\documentclass[prd,aps,showpacs,superscriptaddress,nofootinbib,eqsecnum,amsfonts,amsmath,preprintnumbers,floatfix,floats]{revtex4}

\usepackage{epsfig}
\usepackage{bm}
\usepackage{graphics}
\usepackage{amssymb}
\usepackage{amsmath}
\usepackage{natbib, hyperref}
\usepackage{color}
\definecolor{AliceBlue}{rgb}{0.94,0.97,1.00}
\definecolor{AntiqueWhite1}{rgb}{1.00,0.94,0.86}
\definecolor{AntiqueWhite2}{rgb}{0.93,0.87,0.80}
\definecolor{AntiqueWhite3}{rgb}{0.80,0.75,0.69}
\definecolor{AntiqueWhite4}{rgb}{0.55,0.51,0.47}
\definecolor{AntiqueWhite}{rgb}{0.98,0.92,0.84}
\definecolor{BlanchedAlmond}{rgb}{1.00,0.92,0.80}
\definecolor{BlueViolet}{rgb}{0.54,0.17,0.89}
\definecolor{CadetBlue1}{rgb}{0.60,0.96,1.00}
\definecolor{CadetBlue2}{rgb}{0.56,0.90,0.93}
\definecolor{CadetBlue3}{rgb}{0.48,0.77,0.80}
\definecolor{CadetBlue4}{rgb}{0.33,0.53,0.55}
\definecolor{CadetBlue}{rgb}{0.37,0.62,0.63}
\definecolor{CornflowerBlue}{rgb}{0.39,0.58,0.93}
\definecolor{DarkBlue}{rgb}{0.00,0.00,0.55}
\definecolor{DarkCyan}{rgb}{0.00,0.55,0.55}
\definecolor{DarkGoldenrod1}{rgb}{1.00,0.73,0.06}
\definecolor{DarkGoldenrod2}{rgb}{0.93,0.68,0.05}
\definecolor{DarkGoldenrod3}{rgb}{0.80,0.58,0.05}
\definecolor{DarkGoldenrod4}{rgb}{0.55,0.40,0.03}
\definecolor{DarkGoldenrod}{rgb}{0.72,0.53,0.04}
\definecolor{DarkGray}{rgb}{0.66,0.66,0.66}
\definecolor{DarkGreen}{rgb}{0.00,0.39,0.00}
\definecolor{DarkGrey}{rgb}{0.66,0.66,0.66}
\definecolor{DarkKhaki}{rgb}{0.74,0.72,0.42}
\definecolor{DarkMagenta}{rgb}{0.55,0.00,0.55}
\definecolor{DarkOliveGreen1}{rgb}{0.79,1.00,0.44}
\definecolor{DarkOliveGreen2}{rgb}{0.74,0.93,0.41}
\definecolor{DarkOliveGreen3}{rgb}{0.64,0.80,0.35}
\definecolor{DarkOliveGreen4}{rgb}{0.43,0.55,0.24}
\definecolor{DarkOliveGreen}{rgb}{0.33,0.42,0.18}
\definecolor{DarkOrange1}{rgb}{1.00,0.50,0.00}
\definecolor{DarkOrange2}{rgb}{0.93,0.46,0.00}
\definecolor{DarkOrange3}{rgb}{0.80,0.40,0.00}
\definecolor{DarkOrange4}{rgb}{0.55,0.27,0.00}
\definecolor{DarkOrange}{rgb}{1.00,0.55,0.00}
\definecolor{DarkOrchid1}{rgb}{0.75,0.24,1.00}
\definecolor{DarkOrchid2}{rgb}{0.70,0.23,0.93}
\definecolor{DarkOrchid3}{rgb}{0.60,0.20,0.80}
\definecolor{DarkOrchid4}{rgb}{0.41,0.13,0.55}
\definecolor{DarkOrchid}{rgb}{0.60,0.20,0.80}
\definecolor{DarkRed}{rgb}{0.55,0.00,0.00}
\definecolor{DarkSalmon}{rgb}{0.91,0.59,0.48}
\definecolor{DarkSeaGreen1}{rgb}{0.76,1.00,0.76}
\definecolor{DarkSeaGreen2}{rgb}{0.71,0.93,0.71}
\definecolor{DarkSeaGreen3}{rgb}{0.61,0.80,0.61}
\definecolor{DarkSeaGreen4}{rgb}{0.41,0.55,0.41}
\definecolor{DarkSeaGreen}{rgb}{0.56,0.74,0.56}
\definecolor{DarkSlateBlue}{rgb}{0.28,0.24,0.55}
\definecolor{DarkSlateGray1}{rgb}{0.59,1.00,1.00}
\definecolor{DarkSlateGray2}{rgb}{0.55,0.93,0.93}
\definecolor{DarkSlateGray3}{rgb}{0.47,0.80,0.80}
\definecolor{DarkSlateGray4}{rgb}{0.32,0.55,0.55}
\definecolor{DarkSlateGray}{rgb}{0.18,0.31,0.31}
\definecolor{DarkSlateGrey}{rgb}{0.18,0.31,0.31}
\definecolor{DarkTurquoise}{rgb}{0.00,0.81,0.82}
\definecolor{DarkViolet}{rgb}{0.58,0.00,0.83}
\definecolor{DeepPink1}{rgb}{1.00,0.08,0.58}
\definecolor{DeepPink2}{rgb}{0.93,0.07,0.54}
\definecolor{DeepPink3}{rgb}{0.80,0.06,0.46}
\definecolor{DeepPink4}{rgb}{0.55,0.04,0.31}
\definecolor{DeepPink}{rgb}{1.00,0.08,0.58}
\definecolor{DeepSkyBlue1}{rgb}{0.00,0.75,1.00}
\definecolor{DeepSkyBlue2}{rgb}{0.00,0.70,0.93}
\definecolor{DeepSkyBlue3}{rgb}{0.00,0.60,0.80}
\definecolor{DeepSkyBlue4}{rgb}{0.00,0.41,0.55}
\definecolor{DeepSkyBlue}{rgb}{0.00,0.75,1.00}
\definecolor{DimGray}{rgb}{0.41,0.41,0.41}
\definecolor{DimGrey}{rgb}{0.41,0.41,0.41}
\definecolor{DodgerBlue1}{rgb}{0.12,0.56,1.00}
\definecolor{DodgerBlue2}{rgb}{0.11,0.53,0.93}
\definecolor{DodgerBlue3}{rgb}{0.09,0.45,0.80}
\definecolor{DodgerBlue4}{rgb}{0.06,0.31,0.55}
\definecolor{DodgerBlue}{rgb}{0.12,0.56,1.00}
\definecolor{FloralWhite}{rgb}{1.00,0.98,0.94}
\definecolor{ForestGreen}{rgb}{0.13,0.55,0.13}
\definecolor{GhostWhite}{rgb}{0.97,0.97,1.00}
\definecolor{GreenYellow}{rgb}{0.68,1.00,0.18}
\definecolor{HotPink1}{rgb}{1.00,0.43,0.71}
\definecolor{HotPink2}{rgb}{0.93,0.42,0.65}
\definecolor{HotPink3}{rgb}{0.80,0.38,0.56}
\definecolor{HotPink4}{rgb}{0.55,0.23,0.38}
\definecolor{HotPink}{rgb}{1.00,0.41,0.71}
\definecolor{IndianRed1}{rgb}{1.00,0.42,0.42}
\definecolor{IndianRed2}{rgb}{0.93,0.39,0.39}
\definecolor{IndianRed3}{rgb}{0.80,0.33,0.33}
\definecolor{IndianRed4}{rgb}{0.55,0.23,0.23}
\definecolor{IndianRed}{rgb}{0.80,0.36,0.36}
\definecolor{LavenderBlush1}{rgb}{1.00,0.94,0.96}
\definecolor{LavenderBlush2}{rgb}{0.93,0.88,0.90}
\definecolor{LavenderBlush3}{rgb}{0.80,0.76,0.77}
\definecolor{LavenderBlush4}{rgb}{0.55,0.51,0.53}
\definecolor{LavenderBlush}{rgb}{1.00,0.94,0.96}
\definecolor{LawnGreen}{rgb}{0.49,0.99,0.00}
\definecolor{LemonChiffon1}{rgb}{1.00,0.98,0.80}
\definecolor{LemonChiffon2}{rgb}{0.93,0.91,0.75}
\definecolor{LemonChiffon3}{rgb}{0.80,0.79,0.65}
\definecolor{LemonChiffon4}{rgb}{0.55,0.54,0.44}
\definecolor{LemonChiffon}{rgb}{1.00,0.98,0.80}
\definecolor{LightBlue1}{rgb}{0.75,0.94,1.00}
\definecolor{LightBlue2}{rgb}{0.70,0.87,0.93}
\definecolor{LightBlue3}{rgb}{0.60,0.75,0.80}
\definecolor{LightBlue4}{rgb}{0.41,0.51,0.55}
\definecolor{LightBlue}{rgb}{0.68,0.85,0.90}
\definecolor{LightCoral}{rgb}{0.94,0.50,0.50}
\definecolor{LightCyan1}{rgb}{0.88,1.00,1.00}
\definecolor{LightCyan2}{rgb}{0.82,0.93,0.93}
\definecolor{LightCyan3}{rgb}{0.71,0.80,0.80}
\definecolor{LightCyan4}{rgb}{0.48,0.55,0.55}
\definecolor{LightCyan}{rgb}{0.88,1.00,1.00}
\definecolor{LightGoldenrod1}{rgb}{1.00,0.93,0.55}
\definecolor{LightGoldenrod2}{rgb}{0.93,0.86,0.51}
\definecolor{LightGoldenrod3}{rgb}{0.80,0.75,0.44}
\definecolor{LightGoldenrod4}{rgb}{0.55,0.51,0.30}
\definecolor{LightGoldenrodYellow}{rgb}{0.98,0.98,0.82}
\definecolor{LightGoldenrod}{rgb}{0.93,0.87,0.51}
\definecolor{LightGray}{rgb}{0.83,0.83,0.83}
\definecolor{LightGreen}{rgb}{0.56,0.93,0.56}
\definecolor{LightGrey}{rgb}{0.83,0.83,0.83}
\definecolor{LightPink1}{rgb}{1.00,0.68,0.73}
\definecolor{LightPink2}{rgb}{0.93,0.64,0.68}
\definecolor{LightPink3}{rgb}{0.80,0.55,0.58}
\definecolor{LightPink4}{rgb}{0.55,0.37,0.40}
\definecolor{LightPink}{rgb}{1.00,0.71,0.76}
\definecolor{LightSalmon1}{rgb}{1.00,0.63,0.48}
\definecolor{LightSalmon2}{rgb}{0.93,0.58,0.45}
\definecolor{LightSalmon3}{rgb}{0.80,0.51,0.38}
\definecolor{LightSalmon4}{rgb}{0.55,0.34,0.26}
\definecolor{LightSalmon}{rgb}{1.00,0.63,0.48}
\definecolor{LightSeaGreen}{rgb}{0.13,0.70,0.67}
\definecolor{LightSkyBlue1}{rgb}{0.69,0.89,1.00}
\definecolor{LightSkyBlue2}{rgb}{0.64,0.83,0.93}
\definecolor{LightSkyBlue3}{rgb}{0.55,0.71,0.80}
\definecolor{LightSkyBlue4}{rgb}{0.38,0.48,0.55}
\definecolor{LightSkyBlue}{rgb}{0.53,0.81,0.98}
\definecolor{LightSlateBlue}{rgb}{0.52,0.44,1.00}
\definecolor{LightSlateGray}{rgb}{0.47,0.53,0.60}
\definecolor{LightSlateGrey}{rgb}{0.47,0.53,0.60}
\definecolor{LightSteelBlue1}{rgb}{0.79,0.88,1.00}
\definecolor{LightSteelBlue2}{rgb}{0.74,0.82,0.93}
\definecolor{LightSteelBlue3}{rgb}{0.64,0.71,0.80}
\definecolor{LightSteelBlue4}{rgb}{0.43,0.48,0.55}
\definecolor{LightSteelBlue}{rgb}{0.69,0.77,0.87}
\definecolor{LightYellow1}{rgb}{1.00,1.00,0.88}
\definecolor{LightYellow2}{rgb}{0.93,0.93,0.82}
\definecolor{LightYellow3}{rgb}{0.80,0.80,0.71}
\definecolor{LightYellow4}{rgb}{0.55,0.55,0.48}
\definecolor{LightYellow}{rgb}{1.00,1.00,0.88}
\definecolor{LimeGreen}{rgb}{0.20,0.80,0.20}
\definecolor{MediumAquamarine}{rgb}{0.40,0.80,0.67}
\definecolor{MediumBlue}{rgb}{0.00,0.00,0.80}
\definecolor{MediumOrchid1}{rgb}{0.88,0.40,1.00}
\definecolor{MediumOrchid2}{rgb}{0.82,0.37,0.93}
\definecolor{MediumOrchid3}{rgb}{0.71,0.32,0.80}
\definecolor{MediumOrchid4}{rgb}{0.48,0.22,0.55}
\definecolor{MediumOrchid}{rgb}{0.73,0.33,0.83}
\definecolor{MediumPurple1}{rgb}{0.67,0.51,1.00}
\definecolor{MediumPurple2}{rgb}{0.62,0.47,0.93}
\definecolor{MediumPurple3}{rgb}{0.54,0.41,0.80}
\definecolor{MediumPurple4}{rgb}{0.36,0.28,0.55}
\definecolor{MediumPurple}{rgb}{0.58,0.44,0.86}
\definecolor{MediumSeaGreen}{rgb}{0.24,0.70,0.44}
\definecolor{MediumSlateBlue}{rgb}{0.48,0.41,0.93}
\definecolor{MediumSpringGreen}{rgb}{0.00,0.98,0.60}
\definecolor{MediumTurquoise}{rgb}{0.28,0.82,0.80}
\definecolor{MediumVioletRed}{rgb}{0.78,0.08,0.52}
\definecolor{MidnightBlue}{rgb}{0.10,0.10,0.44}
\definecolor{MintCream}{rgb}{0.96,1.00,0.98}
\definecolor{MistyRose1}{rgb}{1.00,0.89,0.88}
\definecolor{MistyRose2}{rgb}{0.93,0.84,0.82}
\definecolor{MistyRose3}{rgb}{0.80,0.72,0.71}
\definecolor{MistyRose4}{rgb}{0.55,0.49,0.48}
\definecolor{MistyRose}{rgb}{1.00,0.89,0.88}
\definecolor{NavajoWhite1}{rgb}{1.00,0.87,0.68}
\definecolor{NavajoWhite2}{rgb}{0.93,0.81,0.63}
\definecolor{NavajoWhite3}{rgb}{0.80,0.70,0.55}
\definecolor{NavajoWhite4}{rgb}{0.55,0.47,0.37}
\definecolor{NavajoWhite}{rgb}{1.00,0.87,0.68}
\definecolor{NavyBlue}{rgb}{0.00,0.00,0.50}
\definecolor{OldLace}{rgb}{0.99,0.96,0.90}
\definecolor{OliveDrab1}{rgb}{0.75,1.00,0.24}
\definecolor{OliveDrab2}{rgb}{0.70,0.93,0.23}
\definecolor{OliveDrab3}{rgb}{0.60,0.80,0.20}
\definecolor{OliveDrab4}{rgb}{0.41,0.55,0.13}
\definecolor{OliveDrab}{rgb}{0.42,0.56,0.14}
\definecolor{OrangeRed1}{rgb}{1.00,0.27,0.00}
\definecolor{OrangeRed2}{rgb}{0.93,0.25,0.00}
\definecolor{OrangeRed3}{rgb}{0.80,0.22,0.00}
\definecolor{OrangeRed4}{rgb}{0.55,0.15,0.00}
\definecolor{OrangeRed}{rgb}{1.00,0.27,0.00}
\definecolor{PaleGoldenrod}{rgb}{0.93,0.91,0.67}
\definecolor{PaleGreen1}{rgb}{0.60,1.00,0.60}
\definecolor{PaleGreen2}{rgb}{0.56,0.93,0.56}
\definecolor{PaleGreen3}{rgb}{0.49,0.80,0.49}
\definecolor{PaleGreen4}{rgb}{0.33,0.55,0.33}
\definecolor{PaleGreen}{rgb}{0.60,0.98,0.60}
\definecolor{PaleTurquoise1}{rgb}{0.73,1.00,1.00}
\definecolor{PaleTurquoise2}{rgb}{0.68,0.93,0.93}
\definecolor{PaleTurquoise3}{rgb}{0.59,0.80,0.80}
\definecolor{PaleTurquoise4}{rgb}{0.40,0.55,0.55}
\definecolor{PaleTurquoise}{rgb}{0.69,0.93,0.93}
\definecolor{PaleVioletRed1}{rgb}{1.00,0.51,0.67}
\definecolor{PaleVioletRed2}{rgb}{0.93,0.47,0.62}
\definecolor{PaleVioletRed3}{rgb}{0.80,0.41,0.54}
\definecolor{PaleVioletRed4}{rgb}{0.55,0.28,0.36}
\definecolor{PaleVioletRed}{rgb}{0.86,0.44,0.58}
\definecolor{PapayaWhip}{rgb}{1.00,0.94,0.84}
\definecolor{PeachPuff1}{rgb}{1.00,0.85,0.73}
\definecolor{PeachPuff2}{rgb}{0.93,0.80,0.68}
\definecolor{PeachPuff3}{rgb}{0.80,0.69,0.58}
\definecolor{PeachPuff4}{rgb}{0.55,0.47,0.40}
\definecolor{PeachPuff}{rgb}{1.00,0.85,0.73}
\definecolor{PowderBlue}{rgb}{0.69,0.88,0.90}
\definecolor{RosyBrown1}{rgb}{1.00,0.76,0.76}
\definecolor{RosyBrown2}{rgb}{0.93,0.71,0.71}
\definecolor{RosyBrown3}{rgb}{0.80,0.61,0.61}
\definecolor{RosyBrown4}{rgb}{0.55,0.41,0.41}
\definecolor{RosyBrown}{rgb}{0.74,0.56,0.56}
\definecolor{RoyalBlue1}{rgb}{0.28,0.46,1.00}
\definecolor{RoyalBlue2}{rgb}{0.26,0.43,0.93}
\definecolor{RoyalBlue3}{rgb}{0.23,0.37,0.80}
\definecolor{RoyalBlue4}{rgb}{0.15,0.25,0.55}
\definecolor{RoyalBlue}{rgb}{0.25,0.41,0.88}
\definecolor{SaddleBrown}{rgb}{0.55,0.27,0.07}
\definecolor{SandyBrown}{rgb}{0.96,0.64,0.38}
\definecolor{SeaGreen1}{rgb}{0.33,1.00,0.62}
\definecolor{SeaGreen2}{rgb}{0.31,0.93,0.58}
\definecolor{SeaGreen3}{rgb}{0.26,0.80,0.50}
\definecolor{SeaGreen4}{rgb}{0.18,0.55,0.34}
\definecolor{SeaGreen}{rgb}{0.18,0.55,0.34}
\definecolor{SkyBlue1}{rgb}{0.53,0.81,1.00}
\definecolor{SkyBlue2}{rgb}{0.49,0.75,0.93}
\definecolor{SkyBlue3}{rgb}{0.42,0.65,0.80}
\definecolor{SkyBlue4}{rgb}{0.29,0.44,0.55}
\definecolor{SkyBlue}{rgb}{0.53,0.81,0.92}
\definecolor{SlateBlue1}{rgb}{0.51,0.44,1.00}
\definecolor{SlateBlue2}{rgb}{0.48,0.40,0.93}
\definecolor{SlateBlue3}{rgb}{0.41,0.35,0.80}
\definecolor{SlateBlue4}{rgb}{0.28,0.24,0.55}
\definecolor{SlateBlue}{rgb}{0.42,0.35,0.80}
\definecolor{SlateGray1}{rgb}{0.78,0.89,1.00}
\definecolor{SlateGray2}{rgb}{0.73,0.83,0.93}
\definecolor{SlateGray3}{rgb}{0.62,0.71,0.80}
\definecolor{SlateGray4}{rgb}{0.42,0.48,0.55}
\definecolor{SlateGray}{rgb}{0.44,0.50,0.56}
\definecolor{SlateGrey}{rgb}{0.44,0.50,0.56}
\definecolor{SpringGreen1}{rgb}{0.00,1.00,0.50}
\definecolor{SpringGreen2}{rgb}{0.00,0.93,0.46}
\definecolor{SpringGreen3}{rgb}{0.00,0.80,0.40}
\definecolor{SpringGreen4}{rgb}{0.00,0.55,0.27}
\definecolor{SpringGreen}{rgb}{0.00,1.00,0.50}
\definecolor{SteelBlue1}{rgb}{0.39,0.72,1.00}
\definecolor{SteelBlue2}{rgb}{0.36,0.67,0.93}
\definecolor{SteelBlue3}{rgb}{0.31,0.58,0.80}
\definecolor{SteelBlue4}{rgb}{0.21,0.39,0.55}
\definecolor{SteelBlue}{rgb}{0.27,0.51,0.71}
\definecolor{VioletRed1}{rgb}{1.00,0.24,0.59}
\definecolor{VioletRed2}{rgb}{0.93,0.23,0.55}
\definecolor{VioletRed3}{rgb}{0.80,0.20,0.47}
\definecolor{VioletRed4}{rgb}{0.55,0.13,0.32}
\definecolor{VioletRed}{rgb}{0.82,0.13,0.56}
\definecolor{WhiteSmoke}{rgb}{0.96,0.96,0.96}
\definecolor{YellowGreen}{rgb}{0.60,0.80,0.20}
\definecolor{aliceblue}{rgb}{0.94,0.97,1.00}
\definecolor{antiquewhite}{rgb}{0.98,0.92,0.84}
\definecolor{aquamarine1}{rgb}{0.50,1.00,0.83}
\definecolor{aquamarine2}{rgb}{0.46,0.93,0.78}
\definecolor{aquamarine3}{rgb}{0.40,0.80,0.67}
\definecolor{aquamarine4}{rgb}{0.27,0.55,0.45}
\definecolor{aquamarine}{rgb}{0.50,1.00,0.83}
\definecolor{azure1}{rgb}{0.94,1.00,1.00}
\definecolor{azure2}{rgb}{0.88,0.93,0.93}
\definecolor{azure3}{rgb}{0.76,0.80,0.80}
\definecolor{azure4}{rgb}{0.51,0.55,0.55}
\definecolor{azure}{rgb}{0.94,1.00,1.00}
\definecolor{beige}{rgb}{0.96,0.96,0.86}
\definecolor{bisque1}{rgb}{1.00,0.89,0.77}
\definecolor{bisque2}{rgb}{0.93,0.84,0.72}
\definecolor{bisque3}{rgb}{0.80,0.72,0.62}
\definecolor{bisque4}{rgb}{0.55,0.49,0.42}
\definecolor{bisque}{rgb}{1.00,0.89,0.77}
\definecolor{black}{rgb}{0.00,0.00,0.00}
\definecolor{blanchedalmond}{rgb}{1.00,0.92,0.80}
\definecolor{blue1}{rgb}{0.00,0.00,1.00}
\definecolor{blue2}{rgb}{0.00,0.00,0.93}
\definecolor{blue3}{rgb}{0.00,0.00,0.80}
\definecolor{blue4}{rgb}{0.00,0.00,0.55}
\definecolor{blueviolet}{rgb}{0.54,0.17,0.89}
\definecolor{blue}{rgb}{0.00,0.00,1.00}
\definecolor{brown1}{rgb}{1.00,0.25,0.25}
\definecolor{brown2}{rgb}{0.93,0.23,0.23}
\definecolor{brown3}{rgb}{0.80,0.20,0.20}
\definecolor{brown4}{rgb}{0.55,0.14,0.14}
\definecolor{brown}{rgb}{0.65,0.16,0.16}
\definecolor{burlywood1}{rgb}{1.00,0.83,0.61}
\definecolor{burlywood2}{rgb}{0.93,0.77,0.57}
\definecolor{burlywood3}{rgb}{0.80,0.67,0.49}
\definecolor{burlywood4}{rgb}{0.55,0.45,0.33}
\definecolor{burlywood}{rgb}{0.87,0.72,0.53}
\definecolor{cadetblue}{rgb}{0.37,0.62,0.63}
\definecolor{chartreuse1}{rgb}{0.50,1.00,0.00}
\definecolor{chartreuse2}{rgb}{0.46,0.93,0.00}
\definecolor{chartreuse3}{rgb}{0.40,0.80,0.00}
\definecolor{chartreuse4}{rgb}{0.27,0.55,0.00}
\definecolor{chartreuse}{rgb}{0.50,1.00,0.00}
\definecolor{chocolate1}{rgb}{1.00,0.50,0.14}
\definecolor{chocolate2}{rgb}{0.93,0.46,0.13}
\definecolor{chocolate3}{rgb}{0.80,0.40,0.11}
\definecolor{chocolate4}{rgb}{0.55,0.27,0.07}
\definecolor{chocolate}{rgb}{0.82,0.41,0.12}
\definecolor{coral1}{rgb}{1.00,0.45,0.34}
\definecolor{coral2}{rgb}{0.93,0.42,0.31}
\definecolor{coral3}{rgb}{0.80,0.36,0.27}
\definecolor{coral4}{rgb}{0.55,0.24,0.18}
\definecolor{coral}{rgb}{1.00,0.50,0.31}
\definecolor{cornflowerblue}{rgb}{0.39,0.58,0.93}
\definecolor{cornsilk1}{rgb}{1.00,0.97,0.86}
\definecolor{cornsilk2}{rgb}{0.93,0.91,0.80}
\definecolor{cornsilk3}{rgb}{0.80,0.78,0.69}
\definecolor{cornsilk4}{rgb}{0.55,0.53,0.47}
\definecolor{cornsilk}{rgb}{1.00,0.97,0.86}
\definecolor{cyan1}{rgb}{0.00,1.00,1.00}
\definecolor{cyan2}{rgb}{0.00,0.93,0.93}
\definecolor{cyan3}{rgb}{0.00,0.80,0.80}
\definecolor{cyan4}{rgb}{0.00,0.55,0.55}
\definecolor{cyan}{rgb}{0.00,1.00,1.00}
\definecolor{darkblue}{rgb}{0.00,0.00,0.55}
\definecolor{darkcyan}{rgb}{0.00,0.55,0.55}
\definecolor{darkgoldenrod}{rgb}{0.72,0.53,0.04}
\definecolor{darkgray}{rgb}{0.66,0.66,0.66}
\definecolor{darkgreen}{rgb}{0.00,0.39,0.00}
\definecolor{darkgrey}{rgb}{0.66,0.66,0.66}
\definecolor{darkkhaki}{rgb}{0.74,0.72,0.42}
\definecolor{darkmagenta}{rgb}{0.55,0.00,0.55}
\definecolor{darkolive}{rgb}{0.33,0.42,0.18}
\definecolor{darkorange}{rgb}{1.00,0.55,0.00}
\definecolor{darkorchid}{rgb}{0.60,0.20,0.80}
\definecolor{darkred}{rgb}{0.55,0.00,0.00}
\definecolor{darksalmon}{rgb}{0.91,0.59,0.48}
\definecolor{darksea}{rgb}{0.56,0.74,0.56}
\definecolor{darkslate}{rgb}{0.18,0.31,0.31}
\definecolor{darkslate}{rgb}{0.18,0.31,0.31}
\definecolor{darkslate}{rgb}{0.28,0.24,0.55}
\definecolor{darkturquoise}{rgb}{0.00,0.81,0.82}
\definecolor{darkviolet}{rgb}{0.58,0.00,0.83}
\definecolor{deeppink}{rgb}{1.00,0.08,0.58}
\definecolor{deepsky}{rgb}{0.00,0.75,1.00}
\definecolor{dimgray}{rgb}{0.41,0.41,0.41}
\definecolor{dimgrey}{rgb}{0.41,0.41,0.41}
\definecolor{dodgerblue}{rgb}{0.12,0.56,1.00}
\definecolor{firebrick1}{rgb}{1.00,0.19,0.19}
\definecolor{firebrick2}{rgb}{0.93,0.17,0.17}
\definecolor{firebrick3}{rgb}{0.80,0.15,0.15}
\definecolor{firebrick4}{rgb}{0.55,0.10,0.10}
\definecolor{firebrick}{rgb}{0.70,0.13,0.13}
\definecolor{floralwhite}{rgb}{1.00,0.98,0.94}
\definecolor{forestgreen}{rgb}{0.13,0.55,0.13}
\definecolor{gainsboro}{rgb}{0.86,0.86,0.86}
\definecolor{ghostwhite}{rgb}{0.97,0.97,1.00}
\definecolor{gold1}{rgb}{1.00,0.84,0.00}
\definecolor{gold2}{rgb}{0.93,0.79,0.00}
\definecolor{gold3}{rgb}{0.80,0.68,0.00}
\definecolor{gold4}{rgb}{0.55,0.46,0.00}
\definecolor{goldenrod1}{rgb}{1.00,0.76,0.15}
\definecolor{goldenrod2}{rgb}{0.93,0.71,0.13}
\definecolor{goldenrod3}{rgb}{0.80,0.61,0.11}
\definecolor{goldenrod4}{rgb}{0.55,0.41,0.08}
\definecolor{goldenrod}{rgb}{0.85,0.65,0.13}
\definecolor{gold}{rgb}{1.00,0.84,0.00}
\definecolor{gray0}{rgb}{0.00,0.00,0.00}
\definecolor{gray100}{rgb}{1.00,1.00,1.00}
\definecolor{gray10}{rgb}{0.10,0.10,0.10}
\definecolor{gray11}{rgb}{0.11,0.11,0.11}
\definecolor{gray12}{rgb}{0.12,0.12,0.12}
\definecolor{gray13}{rgb}{0.13,0.13,0.13}
\definecolor{gray14}{rgb}{0.14,0.14,0.14}
\definecolor{gray15}{rgb}{0.15,0.15,0.15}
\definecolor{gray16}{rgb}{0.16,0.16,0.16}
\definecolor{gray17}{rgb}{0.17,0.17,0.17}
\definecolor{gray18}{rgb}{0.18,0.18,0.18}
\definecolor{gray19}{rgb}{0.19,0.19,0.19}
\definecolor{gray1}{rgb}{0.01,0.01,0.01}
\definecolor{gray20}{rgb}{0.20,0.20,0.20}
\definecolor{gray21}{rgb}{0.21,0.21,0.21}
\definecolor{gray22}{rgb}{0.22,0.22,0.22}
\definecolor{gray23}{rgb}{0.23,0.23,0.23}
\definecolor{gray24}{rgb}{0.24,0.24,0.24}
\definecolor{gray25}{rgb}{0.25,0.25,0.25}
\definecolor{gray26}{rgb}{0.26,0.26,0.26}
\definecolor{gray27}{rgb}{0.27,0.27,0.27}
\definecolor{gray28}{rgb}{0.28,0.28,0.28}
\definecolor{gray29}{rgb}{0.29,0.29,0.29}
\definecolor{gray2}{rgb}{0.02,0.02,0.02}
\definecolor{gray30}{rgb}{0.30,0.30,0.30}
\definecolor{gray31}{rgb}{0.31,0.31,0.31}
\definecolor{gray32}{rgb}{0.32,0.32,0.32}
\definecolor{gray33}{rgb}{0.33,0.33,0.33}
\definecolor{gray34}{rgb}{0.34,0.34,0.34}
\definecolor{gray35}{rgb}{0.35,0.35,0.35}
\definecolor{gray36}{rgb}{0.36,0.36,0.36}
\definecolor{gray37}{rgb}{0.37,0.37,0.37}
\definecolor{gray38}{rgb}{0.38,0.38,0.38}
\definecolor{gray39}{rgb}{0.39,0.39,0.39}
\definecolor{gray3}{rgb}{0.03,0.03,0.03}
\definecolor{gray40}{rgb}{0.40,0.40,0.40}
\definecolor{gray41}{rgb}{0.41,0.41,0.41}
\definecolor{gray42}{rgb}{0.42,0.42,0.42}
\definecolor{gray43}{rgb}{0.43,0.43,0.43}
\definecolor{gray44}{rgb}{0.44,0.44,0.44}
\definecolor{gray45}{rgb}{0.45,0.45,0.45}
\definecolor{gray46}{rgb}{0.46,0.46,0.46}
\definecolor{gray47}{rgb}{0.47,0.47,0.47}
\definecolor{gray48}{rgb}{0.48,0.48,0.48}
\definecolor{gray49}{rgb}{0.49,0.49,0.49}
\definecolor{gray4}{rgb}{0.04,0.04,0.04}
\definecolor{gray50}{rgb}{0.50,0.50,0.50}
\definecolor{gray51}{rgb}{0.51,0.51,0.51}
\definecolor{gray52}{rgb}{0.52,0.52,0.52}
\definecolor{gray53}{rgb}{0.53,0.53,0.53}
\definecolor{gray54}{rgb}{0.54,0.54,0.54}
\definecolor{gray55}{rgb}{0.55,0.55,0.55}
\definecolor{gray56}{rgb}{0.56,0.56,0.56}
\definecolor{gray57}{rgb}{0.57,0.57,0.57}
\definecolor{gray58}{rgb}{0.58,0.58,0.58}
\definecolor{gray59}{rgb}{0.59,0.59,0.59}
\definecolor{gray5}{rgb}{0.05,0.05,0.05}
\definecolor{gray60}{rgb}{0.60,0.60,0.60}
\definecolor{gray61}{rgb}{0.61,0.61,0.61}
\definecolor{gray62}{rgb}{0.62,0.62,0.62}
\definecolor{gray63}{rgb}{0.63,0.63,0.63}
\definecolor{gray64}{rgb}{0.64,0.64,0.64}
\definecolor{gray65}{rgb}{0.65,0.65,0.65}
\definecolor{gray66}{rgb}{0.66,0.66,0.66}
\definecolor{gray67}{rgb}{0.67,0.67,0.67}
\definecolor{gray68}{rgb}{0.68,0.68,0.68}
\definecolor{gray69}{rgb}{0.69,0.69,0.69}
\definecolor{gray6}{rgb}{0.06,0.06,0.06}
\definecolor{gray70}{rgb}{0.70,0.70,0.70}
\definecolor{gray71}{rgb}{0.71,0.71,0.71}
\definecolor{gray72}{rgb}{0.72,0.72,0.72}
\definecolor{gray73}{rgb}{0.73,0.73,0.73}
\definecolor{gray74}{rgb}{0.74,0.74,0.74}
\definecolor{gray75}{rgb}{0.75,0.75,0.75}
\definecolor{gray76}{rgb}{0.76,0.76,0.76}
\definecolor{gray77}{rgb}{0.77,0.77,0.77}
\definecolor{gray78}{rgb}{0.78,0.78,0.78}
\definecolor{gray79}{rgb}{0.79,0.79,0.79}
\definecolor{gray7}{rgb}{0.07,0.07,0.07}
\definecolor{gray80}{rgb}{0.80,0.80,0.80}
\definecolor{gray81}{rgb}{0.81,0.81,0.81}
\definecolor{gray82}{rgb}{0.82,0.82,0.82}
\definecolor{gray83}{rgb}{0.83,0.83,0.83}
\definecolor{gray84}{rgb}{0.84,0.84,0.84}
\definecolor{gray85}{rgb}{0.85,0.85,0.85}
\definecolor{gray86}{rgb}{0.86,0.86,0.86}
\definecolor{gray87}{rgb}{0.87,0.87,0.87}
\definecolor{gray88}{rgb}{0.88,0.88,0.88}
\definecolor{gray89}{rgb}{0.89,0.89,0.89}
\definecolor{gray8}{rgb}{0.08,0.08,0.08}
\definecolor{gray90}{rgb}{0.90,0.90,0.90}
\definecolor{gray91}{rgb}{0.91,0.91,0.91}
\definecolor{gray92}{rgb}{0.92,0.92,0.92}
\definecolor{gray93}{rgb}{0.93,0.93,0.93}
\definecolor{gray94}{rgb}{0.94,0.94,0.94}
\definecolor{gray95}{rgb}{0.95,0.95,0.95}
\definecolor{gray96}{rgb}{0.96,0.96,0.96}
\definecolor{gray97}{rgb}{0.97,0.97,0.97}
\definecolor{gray98}{rgb}{0.98,0.98,0.98}
\definecolor{gray99}{rgb}{0.99,0.99,0.99}
\definecolor{gray9}{rgb}{0.09,0.09,0.09}
\definecolor{gray}{rgb}{0.75,0.75,0.75}
\definecolor{green1}{rgb}{0.00,1.00,0.00}
\definecolor{green2}{rgb}{0.00,0.93,0.00}
\definecolor{green3}{rgb}{0.00,0.80,0.00}
\definecolor{green4}{rgb}{0.00,0.55,0.00}
\definecolor{greenyellow}{rgb}{0.68,1.00,0.18}
\definecolor{green}{rgb}{0.00,1.00,0.00}
\definecolor{grey0}{rgb}{0.00,0.00,0.00}
\definecolor{grey100}{rgb}{1.00,1.00,1.00}
\definecolor{grey10}{rgb}{0.10,0.10,0.10}
\definecolor{grey11}{rgb}{0.11,0.11,0.11}
\definecolor{grey12}{rgb}{0.12,0.12,0.12}
\definecolor{grey13}{rgb}{0.13,0.13,0.13}
\definecolor{grey14}{rgb}{0.14,0.14,0.14}
\definecolor{grey15}{rgb}{0.15,0.15,0.15}
\definecolor{grey16}{rgb}{0.16,0.16,0.16}
\definecolor{grey17}{rgb}{0.17,0.17,0.17}
\definecolor{grey18}{rgb}{0.18,0.18,0.18}
\definecolor{grey19}{rgb}{0.19,0.19,0.19}
\definecolor{grey1}{rgb}{0.01,0.01,0.01}
\definecolor{grey20}{rgb}{0.20,0.20,0.20}
\definecolor{grey21}{rgb}{0.21,0.21,0.21}
\definecolor{grey22}{rgb}{0.22,0.22,0.22}
\definecolor{grey23}{rgb}{0.23,0.23,0.23}
\definecolor{grey24}{rgb}{0.24,0.24,0.24}
\definecolor{grey25}{rgb}{0.25,0.25,0.25}
\definecolor{grey26}{rgb}{0.26,0.26,0.26}
\definecolor{grey27}{rgb}{0.27,0.27,0.27}
\definecolor{grey28}{rgb}{0.28,0.28,0.28}
\definecolor{grey29}{rgb}{0.29,0.29,0.29}
\definecolor{grey2}{rgb}{0.02,0.02,0.02}
\definecolor{grey30}{rgb}{0.30,0.30,0.30}
\definecolor{grey31}{rgb}{0.31,0.31,0.31}
\definecolor{grey32}{rgb}{0.32,0.32,0.32}
\definecolor{grey33}{rgb}{0.33,0.33,0.33}
\definecolor{grey34}{rgb}{0.34,0.34,0.34}
\definecolor{grey35}{rgb}{0.35,0.35,0.35}
\definecolor{grey36}{rgb}{0.36,0.36,0.36}
\definecolor{grey37}{rgb}{0.37,0.37,0.37}
\definecolor{grey38}{rgb}{0.38,0.38,0.38}
\definecolor{grey39}{rgb}{0.39,0.39,0.39}
\definecolor{grey3}{rgb}{0.03,0.03,0.03}
\definecolor{grey40}{rgb}{0.40,0.40,0.40}
\definecolor{grey41}{rgb}{0.41,0.41,0.41}
\definecolor{grey42}{rgb}{0.42,0.42,0.42}
\definecolor{grey43}{rgb}{0.43,0.43,0.43}
\definecolor{grey44}{rgb}{0.44,0.44,0.44}
\definecolor{grey45}{rgb}{0.45,0.45,0.45}
\definecolor{grey46}{rgb}{0.46,0.46,0.46}
\definecolor{grey47}{rgb}{0.47,0.47,0.47}
\definecolor{grey48}{rgb}{0.48,0.48,0.48}
\definecolor{grey49}{rgb}{0.49,0.49,0.49}
\definecolor{grey4}{rgb}{0.04,0.04,0.04}
\definecolor{grey50}{rgb}{0.50,0.50,0.50}
\definecolor{grey51}{rgb}{0.51,0.51,0.51}
\definecolor{grey52}{rgb}{0.52,0.52,0.52}
\definecolor{grey53}{rgb}{0.53,0.53,0.53}
\definecolor{grey54}{rgb}{0.54,0.54,0.54}
\definecolor{grey55}{rgb}{0.55,0.55,0.55}
\definecolor{grey56}{rgb}{0.56,0.56,0.56}
\definecolor{grey57}{rgb}{0.57,0.57,0.57}
\definecolor{grey58}{rgb}{0.58,0.58,0.58}
\definecolor{grey59}{rgb}{0.59,0.59,0.59}
\definecolor{grey5}{rgb}{0.05,0.05,0.05}
\definecolor{grey60}{rgb}{0.60,0.60,0.60}
\definecolor{grey61}{rgb}{0.61,0.61,0.61}
\definecolor{grey62}{rgb}{0.62,0.62,0.62}
\definecolor{grey63}{rgb}{0.63,0.63,0.63}
\definecolor{grey64}{rgb}{0.64,0.64,0.64}
\definecolor{grey65}{rgb}{0.65,0.65,0.65}
\definecolor{grey66}{rgb}{0.66,0.66,0.66}
\definecolor{grey67}{rgb}{0.67,0.67,0.67}
\definecolor{grey68}{rgb}{0.68,0.68,0.68}
\definecolor{grey69}{rgb}{0.69,0.69,0.69}
\definecolor{grey6}{rgb}{0.06,0.06,0.06}
\definecolor{grey70}{rgb}{0.70,0.70,0.70}
\definecolor{grey71}{rgb}{0.71,0.71,0.71}
\definecolor{grey72}{rgb}{0.72,0.72,0.72}
\definecolor{grey73}{rgb}{0.73,0.73,0.73}
\definecolor{grey74}{rgb}{0.74,0.74,0.74}
\definecolor{grey75}{rgb}{0.75,0.75,0.75}
\definecolor{grey76}{rgb}{0.76,0.76,0.76}
\definecolor{grey77}{rgb}{0.77,0.77,0.77}
\definecolor{grey78}{rgb}{0.78,0.78,0.78}
\definecolor{grey79}{rgb}{0.79,0.79,0.79}
\definecolor{grey7}{rgb}{0.07,0.07,0.07}
\definecolor{grey80}{rgb}{0.80,0.80,0.80}
\definecolor{grey81}{rgb}{0.81,0.81,0.81}
\definecolor{grey82}{rgb}{0.82,0.82,0.82}
\definecolor{grey83}{rgb}{0.83,0.83,0.83}
\definecolor{grey84}{rgb}{0.84,0.84,0.84}
\definecolor{grey85}{rgb}{0.85,0.85,0.85}
\definecolor{grey86}{rgb}{0.86,0.86,0.86}
\definecolor{grey87}{rgb}{0.87,0.87,0.87}
\definecolor{grey88}{rgb}{0.88,0.88,0.88}
\definecolor{grey89}{rgb}{0.89,0.89,0.89}
\definecolor{grey8}{rgb}{0.08,0.08,0.08}
\definecolor{grey90}{rgb}{0.90,0.90,0.90}
\definecolor{grey91}{rgb}{0.91,0.91,0.91}
\definecolor{grey92}{rgb}{0.92,0.92,0.92}
\definecolor{grey93}{rgb}{0.93,0.93,0.93}
\definecolor{grey94}{rgb}{0.94,0.94,0.94}
\definecolor{grey95}{rgb}{0.95,0.95,0.95}
\definecolor{grey96}{rgb}{0.96,0.96,0.96}
\definecolor{grey97}{rgb}{0.97,0.97,0.97}
\definecolor{grey98}{rgb}{0.98,0.98,0.98}
\definecolor{grey99}{rgb}{0.99,0.99,0.99}
\definecolor{grey9}{rgb}{0.09,0.09,0.09}
\definecolor{grey}{rgb}{0.75,0.75,0.75}
\definecolor{honeydew1}{rgb}{0.94,1.00,0.94}
\definecolor{honeydew2}{rgb}{0.88,0.93,0.88}
\definecolor{honeydew3}{rgb}{0.76,0.80,0.76}
\definecolor{honeydew4}{rgb}{0.51,0.55,0.51}
\definecolor{honeydew}{rgb}{0.94,1.00,0.94}
\definecolor{hotpink}{rgb}{1.00,0.41,0.71}
\definecolor{indianred}{rgb}{0.80,0.36,0.36}
\definecolor{ivory1}{rgb}{1.00,1.00,0.94}
\definecolor{ivory2}{rgb}{0.93,0.93,0.88}
\definecolor{ivory3}{rgb}{0.80,0.80,0.76}
\definecolor{ivory4}{rgb}{0.55,0.55,0.51}
\definecolor{ivory}{rgb}{1.00,1.00,0.94}
\definecolor{khaki1}{rgb}{1.00,0.96,0.56}
\definecolor{khaki2}{rgb}{0.93,0.90,0.52}
\definecolor{khaki3}{rgb}{0.80,0.78,0.45}
\definecolor{khaki4}{rgb}{0.55,0.53,0.31}
\definecolor{khaki}{rgb}{0.94,0.90,0.55}
\definecolor{lavenderblush}{rgb}{1.00,0.94,0.96}
\definecolor{lavender}{rgb}{0.90,0.90,0.98}
\definecolor{lawngreen}{rgb}{0.49,0.99,0.00}
\definecolor{lemonchiffon}{rgb}{1.00,0.98,0.80}
\definecolor{lightblue}{rgb}{0.68,0.85,0.90}
\definecolor{lightcoral}{rgb}{0.94,0.50,0.50}
\definecolor{lightcyan}{rgb}{0.88,1.00,1.00}
\definecolor{lightgoldenrod}{rgb}{0.93,0.87,0.51}
\definecolor{lightgoldenrod}{rgb}{0.98,0.98,0.82}
\definecolor{lightgray}{rgb}{0.83,0.83,0.83}
\definecolor{lightgreen}{rgb}{0.56,0.93,0.56}
\definecolor{lightgrey}{rgb}{0.83,0.83,0.83}
\definecolor{lightpink}{rgb}{1.00,0.71,0.76}
\definecolor{lightsalmon}{rgb}{1.00,0.63,0.48}
\definecolor{lightsea}{rgb}{0.13,0.70,0.67}
\definecolor{lightsky}{rgb}{0.53,0.81,0.98}
\definecolor{lightslate}{rgb}{0.47,0.53,0.60}
\definecolor{lightslate}{rgb}{0.47,0.53,0.60}
\definecolor{lightslate}{rgb}{0.52,0.44,1.00}
\definecolor{lightsteel}{rgb}{0.69,0.77,0.87}
\definecolor{lightyellow}{rgb}{1.00,1.00,0.88}
\definecolor{limegreen}{rgb}{0.20,0.80,0.20}
\definecolor{linen}{rgb}{0.98,0.94,0.90}
\definecolor{magenta1}{rgb}{1.00,0.00,1.00}
\definecolor{magenta2}{rgb}{0.93,0.00,0.93}
\definecolor{magenta3}{rgb}{0.80,0.00,0.80}
\definecolor{magenta4}{rgb}{0.55,0.00,0.55}
\definecolor{magenta}{rgb}{1.00,0.00,1.00}
\definecolor{maroon1}{rgb}{1.00,0.20,0.70}
\definecolor{maroon2}{rgb}{0.93,0.19,0.65}
\definecolor{maroon3}{rgb}{0.80,0.16,0.56}
\definecolor{maroon4}{rgb}{0.55,0.11,0.38}
\definecolor{maroon}{rgb}{0.69,0.19,0.38}
\definecolor{mediumaquamarine}{rgb}{0.40,0.80,0.67}
\definecolor{mediumblue}{rgb}{0.00,0.00,0.80}
\definecolor{mediumorchid}{rgb}{0.73,0.33,0.83}
\definecolor{mediumpurple}{rgb}{0.58,0.44,0.86}
\definecolor{mediumsea}{rgb}{0.24,0.70,0.44}
\definecolor{mediumslate}{rgb}{0.48,0.41,0.93}
\definecolor{mediumspring}{rgb}{0.00,0.98,0.60}
\definecolor{mediumturquoise}{rgb}{0.28,0.82,0.80}
\definecolor{mediumviolet}{rgb}{0.78,0.08,0.52}
\definecolor{midnightblue}{rgb}{0.10,0.10,0.44}
\definecolor{mintcream}{rgb}{0.96,1.00,0.98}
\definecolor{mistyrose}{rgb}{1.00,0.89,0.88}
\definecolor{moccasin}{rgb}{1.00,0.89,0.71}
\definecolor{navajowhite}{rgb}{1.00,0.87,0.68}
\definecolor{navyblue}{rgb}{0.00,0.00,0.50}
\definecolor{navy}{rgb}{0.00,0.00,0.50}
\definecolor{oldlace}{rgb}{0.99,0.96,0.90}
\definecolor{olivedrab}{rgb}{0.42,0.56,0.14}
\definecolor{orange1}{rgb}{1.00,0.65,0.00}
\definecolor{orange2}{rgb}{0.93,0.60,0.00}
\definecolor{orange3}{rgb}{0.80,0.52,0.00}
\definecolor{orange4}{rgb}{0.55,0.35,0.00}
\definecolor{orangered}{rgb}{1.00,0.27,0.00}
\definecolor{orange}{rgb}{1.00,0.65,0.00}
\definecolor{orchid1}{rgb}{1.00,0.51,0.98}
\definecolor{orchid2}{rgb}{0.93,0.48,0.91}
\definecolor{orchid3}{rgb}{0.80,0.41,0.79}
\definecolor{orchid4}{rgb}{0.55,0.28,0.54}
\definecolor{orchid}{rgb}{0.85,0.44,0.84}
\definecolor{palegoldenrod}{rgb}{0.93,0.91,0.67}
\definecolor{palegreen}{rgb}{0.60,0.98,0.60}
\definecolor{paleturquoise}{rgb}{0.69,0.93,0.93}
\definecolor{paleviolet}{rgb}{0.86,0.44,0.58}
\definecolor{papayawhip}{rgb}{1.00,0.94,0.84}
\definecolor{peachpuff}{rgb}{1.00,0.85,0.73}
\definecolor{peru}{rgb}{0.80,0.52,0.25}
\definecolor{pink1}{rgb}{1.00,0.71,0.77}
\definecolor{pink2}{rgb}{0.93,0.66,0.72}
\definecolor{pink3}{rgb}{0.80,0.57,0.62}
\definecolor{pink4}{rgb}{0.55,0.39,0.42}
\definecolor{pink}{rgb}{1.00,0.75,0.80}
\definecolor{plum1}{rgb}{1.00,0.73,1.00}
\definecolor{plum2}{rgb}{0.93,0.68,0.93}
\definecolor{plum3}{rgb}{0.80,0.59,0.80}
\definecolor{plum4}{rgb}{0.55,0.40,0.55}
\definecolor{plum}{rgb}{0.87,0.63,0.87}
\definecolor{powderblue}{rgb}{0.69,0.88,0.90}
\definecolor{purple1}{rgb}{0.61,0.19,1.00}
\definecolor{purple2}{rgb}{0.57,0.17,0.93}
\definecolor{purple3}{rgb}{0.49,0.15,0.80}
\definecolor{purple4}{rgb}{0.33,0.10,0.55}
\definecolor{purple}{rgb}{0.63,0.13,0.94}
\definecolor{red1}{rgb}{1.00,0.00,0.00}
\definecolor{red2}{rgb}{0.93,0.00,0.00}
\definecolor{red3}{rgb}{0.80,0.00,0.00}
\definecolor{red4}{rgb}{0.55,0.00,0.00}
\definecolor{red}{rgb}{1.00,0.00,0.00}
\definecolor{rosybrown}{rgb}{0.74,0.56,0.56}
\definecolor{royalblue}{rgb}{0.25,0.41,0.88}
\definecolor{saddlebrown}{rgb}{0.55,0.27,0.07}
\definecolor{salmon1}{rgb}{1.00,0.55,0.41}
\definecolor{salmon2}{rgb}{0.93,0.51,0.38}
\definecolor{salmon3}{rgb}{0.80,0.44,0.33}
\definecolor{salmon4}{rgb}{0.55,0.30,0.22}
\definecolor{salmon}{rgb}{0.98,0.50,0.45}
\definecolor{sandybrown}{rgb}{0.96,0.64,0.38}
\definecolor{seagreen}{rgb}{0.18,0.55,0.34}
\definecolor{seashell1}{rgb}{1.00,0.96,0.93}
\definecolor{seashell2}{rgb}{0.93,0.90,0.87}
\definecolor{seashell3}{rgb}{0.80,0.77,0.75}
\definecolor{seashell4}{rgb}{0.55,0.53,0.51}
\definecolor{seashell}{rgb}{1.00,0.96,0.93}
\definecolor{sienna1}{rgb}{1.00,0.51,0.28}
\definecolor{sienna2}{rgb}{0.93,0.47,0.26}
\definecolor{sienna3}{rgb}{0.80,0.41,0.22}
\definecolor{sienna4}{rgb}{0.55,0.28,0.15}
\definecolor{sienna}{rgb}{0.63,0.32,0.18}
\definecolor{skyblue}{rgb}{0.53,0.81,0.92}
\definecolor{slateblue}{rgb}{0.42,0.35,0.80}
\definecolor{slategray}{rgb}{0.44,0.50,0.56}
\definecolor{slategrey}{rgb}{0.44,0.50,0.56}
\definecolor{snow1}{rgb}{1.00,0.98,0.98}
\definecolor{snow2}{rgb}{0.93,0.91,0.91}
\definecolor{snow3}{rgb}{0.80,0.79,0.79}
\definecolor{snow4}{rgb}{0.55,0.54,0.54}
\definecolor{snow}{rgb}{1.00,0.98,0.98}
\definecolor{springgreen}{rgb}{0.00,1.00,0.50}
\definecolor{steelblue}{rgb}{0.27,0.51,0.71}
\definecolor{tan1}{rgb}{1.00,0.65,0.31}
\definecolor{tan2}{rgb}{0.93,0.60,0.29}
\definecolor{tan3}{rgb}{0.80,0.52,0.25}
\definecolor{tan4}{rgb}{0.55,0.35,0.17}
\definecolor{tan}{rgb}{0.82,0.71,0.55}
\definecolor{thistle1}{rgb}{1.00,0.88,1.00}
\definecolor{thistle2}{rgb}{0.93,0.82,0.93}
\definecolor{thistle3}{rgb}{0.80,0.71,0.80}
\definecolor{thistle4}{rgb}{0.55,0.48,0.55}
\definecolor{thistle}{rgb}{0.85,0.75,0.85}
\definecolor{tomato1}{rgb}{1.00,0.39,0.28}
\definecolor{tomato2}{rgb}{0.93,0.36,0.26}
\definecolor{tomato3}{rgb}{0.80,0.31,0.22}
\definecolor{tomato4}{rgb}{0.55,0.21,0.15}
\definecolor{tomato}{rgb}{1.00,0.39,0.28}
\definecolor{turquoise1}{rgb}{0.00,0.96,1.00}
\definecolor{turquoise2}{rgb}{0.00,0.90,0.93}
\definecolor{turquoise3}{rgb}{0.00,0.77,0.80}
\definecolor{turquoise4}{rgb}{0.00,0.53,0.55}
\definecolor{turquoise}{rgb}{0.25,0.88,0.82}
\definecolor{violetred}{rgb}{0.82,0.13,0.56}
\definecolor{violet}{rgb}{0.93,0.51,0.93}
\definecolor{wheat1}{rgb}{1.00,0.91,0.73}
\definecolor{wheat2}{rgb}{0.93,0.85,0.68}
\definecolor{wheat3}{rgb}{0.80,0.73,0.59}
\definecolor{wheat4}{rgb}{0.55,0.49,0.40}
\definecolor{wheat}{rgb}{0.96,0.87,0.70}
\definecolor{whitesmoke}{rgb}{0.96,0.96,0.96}
\definecolor{white}{rgb}{1.00,1.00,1.00}
\definecolor{yellow1}{rgb}{1.00,1.00,0.00}
\definecolor{yellow2}{rgb}{0.93,0.93,0.00}
\definecolor{yellow3}{rgb}{0.80,0.80,0.00}
\definecolor{yellow4}{rgb}{0.55,0.55,0.00}
\definecolor{yellowgreen}{rgb}{0.60,0.80,0.20}
\definecolor{yellow}{rgb}{1.00,1.00,0.00}

\def\bea{\begin{eqnarray}}
\def\eea{\end{eqnarray}}
\def\vt{\vartheta}

\begin{document}

\newcommand{\rhat}{\hat{r}}
\newcommand{\iotahat}{\hat{\iota}}
\newcommand{\phihat}{\hat{\phi}}
\newcommand{\h}{\mathfrak{h}}
\newcommand{\be}{\begin{equation}}
\newcommand{\ee}{\end{equation}}
\newcommand{\ber}{\begin{eqnarray}}
\newcommand{\eer}{\end{eqnarray}}
\newcommand{\fmerg}{f_{\rm merg}}
\newcommand{\fcut}{f_{\rm cut}}
\newcommand{\fring}{f_{\rm ring}}
\newcommand{\cA}{\mathcal{A}}
\newcommand{\ie}{i.e.}
\newcommand{\df}{{\mathrm{d}f}}
\newcommand{\rmi}{\mathrm{i}}
\newcommand{\rmd}{\mathrm{d}}
\newcommand{\rme}{\mathrm{e}}
\newcommand{\dt}{{\mathrm{d}t}}
\newcommand{\pj}{\partial_j}
\newcommand{\pk}{\partial_k}
\newcommand{\psifl}{\Psi(f; {\bm \lambda})}
\newcommand{\hp}{h_+(t)}
\newcommand{\hc}{h_\times(t)}
\newcommand{\Fp}{F_+}
\newcommand{\Fc}{F_\times}
\newcommand{\Ylm}{Y_{\ell m}^{-2}}
\def\no{\nonumber \\ & \quad}
\def\noQ{\nonumber \\}
\newcommand{\mc}{M_c}
\newcommand{\vek}[1]{\boldsymbol{#1}}
\newcommand{\vdag}{(v)^\dagger}
\newcommand{\bvtheta}{{\bm \vartheta}}
\newcommand{\btheta}{{\bm \theta}}
\newcommand{\brho}{{\bm \rho}}
\newcommand{\pa}{\partial_a}
\newcommand{\pb}{\partial_b}
\newcommand{\Psieff}{\Psi_{\rm eff}}
\newcommand{\Aeff}{A_{\rm eff}}
\newcommand{\deff}{d_{\rm eff}}
\newcommand{\corr}{\mathcal{C}}
\newcommand{\bvthat}{\hat{\mbox{\boldmath $\vt$}}}
\newcommand{\bvt}{\mbox{\boldmath $\vt$}}

\newcommand{\comment}[1]{{\textsf{#1}}}
\newcommand{\ajith}[1]{\textcolor{magenta}{\textit{Ajith: #1}}}
\newcommand{\sukanta}[1]{\textcolor{blue}{\textit{Sukanta: #1}}}

\newcommand{\AEIHann}{Max-Planck-Institut f\"ur Gravitationsphysik 
(Albert-Einstein-Institut) and Leibniz Universit\"at Hannover, 
Callinstr.~38, 30167~Hannover, Germany}
\newcommand{\WSU}{Department of Physics \& Astronomy, Washington State University,
1245 Webster, Pullman, WA 99164-2814, U.S.A. \\
}
\newcommand{\IUCAA}{Inter-University Centre for Astronomy and Astrophysics, Post Bag 4, Ganeshkhind, Pune 411 007, India \\
}
\newcommand{\LIGOCaltech}{LIGO Laboratory, California Institute of Technology, 
Pasadena, CA 91125, U.S.A.}
\newcommand{\TAPIR}{Theoretical Astrophysics, California Institute of Technology, 
Pasadena, CA 91125, U.S.A.}


\title{Preparations for detecting and characterizing gravitational-wave signals from binary black hole coalescences}


\preprint{LIGO-P1200168}

\author{Thilina Dayanga}
\email{wdayanga@wsu.edu}
\affiliation{\WSU}

\author{Sukanta Bose}
\email{sukanta@wsu.edu}
\affiliation{\WSU}
\affiliation{\IUCAA}

\pacs{04.30.Tv,04.30.-w,04.80.Nn,97.60.Lf}

\begin{abstract}


We evaluate how well ``EOBNR'' waveforms, obtained from the effective one-body formalism, perform in detecting gravitational wave (GW) signals from binary black hole (BBH) coalescences modelled by numerical relativity (NR) groups participating in the second edition of the numerical injection analysis (NINJA-2).
In this study, NINJA-2 NR-based signals that are available in the public domain were injected in simulated Gaussian, stationary data prepared for three LIGO-Virgo detectors with early Advanced LIGO sensitivities.
Here we studied only non-spinning BBH signals. A total of 2000 such signals from 20 NR-based signal families were injected in a two-month long data set. 
The all-sky, all-time compact binary coalescence (CBC) search pipeline was run along with an added coherent stage to search for those signals. 
We find that the EOBNR templates are only slightly less efficient (by a few percent) in detecting non-spinning NR-based signals than in detecting EOBNR injections. On the other hand, the coherent stage improves the signal detectability by a few percent over a coincident search. In regards to signal parameters, such as the binary component masses and signal end-time, the magnitude and nature of the systematic errors in their measurement 
show some interesting but limited variations. In particular, a very small fraction of signals are systematically detected with templates of slightly more massive systems and, therefore, have a measured end-time that is earlier than the true one. The same signals have a worse match than other signals with EOBNR templates of the same parameters.
We compare these observations with the results of a study where EOBNR templates were used to find EOBNR signal injections to account for any biases that might arise from the data analysis pipeline itself. 
We discuss how our results can be utilized by various source modelling groups and data analysis pipelines to explore ways of improving the detectability of BBH signals. 

\end{abstract}
\maketitle

\section{Introduction}

Inspiraling binary black holes (BBHs) are one of the most promising gravitational-wave sources that the second generation ground-based detectors, such as the Advanced Laser Interferometer Gravitational-wave Observatory (aLIGO) \cite{aLIGO} and Advanced Virgo (AdV) \cite{Adv} detectors, are likely to detect. Currently these laser interferometric detectors are being upgraded 
and will start collecting data in a few years' time with a sensitivity improvement of about an order of magnitude. The new detector KAGRA \cite{Somiya:2011np} is also expected to take data with a similar design sensitivity later this decade.
These detectors will define what is being termed as the Advanced Detector Era (ADE).
Second generation instruments will have sensitivity in a broader frequency band compared to the initial detectors and will observe gravitational waves (GWs) from compact binary coalescence (CBC) signals starting at a lower frequency.
Observations of X-ray binaries IC10 X-1 \cite{Silverman:2008} and NGC 300 X-1 \cite{Crowther:2010} indicate that the masses of the stellar-mass components of a BBH can be as high as $20 - 30 M_\odot$. The discovery of HLX-1 in ESO 243-49 that has a lower mass limit of approximately $500~\mbox{M}_{\odot}$ presents strong evidence for the existence of intermediate-mass black holes ~\cite{Farrell:2010bf}. As described in Refs. \cite{Abadie:2011} and \cite{Aasi:2012}, binary black holes with component masses that high or higher will be detectable in the ADE detectors only through the merger and ringdown signals. This paper presents a search for BBH systems with total mass $25M_\odot \leq M \leq 100M_\odot$ and component masses $3M_\odot \leq m_1,m_2 \leq 97M_\odot$. 
Searches in real LIGO-Virgo data for BBHs in the same mass range were conducted with Inspiral-Merger-Ringdown (IMR) templates in Refs. \cite{Abadie:2011,Aasi:2012}.

\begin{figure*}[tb]
\centering
\includegraphics[width=8.5cm]{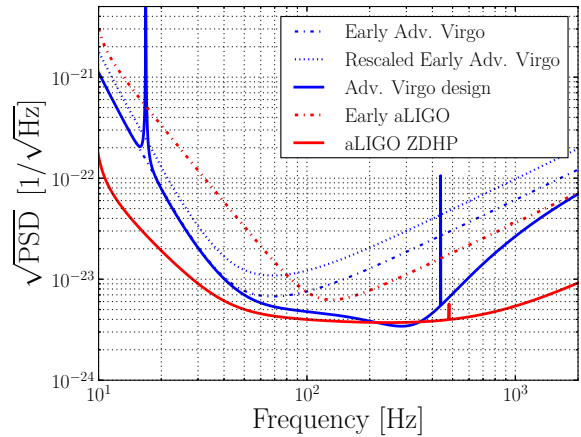}
\caption{
Comparison of early and final Advanced LIGO (aLIGO) and Advanced Virgo (Adv. Virgo) design amplitude spectral densities (ASDs) \cite{Lynn}. The ASD is the square-root of the power spectral density (PSD) \cite{Helstrom}. The red dotted and solid lines represent early and zero-detuned high-power (ZDHP) aLIGO design ASDs, respectively \cite{aLIGOASD}. The solid blue curve shows the Adv. Virgo design ASD \cite{AdVASD}. The blue dotted curve is obtained by rescaling it so that its horizon distance \cite{Mandel:2007hi} is similar to that corresponding to the early aLIGO ASD.
}
\label{fig:psd}
\end{figure*}

The ability to detect GW signals arising from BBH coalescences crucially depends on the accuracy of the waveform models used in designing search templates for detection pipelines. 
Estimation of parameters of BBH signals also demands accurate knowledge of the inspiral, merger and ringdown phases of the waveforms \cite{Bose:2010zz}. In 2008, the Numerical Injection Analysis (NINJA) collaboration was formed to facilitate the interaction of the Numerical Relativity and the GW Data Analysis communities with the objective of modelling CBC signals and using them to perfect GW search pipelines and parameter estimation algorithms. 
The main purpose of the first NINJA project (NINJA-1) \cite{Aylott:2009ya} was to foster the exchange of numerical-relativity waveforms to evaluate the performance of a variety of data analysis pipelines in detecting them in simulated initial LIGO and Virgo detector noise. 
The NINJA-2 exercise, on the other hand, creates the opportunity to test and compare detection pipelines more meaningfully than in NINJA-1 owing to the strict requirements imposed on the accuracy and length of the 
NR-based waveforms employed by it. 
The NINJA-2 project required each BBH waveform to include at least
five orbits of usable data before merger, i.e. neglecting the initial burst of junk radiation. Additionally, NR waveform amplitude had to be accurate to within 5
of gravitational-wave frequency) should have an accumulated uncertainty over the entire inspiral, merger and ringdown (of the numerical simulation) of no more than 0.5 rad. Following these restrictions eight NR groups have contributed 56 waveforms to NINJA-2 project \cite{Ninja2Catalog:2012}. 
Since the typical signal will have more cycles in band than what NR alone can produce at the desired accuracy with the computational resources at hand, post-Newtonian theory is used to model the remaining cycles to produce long {\em hybrid} signals, which we will refer to here as pN-NR hybrid signals or simply NR or NR-based signals. The NINJA-2 waveform-catalog paper \cite{Ninja2Catalog:2012} describes in detail the waveform requirements and method used to construct the pN-NR hybrid waveforms.

In this paper, we quantify how well ``EOBNR'' waveforms, obtained from the effective one-body formalism \cite{Buonanno:2007pf} by calibrating against a specific family of NR waveforms, namely, the ones obtained by the NASA-Goddard group \cite{Buonanno:2007pf},
perform in detecting pN-NR BBH signals modelled for NINJA-2. In NINJA-2, a variety of NR-based signals were injected in simulated Gaussian, stationary data from three LIGO-Virgo detectors with early ADE sensitivities (see Fig. \ref{fig:psd}). These signals were constructed with contributions from various numerical relativity groups (see Table 1 in Ref. \cite{Ninja2Catalog:2012}), and are available in the public domain \cite{Ninja2Catalogpublic:2012}. 
For the study reported here we focus on non-spinning BBH signals. A total of 2000 such signals from 20 pN-NR signal families were injected in a two-month long data set \cite{Ninja2Catalogpublic:2012}.
The coincident all-sky, all-time compact binary coalescence (CBC) search pipeline \cite{Babak:2012zx} was run along with an added coherent stage \cite{Bose:2011km} to search for those signals. 
We also compare these observations with the results of a study where EOBNR templates were used to find EOBNR signal injections to account for any biases that might arise from the data analysis pipeline itself. 
We find that the EOBNR templates are slightly less efficient, by about a percent, in detecting non-spinning NR-based signals than in detecting EOBNR injections. 
Also, the magnitude and nature of the systematic error in the measurement of signal parameters show some interesting but limited variations. In particular, a very small fraction of signals are systematically detected with more massive templates and, therefore, have a measured end-time that is earlier than the true one. The same signals have a worse match than other signals with EOBNR templates of the same parameters 
\cite{Ninja2Catalog:2012}.
We show how a coherent all-sky, all-time search can improve CBC detection efficiencies over a coincident analysis by improving the detection probability for any given false-alarm probability. The coherent method requires coincidence of the signal arrival times and other signal parameters, just like a coincident search method \cite{Babak:2012zx} does. But the former also checks for the consistency of the signal phases and amplitudes in the various detectors in a network with a physical value for the signal time-delays across the detector baselines. It also provides multi-baseline signal-based tests like the null-stream test that are effective in discriminating real signals from noise artifacts, especially, when the sensitivities of the detectors are comparable \cite{Guersel:1989th,Bose:2011km,Harry-2010PhysRevD,Talukder:2013}.

Owing to the high computational cost of employing a fully coherent all-sky, all-time CBC search \cite{Pai:2000zt},
here we use the {\em hierarchical} coherent algorithm \cite{Bose:2011km} to search for BBH signals. 
This algorithm and the detection statistic are described in Sec. \ref{sec:stats}.
Coherent data analysis methods have been formulated for searching GW signals from a variety of modelled and unmodelled sources \cite{Pai:2000zt,Arnaud-2003,Schutz-Merit:2011,Klimenko-2011,Bose:2011km,Harry-2010PhysRevD}. It is the optimal method in stationary, Gaussian noise under the Neyman-Pearson criterion \cite{Helstrom}. In CBC searches, it has been used recently in targeted searches \cite{Harry-2010PhysRevD} where the time of occurrence of the signal and the sky-position of the source are known, e.g., from the observation of an electromagnetic (EM) counterpart, such as a gamma-ray burst (GRB). It was demonstrated to perform better than the targeted all-sky, all-time coincident search.

In Sec. \ref{sec:detection}, we compare the performances of the coincident and the hierarchical coherent searches. We find the latter to be somewhat better than the former and explain what factors contribute to this improvement. We argue that the level of improvement is as per expectations for a hierarchical method and that a fully coherent method should be able to yield a much better detection efficiency. We present results from multiple sanity tests that check if the signal injection recovery is consistent with our expectations of a BBH search. Moreover, the effect of signal-based discriminatory tests, such as the chi-square and the null-stream tests, on the performance of that search is also analyzed.

In Sec. \ref{sec:estimation}, the accuracy with which various signal parameters are recovered is described. Those results are found to be mostly devoid of systematic errors when compared with measurements of EOBNR injection parameters by using EOBNR templates. The very few cases where a bias was found, the explanation lies in the mismatch of the pN-NR hybrid signal in those cases with the EOBNR templates, as was seen in Ref. \cite{Ninja2Catalog:2012}.



\section{Coincident and coherent searches for binary black hole systems}
\label{sec:stats}

\subsection{Search algorithm}
\label{subsec:algo}

The all-sky, all-time coincident search pipeline that was used to detect injected pN-NR hybrid signals is described in detail in Ref. \cite{Babak:2012zx}. On the other hand, the hierarchical coherent search pipeline used here was introduced in Ref. \cite{Bose:2011km}. The latter pipeline consists of the former with an additional stage that computes the coherent SNR of all coincident triggers found by the former.\footnote{That stage also computes the null stream \cite{Guersel:1989th,Harry-2010PhysRevD}, which will be described in Sec. \ref{subsec:null}.} We will variously refer to these two stages as the coincident (or the first) and the coherent (or the second) stage, respectively.
Both these stages comprise multiple steps. The coincident stage first splits the strain data time-series from every detector in the network into 2048 sec chunks. The noise power spectral density (PSD) \cite{Helstrom} is estimated for each chunk and is used to construct a template bank for matched filtering. 
Whenever the signal-to-noise (SNR) of the filtered output crosses a preset threshold, which was chosen to be 5.5 in Refs. \cite{Abadie:2011,Aasi:2012} and in this work, the template parameters and the time of the trigger are saved for each detector.
Next the triggers from individual detectors
are compared for coincidences in mass and end-time in two or more detectors to identify multi-detector coincident GW candidate events. Triggers in a detector that do not find any coincidence with a trigger in another detector are dropped from further analysis since currently we do not have a method for assessing the noise background for single detector events.
A candidate event is termed as double-coincident (triple-coincident) if the masses and end-times of the triggers in two (three) detectors are found to be very similar \cite{Robinson:2008un}, as stipulated in advance of the searches. Since the maximum number of detectors available to this search is three, these are the only two types of coincident candidate events possible here. In a real search, a candidate event is subjected to further checks before it is announced as a GW event to ensure that it was not caused by an environmental or instrumental artifact.

The coherent stage requires coincident trigger times in order to begin the coherent analysis. This analysis includes multiple steps, which are similar to the coincident counterpart but include some important modifications. First, a template bank, termed as the coherent bank, is constructed for the coherent analysis using the parameters of the coincident triggers. The triggers identified by the coincident stage can have different mass parameters in different individual detectors due to the different noise in each detector. By definition, however, the same template must be used in every detector for computing the coherent SNR. Here, we choose the mass-template in the loudest detector for that computation. Reference \cite{Bose:2011km} explains this process in detail.
That common template is next used to compute the matched-filter output, which is in the form of an amplitude and phase time-series, for every detector in the coincidence.
That step also computes the template normalization factors and implements signal-based vetoes. 

Similar to the matched-filtering step in the coincident stage, the coherent matched-filtering step used a thresholding criterion in previous studies \cite{Bose:2011km} to reduce the computational cost. This threshold value is set for the individual detectors as in the coincident stage but the value is lowered to allow more triple coincident triggers compared to the coincident counterpart. Typically, the values of previous studies used to be 5.5 for the coincident matched filtering step and 5.0 for the coherent one. One of the main goals of the coherent analysis is to convert all double-coincident triggers in the coincident stage to triples, whenever data are available from all three detectors, to improve the significance of real signals. 
Although a lower threshold in the coherent stage on individual detector SNRs helps to convert more double coincident events into triples many more remain as doubles when that threshold is non-zero. 
To improve upon earlier studies, we devised a method to handle these additional background triggers more efficiently, essentially, by reducing the maximum coincidence duration analyzed at a time. 
Lowering that threshold to zero allows converting all triggers that were double-coincident in the first stage to triple-coincident ones \footnote{Results from the new pipeline show that a few signals that show up as doubles in the coincident stage are still not converted into triples in the coherent stage. This is because those (simulated) signals are very weak in the third detector, thereby, resulting in unphysical time-delays between that and the other two detectors.}. This provides additional information about every single trigger due to the phase consistency check we can impose on the signal in every participating detector. The matched-filtering step in the coherent stage is followed by the computation of the coherent SNR and the null stream for every trigger.
Since we use large number of templates in our template bank, same feature in detector data can be picked by different template waveforms. This leads to have multiple GW triggers at same end-time. Additionally, for coherent searches multiple sky positions can give triggers for same feature in data. These set of triggers is also know as clusters. The last step of this pipeline clusters the gravitational-wave triggers in time and sky position so that only the most significant of them is retained per cluster.

\subsection{Detection Statistics}


The BBH search algorithms exploit the knowledge of their GW signals to define the templates used for matched filtering. Nine parameters describe the GW signal from nonspinning BBH sources studied here. These are the two component masses $m_1$ and $m_2$, the luminosity distance to the source $d$, the right ascension and declination angles $(\alpha,\delta)$ specifying its sky position, its orbital inclination angle $\iota$ to the line-of-sight, the angle $\psi$ describing the orientation of its signal polarization ellipse, the signal coalescence time $t_c$, and the signal coalescence phase $\phi_c$. Alternative mass parameters, in the form of the total mass $M$, the symmetrized mass-ratio $\eta \equiv m_1m_2/M^2$, and the chirp mass $M_{\rm chirp}\equiv \eta^{3/5}M$ are also often used to describe BBH signal parameters.

\begin{figure*}[tb]
\centering
\includegraphics[width=8.5cm]{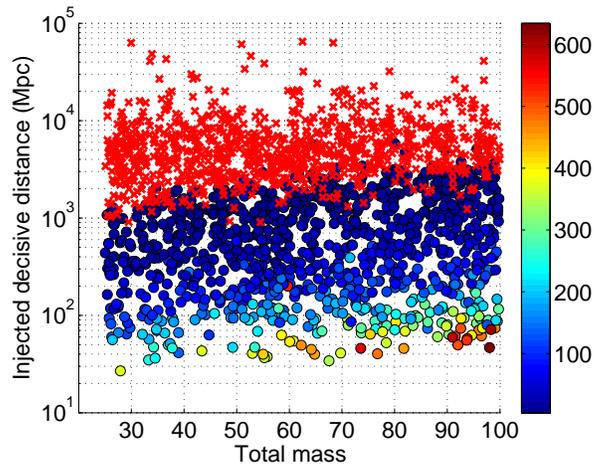}
\caption{
The injected decisive distance of found and missed pN-NR
simulated signals plotted as a function of total mass of the binary black hole system. The colorbar shows the coherent SNR. A total of 1033 injections were found out of 2000 injections by the search pipeline. Red crosses represent missed simulated signals. 
}
\label{fig:foundmissed}
\end{figure*}

For the coincident analysis, the detection statistic used here is the one that was introduced in Ref. \cite{Aasi:2012} for the {\em high-mass} search in LIGO-Virgo data from the sixth LIGO Science Run (S6) and the second and third Virgo Science Runs (VSR2/3). That search targeted GW signals from BBH sources with each component mass between $3-97 M_\odot$ and total binary mass between $25-100 M_\odot$. If the matched-filter output of a unit-norm mass template from a stretch of single-detector data is denoted by $\rho$, then the detection statistic used for long-duration templates, with a duration $t_{\rm dur} \geq 0.2$ sec,
is the {\em new} SNR defined as follows:
\begin{equation}\label{eq:new_snr}
 {\rho}_{\rm new} = \begin{cases} 
 {\displaystyle \frac{\rho}{[(1+(\chi^2_r)^3)/2]^{1/6}}} & \mbox{for } \chi^2_r > 1, \\
 \rho & \mbox{for } \chi^2_r \leq 1,
 \end{cases}  
\end{equation}
where $\chi^2_r \equiv \chi^2/(2p-2)$, and $\chi^2$ is the signal-based time-frequency discriminator studied in Ref. \cite{Allen:2004gu} 
with a chi-squared distribution of $p$ degrees of freedom.
On the other hand, for 
short-duration templates, with a duration smaller than 0.2 sec, the detection statistic used is
\begin{equation} \label{eq:effsnr}
 \rho_{\rm eff} = \frac{\rho}{[\chi^2_r(1+\rho^2/50)]^{1/4}} \,,
\end{equation}
which is termed as the {\em effective} SNR.
These choices were arrived at after comparing their detection efficiencies with those of $\rho$ and other alternative statistics. 
In a network with $M$ detectors, the coincident detection statistic employed here is
\be
\label{eq:coincstat}
 {\rho}_{\rm coinc} = \begin{cases} 
{\displaystyle  \left[ \sum_{I=1}^M \left(\rho_{\rm eff}^I\right)^2 \right]^{1/2}}
& \mbox{if } t^I_{\rm dur} < 0.2 {\rm sec}, \mbox{for any } I \,, \\
{\displaystyle  \left[ \sum_{I=1}^M \left(\rho_{\rm new}^I\right)^2 \right]^{1/2}}
& \mbox{for all other cases} \,, 
 \end{cases}  
\ee
where $x^I$ denotes the value of $x$ for the $I$th detector. 
Employing $\chi^2$ 
in the detection statistic was found to improve the performance of a search, especially, in real data, which is neither Gaussian nor stationary. Contrastingly, a network detection statistic formed from only the $\rho^I$ is the {\em combined} SNR, which is just $\left[ \sum_{I=1}^M \left(\rho^I\right)^2 \right]^{1/2}$.

The coherent statistic used here is the same as the one defined in 
Eq. (2.30) in Ref. \cite{Bose:2011km}.



When describing how effective a search is in finding signals, one of the parameters used is how distant their sources are. However, from the measured strength of a signal, it is not always possible to deduce the source luminosity distance. This is because
the strength of a signal in a detector is determined not only by the proximity of a source but also its location,
its orbital inclination angle to the line-of-sight,
and the angle $\psi$ describing the orientation of its polarization ellipse. In fact, a source at a luminosity distance $d$ appears to be at an {\em effective} distance of 
\be\label{eq:effdist}
d_{\rm eff} = \frac{d}{\sqrt{F_+^2(\alpha,\delta,\psi)\left(1+\cos^2\iota\right)^2/4
  +F_\times^2(\alpha,\delta,\psi)\cos^2\iota}}\,,
\ee
where $F_{+,\times}$ are the detector antenna-pattern functions. Due to the different orientations and, therefore, $F_{+,\times}$,
the effective distance of a source can vary from one detector to another. Since for a detection we require coincidence in at least two detectors, it is essential that the larger of the two corresponding effective distances not be too large for the signal to fall below the detection threshold of the weaker detector. This is why it is useful to define the {\em injected decisive} distance as the injected effective distance in the second loudest detector in a coincidence. Indeed, in Fig. \ref{fig:foundmissed} we show how this quantity and the coherent SNR vary as a function of the total-mass of the injected sources. It is manifest that the closer sources are found with a higher coherent SNR whereas many of the very distant sources are missed.



\begin{figure*}[tb]
\centering
\includegraphics[width=8.5cm]{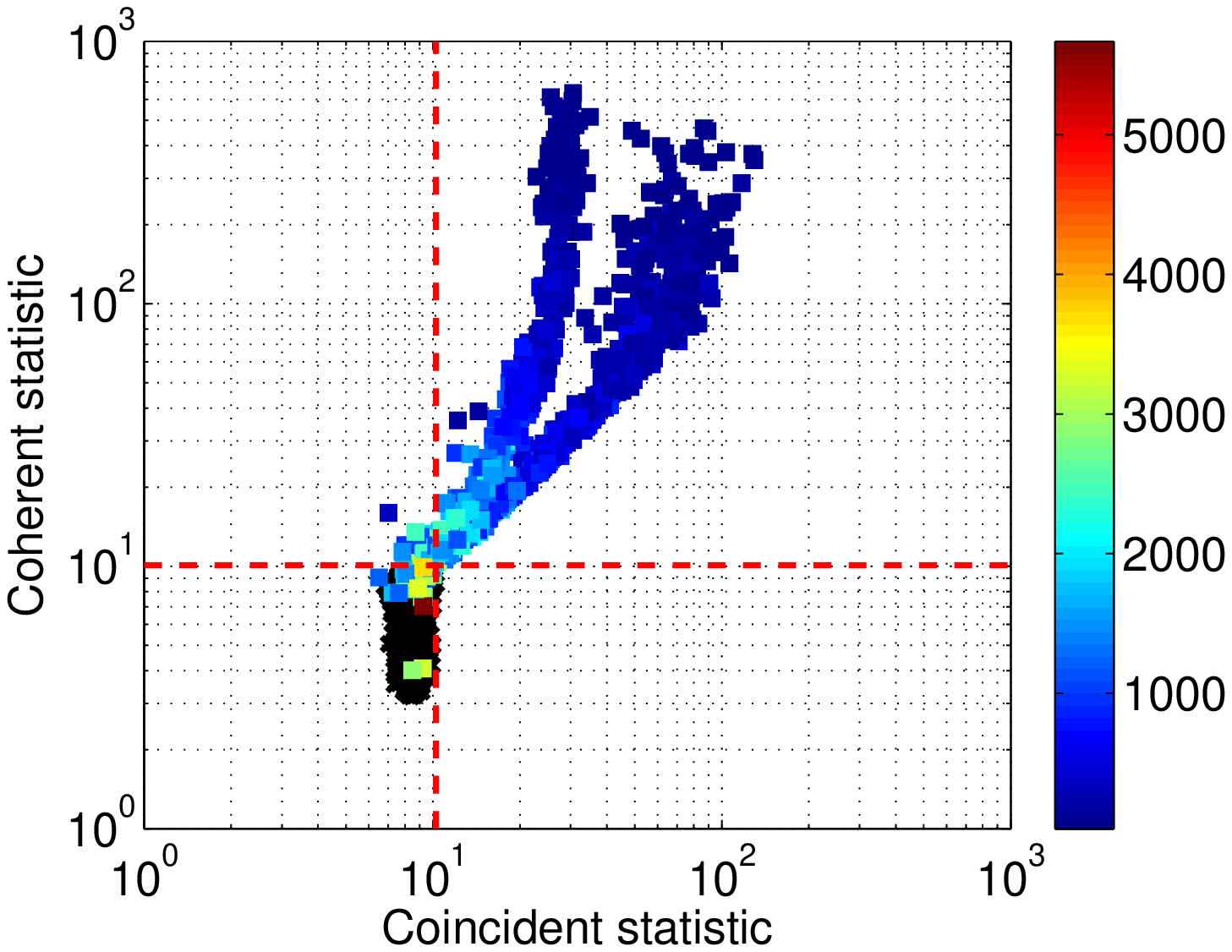}
\includegraphics[width=8.5cm]{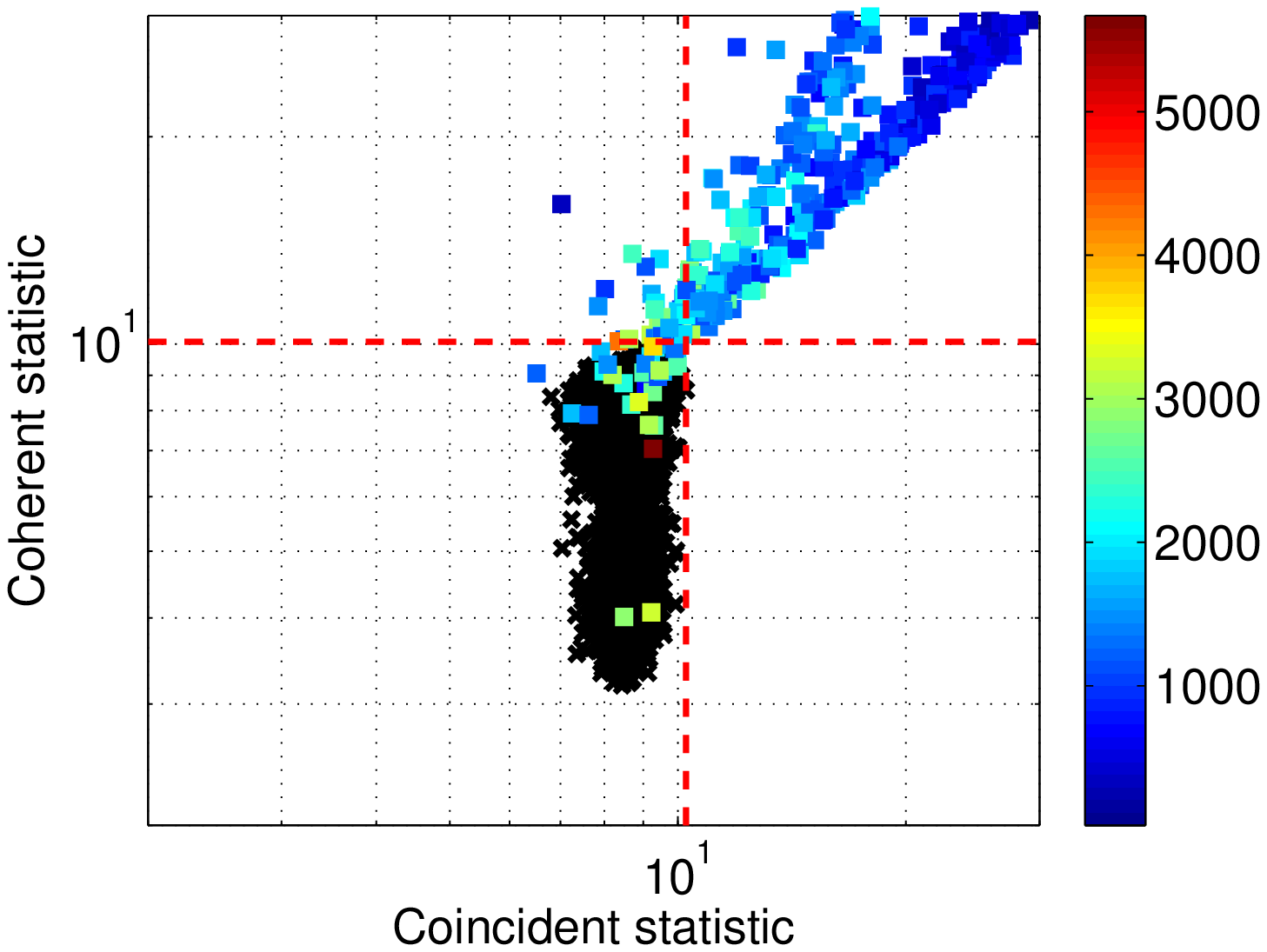}
\includegraphics[width=8.5cm]{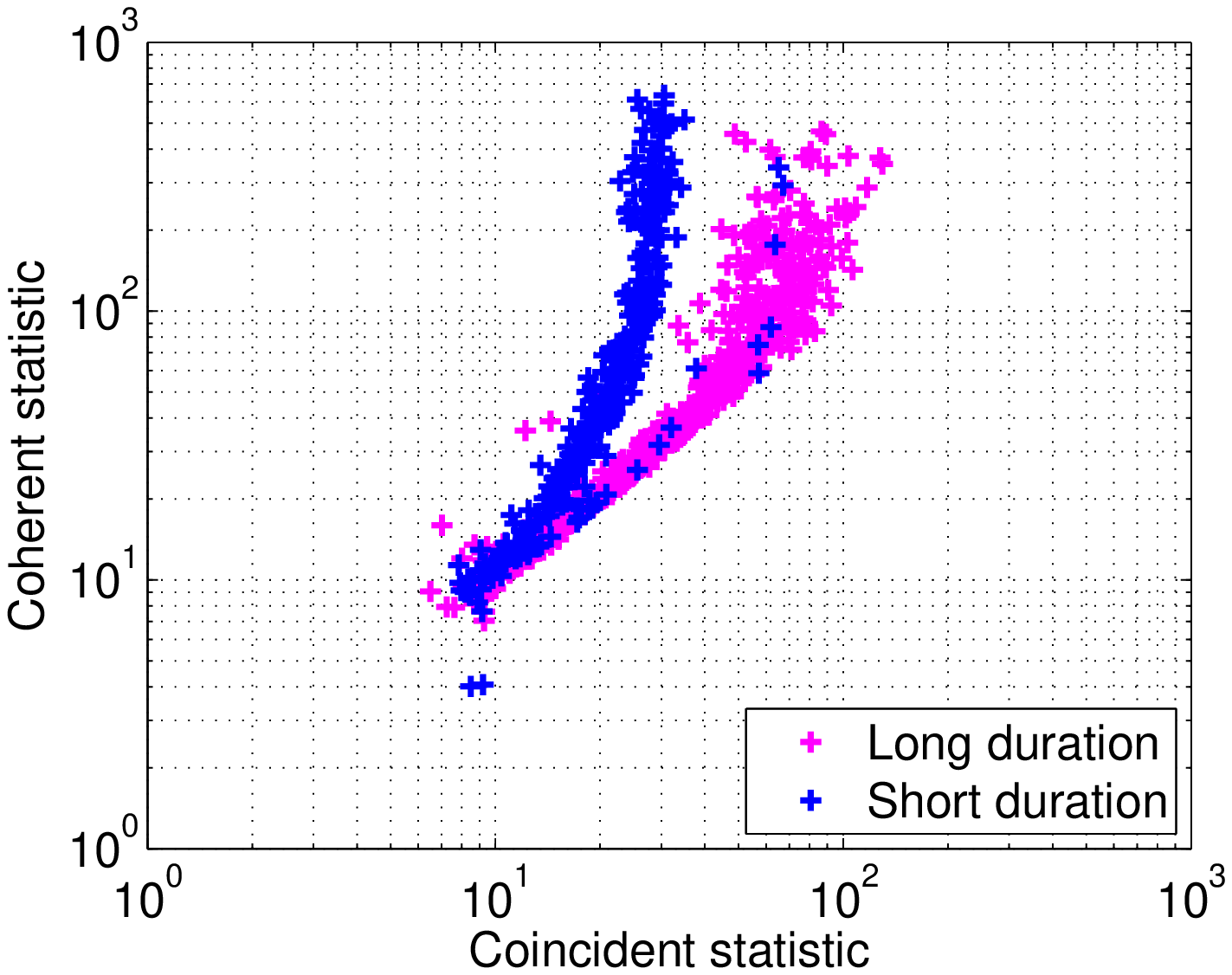}
\caption{
{\bf Top left:} The coherent and coincident detection statistic values of found simulated signals (colored squares) and background events (black crosses). The colorbars show the injected decisive distance in Mpc. The coherent SNR is used as the coherent detection statistic. The coincident detection statistic is ${\rho}_{\rm coinc}$, as defined in Eq. (\ref{eq:coincstat}), which is the one that was used in the high-mass search of LIGO's S6 and Virgo's VSR 2/3 data \cite{Aasi:2012}. {\bf Top Right:} Zoomed version of the left diagram shows better the distribution of the weak injections in the two statistics. {\bf Bottom:} Distribution of pN-NR simulated injections can be mostly separated into two groups characterized by the duration, long or short, of the template they triggered in the coherent search. This plot shows that a small subset of the triggers with a long-duration template fall in the same cluster as the ones with a short-duration template, and {\it vice versa}. This is because a signal can trigger multiple templates; the loudest trigger in coherent statistic can correspond to a different template than that of the loudest trigger in a coincident statistic, which penalizes short-duration ones more than the long-duration ones.
}
\label{fig:scatter}
\end{figure*}

\section{Detecting pN-NR hybrid signals with EOBNR templates}
\label{sec:detection}

In this study we analyzed the performance of the hierarchical coherent pipeline in simulated Gaussian noise. Strain noise time-series were produced for two LIGO detectors, H1 and L1, and the Virgo detector, V1,
using the design noise curves of early advanced LIGO and advanced Virgo detectors \cite{Aasi:2013wya}, respectively.
Each time-series is continuous and is of two months' duration.
Performance of the search pipelines was evaluated by injecting pN-NR waveforms in the simulated data and recovering them with EOBNR templates \cite{Buonanno:2000}. As with the LIGO-Virgo highmass search in S6-VSR2/3 data, here too the template bank used for matched-filtering spans $3-97 M_\odot$ for each of the two component masses and $25-100 M_\odot$ for the total mass.
The noise curves used to simulate noise in this study is the same as the one used for NINJA-2 blind injection challenge \cite{Ninja2BIC:2013}. Unlike in blind injection studies, it is Gaussian and stationary and has early aLIGO noise amplitude spectral density (ASD) for H1 and L1 and rescaled early advanced Virgo ASD for V1, as depicted in Fig. \ref{fig:psd}. As the name suggests, the latter is obtained by rescaling early AdV ASD to match the noise expected in V1 when H1 and L1 will be taking early aLIGO data in the future. Rescaling Virgo noise curve creates the opportunity to have three equally sensitive detectors and it benefits parameter estimation studies and multi-detector signal based vetoes in coherent analysis. It is important to note that these curves are not the 2015-predicted curves, but a best guess at them from over a year ago when NINJA-2 data sets were created.
In the past, real data from the LIGO and Virgo detectors had more dissimilar ASDs than the one used in NINJA-2.
This aspect of our search assumes special significance since
we later study the performance of the null stream,
which is more powerful for detectors in a network with similar rather than disparate sensitivities. 

Since the hierarchical coherent pipeline is not a fully coherent all-sky search, it depends on some other search method to identify the GW event times. In our study we analyze GW events identified by the coincident CBC pipeline. The coherent pipeline takes GW event times found by the latter as its input and does a coherent analysis of the individual detector data from around those trigger times. Compared to a fully coherent all-sky search, which combines complete time series coherently to identify GW events, the hierarchical coherent search requires less computational power. 
One of the main goals of this paper is to draw lessons from the hierarchical coherent search that can benefit the development of a fully coherent search for the ADE.

One major difference between the hierarchical coherent analysis used in this paper and those employed in the past \cite{Bose:2011km} is that after the coincidence stage identifies a part of the data that offers an interesting trigger, its coherent SNR is computed by involving every detector that was active at the time, even if it did not contribute to the coincidence in that stage (e.g., because the SNR in the third detector was below threshold around that event time).
Previous searches used thresholds for individual matched-filter outputs to reduce the computational cost. Typical value for this SNR threshold was 5.0, which was chosen to be somewhat below the threshold in the first stage, of 5.5, chosen for the matched filtering step.
By placing an SNR threshold of 5.0 the coherent pipeline allowed the conversion of some double coincident triggers 
(namely, H1L1, L1V1 and H1V1) to triple-coincident H1L1V1 triggers at times when all three detectors had science data, often termed as {\em triple time}. In this study we lower the individual detector thresholds in the coherent stage to zero. This allows all double and triple coincident triggers, in tripe time, in the coincident stage to be analyzed as triple-coincident triggers in the coherent stage.
By doing so for double coincident triggers, the coherent stage uses new information available from the third detector data for constructing the coherent statistic that was not used in constructing the coincident statistic in the first stage (apart from the fact that the third detector is weaker than the other two). This is an important change vis-a-vis the previous hierarchical coherent searches.

\begin{figure*}[tb]
\centering
\includegraphics[width=8.5cm]{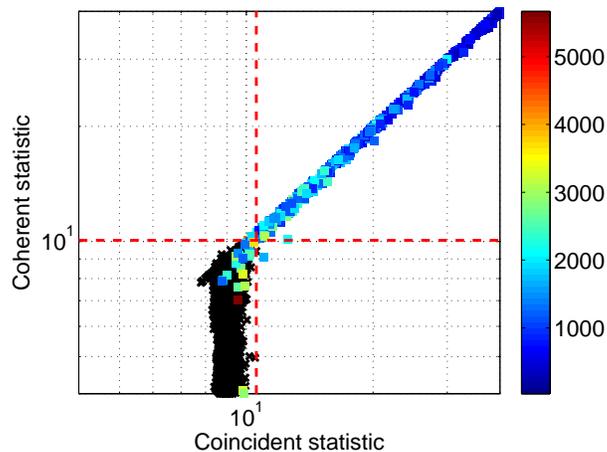}
\caption{
The coherent and coincident detection statistic values of found simulated signals (colored squares) and background events (black crosses). The colorbar shows injected decisive distance in Mpc. The coincident detection statistic used here is the combined SNR. 
This plot confirms that the maximum value the coherent SNR of a trigger can assume is the combined SNR.
}
\label{fig:scatternew}
\end{figure*}

We analyzed the performance of the hierarchical coherent pipeline by injecting and recovering the different NR wave-forms submitted to the NINJA-2 project by various NR groups. For this study we considered only non-spinning waveforms with different mass ratios. Two sets of simulated GW signals were injected with distances between $20$ Mpc to $2900$ Mpc, distributed uniformly in linear or logarithmic distance. Just as for the template back, for the injections as well the total mass was chosen to be in the range $25-100 M_\odot$ and the component masses between $3-97 M_\odot$,
All the injections recovered by the coincident algorithm were found by the hierarchical coherent pipeline as well, which is expected. On the other hand, although one might expect to find only triple-coincident events after the coherent stage, owing to the removal of the individual detector thresholds there, some events were found to be double-coincident. This is because these injections have very large effective distance in one of the detectors compared to the other two; the event time registered in the detector with the large effective distance (and, therefore, low SNR) typically had a large error
and often did not fall within the light travel-time window corresponding to the two baselines formed with the other two detectors. This is why the hierarchical coherent pipeline did not promote them to be triple coincidences but retained them as double coincidences. 

Figure \ref{fig:foundmissed} presents the injected decisive distance of found and missed injected BBH signals as a function of their total mass. As introduced earlier, the injected decisive distance is defined as the effective distance of a source in the detector where it was the second loudest, and is termed so because at least two detectors are required to hear the signal louder than the threshold to produce a coincident event.
The closest missed injection had an injected decisive distance of 907.34 Mpc. 
Figure \ref{fig:scatter} compares the distribution of injected signals and background triggers 
using both coincident and coherent detection statistics. The coherent analysis used the coherent network SNR as the detection statistic whereas the coincident analysis employed the same detection statistic as the one used for LIGO-Virgo BBH search of the S6-VSR2/3 data
\cite{Aasi:2012}, namely, the one defined in Eq. (\ref{eq:coincstat}). 
Due to the relatively poor performance of the $\chi^2$ test for short-duration templates compared to long-duration ones, this statistic utilizes that information differently for GW triggers associated with those two categories of templates. This is reflected in its definition in Eq. (\ref{eq:coincstat}), and the phenomenological case for it is explained in Ref. \cite{Aasi:2012}.

Figure \ref{fig:scatternew} plots the coherent SNR of every found injection trigger and background trigger versus the coincident statistic, which in this case is the combined SNR.
The color-filled squares denote found simulated signals and the black crosses represent background triggers. While the loudest background trigger in the coincident analysis has a coincident detection statistic value of 10.3, its counterpart in the coherent analysis has a coherent SNR of 10.0. 
The right plot in Fig. \ref{fig:scatternew} is the zoomed version of the left plot, focusing on the most interesting region of the plot where the simulated signals and background events start mixing. 
The two plots
show two red dashed lines, intersecting at right angles, that divide each plot into four quadrants. These lines cut across the loudest background triggers for each analysis, respectively. For instance, colored squares in the region $\rho_{\rm coinc} > 10.25$ are simulated signals found louder than the loudest background event, or simulated signals found with zero False Alarm Rate (FAR), in the coincident analysis. Similarly, the colored squares in the region $\rho_{\rm coh} > 10.0$ are simulated signals found with zero FAR in the coherent analysis. 
All the background triggers are confined to the bottom-left quadrant. The top-right quadrant has only injection triggers; these signals are found with a zero FAR in both coincident and coherent searches. The bottom-right quadrant has only injection triggers that have a zero FAR in the coincident search; note that these injection triggers have a non-zero FAR in the coherent search because the loudest background in the coherent search has an SNR (shown by the horizontal red line that is) greater than that of any of these inection triggers. On the other hand, the top-left quadrant has only injection triggers that have a zero FAR in the coherent search. When one counts the injection triggers in these quadrants, one finds that there are more of them in the top-left quadrant than in the bottom-right quadrant. This leads to the inference that in this study the coherent statistic performed better than the coincident one, at zero FAR.

It is also interesting to find that in Fig. \ref{fig:scatter} most of the simulated signals lie above the diagonal; this is due to the re-weighting of coincident SNR based on their $\chi^{2}$ values.  
One also finds that the injection triggers in these plots are distributed in two branches. To investigate the reason behind this feature, we plotted the same set of simulated signals after binning them into two different sets based on the duration of the templates that detected them. 
Events are termed to be of a short duration if at least one of the detectors finding it has a template duration less than 0.2 sec. To be categorized as a long duration event all the participating detectors should have their template durations greater than or equal to 0.2 sec for that event. The bottom plot in that figure indeed shows that these two branches of simulated signals were formed mostly from long and short duration events. That plot also shows that a few triggers of each kind are picked by templates of the opposite kind.
This is because a signal can trigger multiple templates; the loudest coherent SNR trigger in the second stage can correspond to a different template than the loudest coincident SNR trigger in the first stage. Note that the coincident statistic penalizes short-duration templates more than the long-duration ones. 


\begin{figure*}[tb]
\centering
\includegraphics[width=8.5cm]{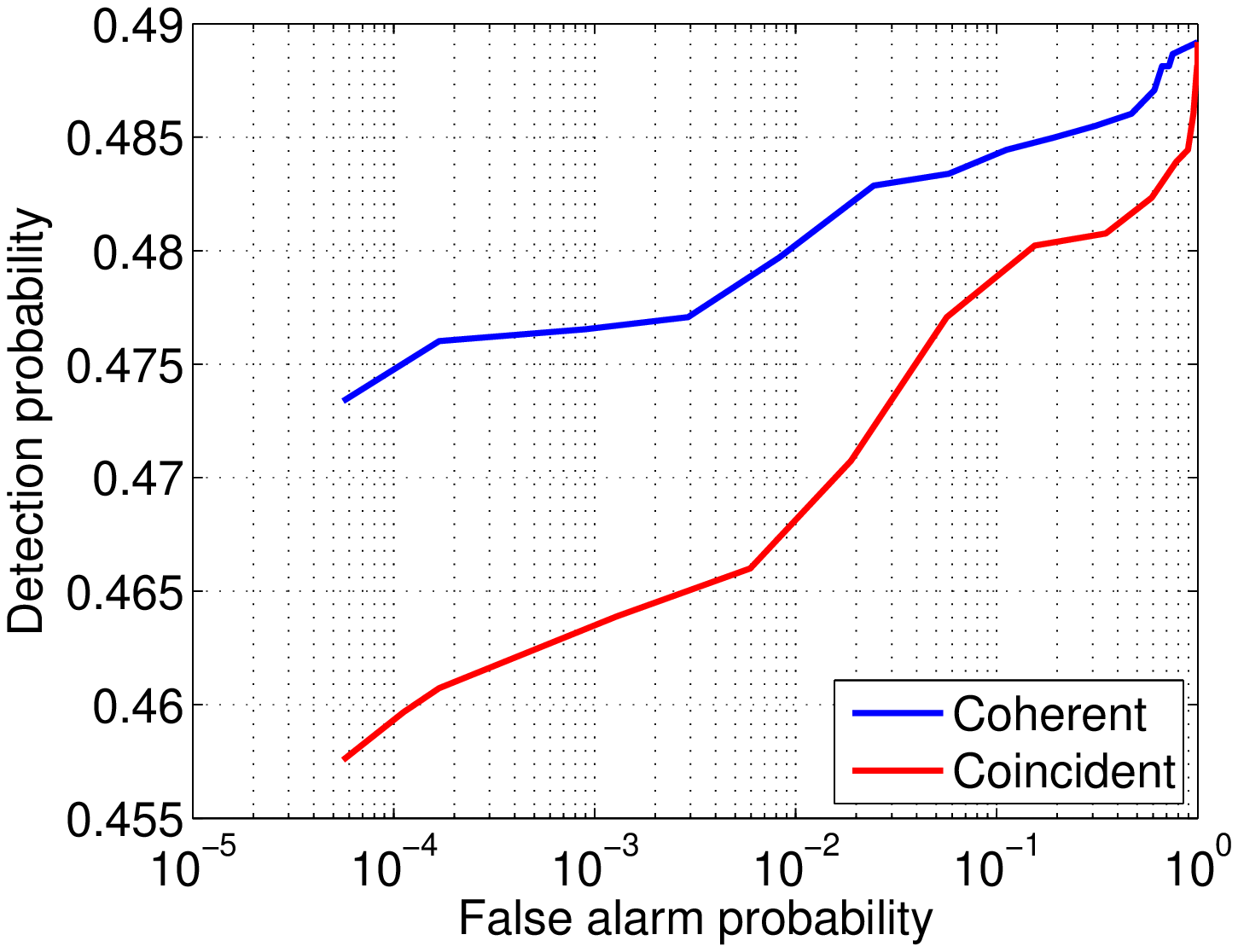}
\includegraphics[width=8.5cm]{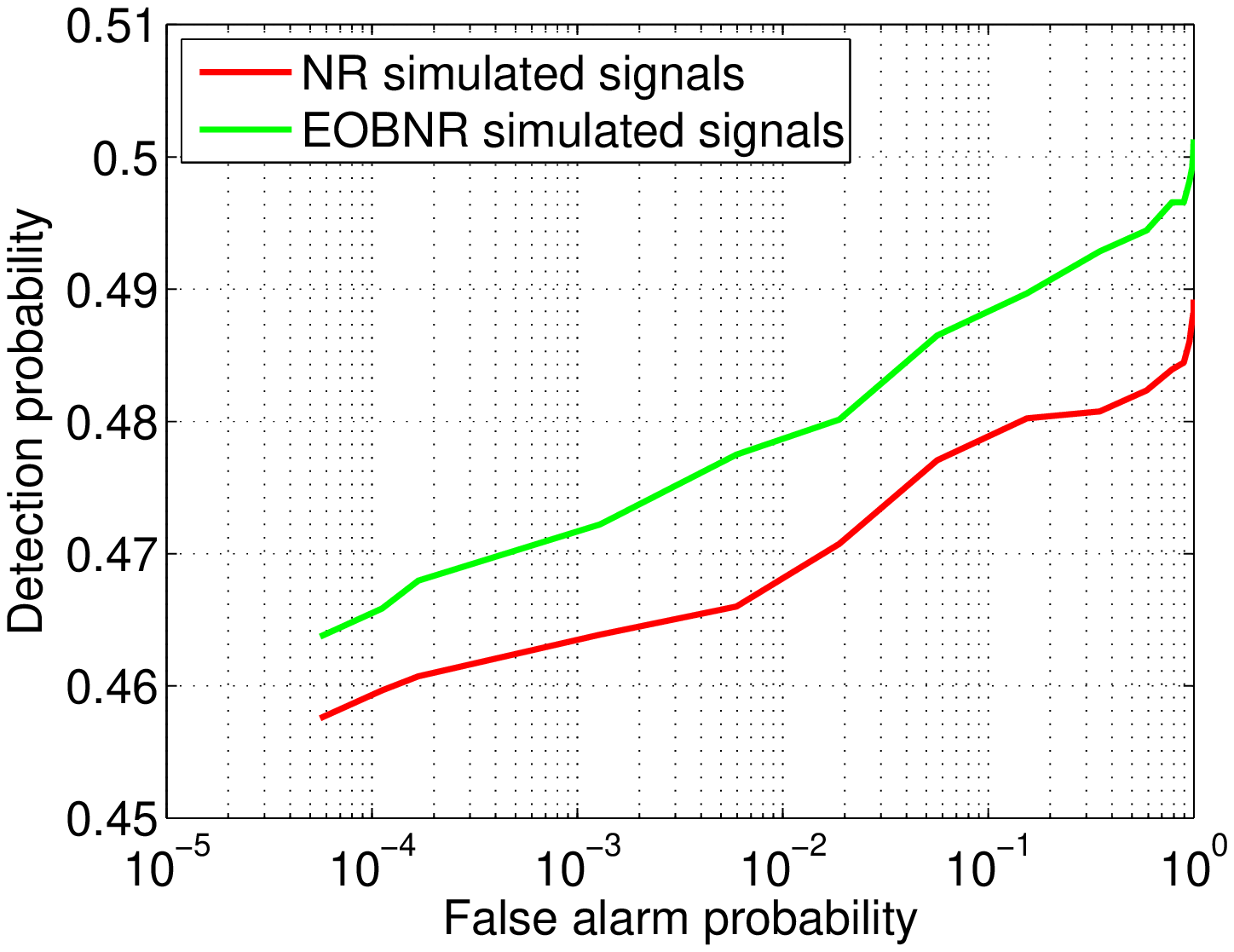}
\caption{
{\bf Left:} Receiver Operating Characteristic (ROC) curves from the injection and (partial) recovery of 2000 numerical relativity simulated signals. The two curves here compare the performances of the coherent and coincident searches. {\bf Right:} The ROC curves for two different simulated signal families, namely, EOBNR and pN-NR, recovered with the same EOBNR template bank. 
}
\label{fig:roc}
\end{figure*}

The maximum value that the coherent SNR can take for any trigger, from background or injection, is the combined SNR.
The combined SNR is the square root of the quadrature sum of the individual detector SNRs. Figure \ref{fig:scatternew} confirms this statement for both simulated signals and background events. 
It is clear that most of the simulated signals fall on diagonal. A relative few, especially of the weak kind, 
fall below the diagonal. This is because the contribution from their cross detector terms to the coherent SNR is less than maximal. 
As expected most of background events can be found below the diagonal due to the incoherence of their individual detector triggers. 
However, these is a small fraction of background triggers that lie on the diagonal. A majority of these are triggers were associated with sky positions that created issues for polarisation matrix inversion. Due to ill-posed polarisation matrix, detection statistic of these triggers had un-physical values larger than the combined SNR. Therefore we assigned combined SNR for detection statistic of such triggers allowing them to be on diagonal.  


\subsection{Comparing ROC curves}

In this section we address the following two issues. First, we enquire if the coherent stage improves the detectability of a CBC signal, which in NINJA-2 is limited to the pN-NR hybrid kind. If an improvement is found then it will make the case for developing computationally viable fully coherent CBC searches. Second, we ask how much worse does an EOBNR template-bank perform in detecting pN-NR signals than those modelled by the EOBNR formalism itself. This exercise probes, in a limited way, if the differences in the deduction of the pN-NR waveforms, on the one hand, and the EOBNR waveforms, on the other hand, are consequential enough to affect the ability of our search pipelines to make a detection. The limitation of this exercise is that it does not address how different EOBNR and pN-NR hybrid waveforms are from a fully accurate waveform solution from General Relativity. For readers interested in that subject, we refer them to Refs. \cite{Hannam:2010ky,Boyle:2011dy,Bernuzzi:2011aq,MacDonald:2011ne,Ohme:2013nsa}, and the references therein. We make both types of comparisons by computing the Receiver Operating Characteristic (ROC) curves \cite{Helstrom} for each case.

The ROC curves in the left panel of Fig. \ref{fig:roc} compare the performances of the hierarchical coherent and coincident searches for the same set of simulated pN-NR hybrid signals. The detection probability of the 
former is higher than the latter for all values of the false alarm probability (FAP) that could be computed in this exercise. But the region in this figure where this observation matters the most is the low FAP region around $10^{-5}-10^{-4}$, where the first GW detections are expected to be made.
Note, however, that the improvement obtained by running the additional coherent step in this coincident pipeline is limited by its hierarchical nature and, therefore, the utility of this study is that it suggests that
a fully-coherent, all-sky, all-time pipeline should perform even better.

The FAP at a given value of the detection statistic, $\rho_{\rm threshold}$, is obtained as follows
\begin{equation}
{\rm FAP}(\rho_{\rm threshold}) = \frac{N_{\rm background}(\rho > \rho_{\rm threshold})}{N_{\rm total \; background}} \,,
\end{equation}
where $N_{\rm background}(\rho > \rho_{\rm threshold})$ is the number of background triggers that have the value of their chosen detection statistic $\rho$ greater than $\rho_{\rm threshold}$, and $N_{\rm total \; background}$ is the total number of background triggers, with any value of $\rho$, found by the search.
On the other hand, the detection probability 
is given by
\begin{equation}
\label{eq:detp}
P_{\rm detection}(\rho > \rho_{\rm threshold}) = \frac{N_{\rm recovered} (\rho > \rho_{\rm threshold})} {N_{\rm recovered}(\rho > \rho_{\rm threshold}) + N_{\rm missed}} \,,
\end{equation} 
where 
$N_{\rm recovered}(\rho > \rho_{\rm threshold})$ is the number of recovered simulated signals with the detection statistic value greater than the value corresponding to a given false alarm probability, and $N_{\rm missed}$ 
denotes the number of simulated signals either with $\rho < \rho_{\rm threshold}$ or totally missed by the detection pipeline.


To get some insight into the relative behavior of the ROC curves in Fig. \ref{fig:roc} note that  the rate of change of FAP with respect to the detection statistic is faster in the case of the coincident statistic than in the case of the coherent statistic. However, the value of the detection statistic for an injection trigger scales inversely with its distance. Also, the detection probability decreases with increasing source distance.
Thus, it follows that the rate of change of FAP with respect to the detection {\em probability} is faster in the case of the coincident statistic than the coherent statistic. This is exactly what is found in the left plot in Fig. \ref{fig:roc}. The fact that the two ROC curves there should meet at a high enough FAP, where the detection threshold is very small, then implies that the ROC curve of the coherent search should be above that of the coincident search. The detection probability does not go to unity in that figure because a hard cut-off of $\rho = 5.5$ was placed in each detector in the first stage, thereby, causing a fraction of the triggers to be missed. In this experiment that fraction happened to be about 51\%. Figure \ref{fig:background} shows how the loudness of the background triggers is distributed for the two searches. 

Figure \ref{fig:roc} shows that at a FAP of $6\times 10^{-5}$ the difference between the detection probabilities of the coherent and coincident searches is approximately $(47.4 - 45.8)$\% $= 1.6$\%. This implies that for a total of 2000 injections the former search found 32 more than the latter, at that FAP.\footnote{Note that these two searches are correlated in the sense that a strong (weak) signal in one is highly likely to produce a strong (weak) signal in the other.} One must bear care, however, in drawing conclusions for astrophysical searches from this simulation study because the source rate for aLIGO is estimated to be in the several tens and not thousands \cite{Abadie:2010cf}, and the aforementioned improvement will affect the detection of very few sources. Instead, the main conclusion is that while hierarchical coherent searches may not be worth investing resources into, they demonstrate the (small) improvement we expect of them and, in turn, suggest that it may be worthwhile to explore how much more gain in detection rate one can achieve with fully coherent all-sky, all-time algorithms in realistic search pipelines.


We next enquire how much the detectability of a signal suffers owing to the fact the pN-NR hybrid waveforms that are used to model the signals are not quite the same as the EOBNR waveforms used to model the templates \cite{Ninja2Catalog:2012}. The right panel of Fig. \ref{fig:roc} compares the performance of two sets, with the same number of simulated signals injected at the same sky locations, but with two different signal families: The green curve corresponds to signals modelled with the EOBNR family and the red curve represents simulated signals made using the NINJA-2 pN-NR hybrid waveforms. Both sets of simulated signals were searched with EOBNR templates. As expected, simulated signals from the same family as the templates, namely EOBNR, register a better performance, at all FAP values. That figure also shows that the mismatch between the template and the signal families result in a small effect on the detection probability, which is of the order of a few percent at the most.

\begin{figure*}[tb]
\centering
\includegraphics[width=8.5cm]{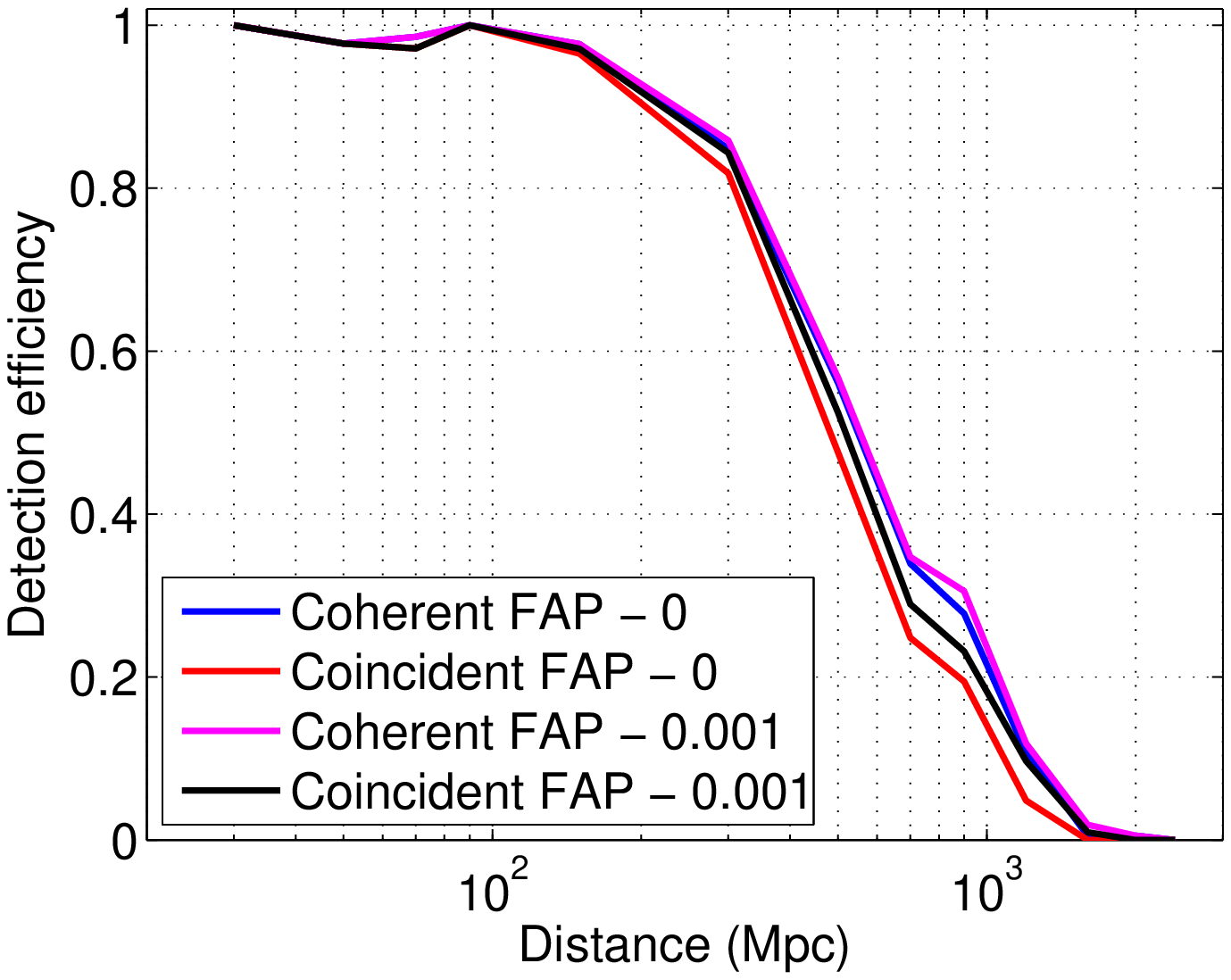}
\includegraphics[width=8.5cm]{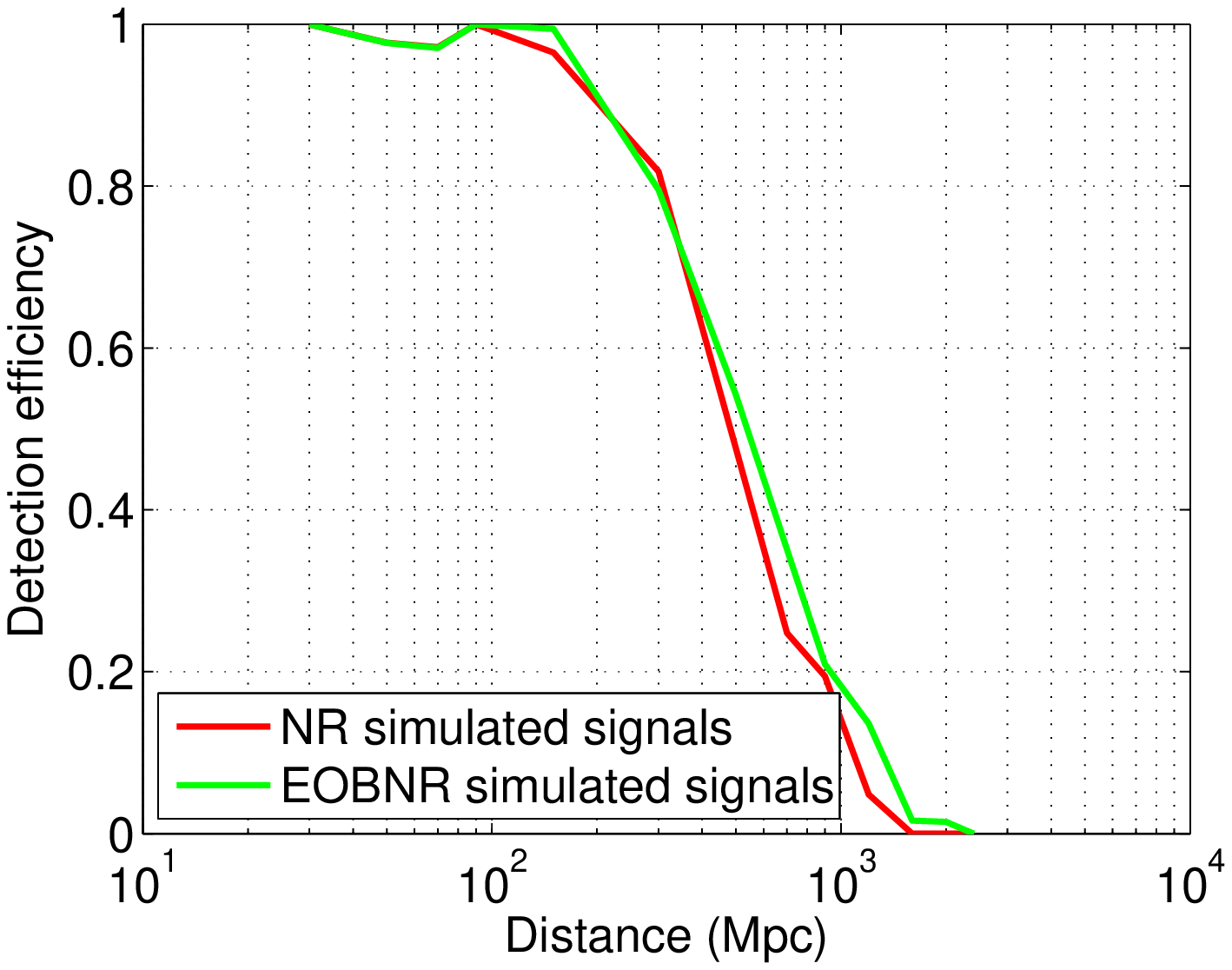}
\caption{
{\bf Left:} The detection efficiency is plotted as a function of the 
injected decisive distance (Mpc), for the hierarchical coherent and coincident searches, at false-alarm rates (FAPs) of 0.000 and 0.001. {\bf Right:} Detection efficiency comparison of two different sets of simulated signals, namely, EOBNR and pN-NR.
}
\label{fig:eff}
\end{figure*}

The variation of detection efficiency with distance also shows the expected behavior. We define it to have the same expression as $P_{\rm detection}$ in Eq. (\ref{eq:detp}), but now $P_{\rm detection}$, $N_{\rm recovered} (\rho > \rho_{\rm threshold})$, and $N_{\rm missed}$ are all computed in multiple distance bins. Figure \ref{fig:eff} shows the detection efficiencies of the hierarchical coherent and coincident searches at $\rho_{\rm threshold}$ corresponding to FAP=0 and $10^{-3}$. The detection efficiencies for all of the searches are very similar; they begin at 100\% for nearby sources and fall off to zero near 2 Gpc, which is the greatest horizon distance \cite{Schutz-Merit:2011} of the set of BBH sources simulated here.

\subsection{The null-stream}
\label{subsec:null}

Real data is neither Gaussian nor stationary and a statistic that is optimal in Gaussian and stationary noise may not remain so in real data.
This makes the case for seeking a more effective detection statistic and signal-based vetoes for LIGO/Virgo science data. An advantage of multi-detector coherent searches is that a detector network with three or more baselines can (over-)determine the two waveform polarizations, in addition to identifying the source location with time-delay triangulation \cite{Schutz-Merit:2011}. In such cases, one can form a linear combination of detector time-series outputs that contains no GW signal \cite{Guersel:1989th}. Such a combination is called the ``null stream". The null stream is consistent with the noise in a detector network, and a noise artifact or a glitch in a detector that is uncorrelated with noise in the other detectors is expected to leave a residue in the null stream. Therefore, its presence can be used to veto a candidate event. The simplest example of the null-stream is the one for a network consisting two co-located, co-aligned detectors. In the simple case where both detectors in such a pair have the same sensitivity, the null stream corresponding to any search template is proportional to the difference of its matched-filter outputs from the data of the two individual detectors. 
When the two detectors in the pair have different sensitivities, the matched-filter outputs are inversely weighted with their respective template-norms before calculating the difference \cite{Bose:2011km}. The value of the null-stream time-series at the trigger time is called the null statistic.

\begin{figure*}[tb]
\centering
\includegraphics[width=8.5cm]{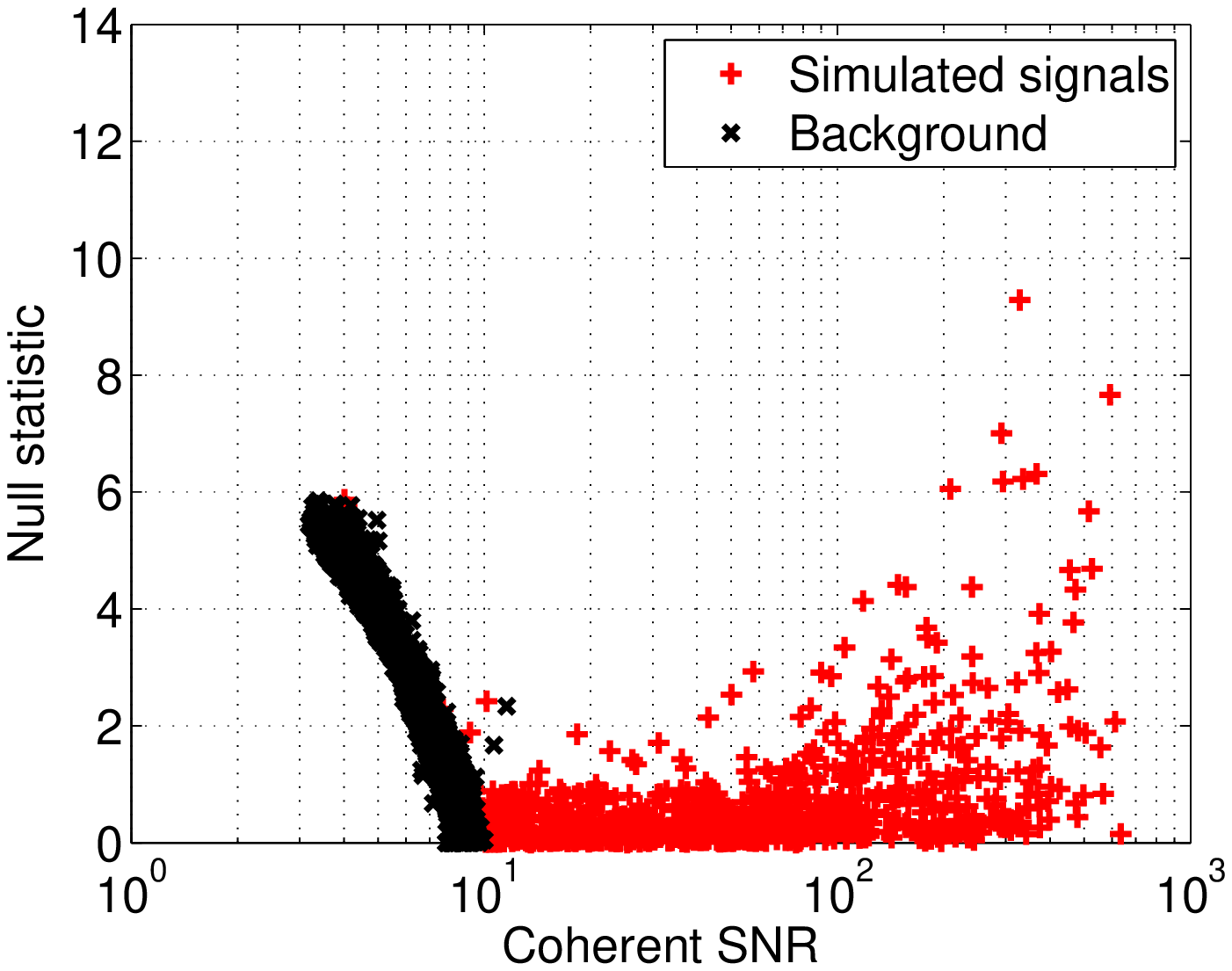}
\includegraphics[width=8.5cm]{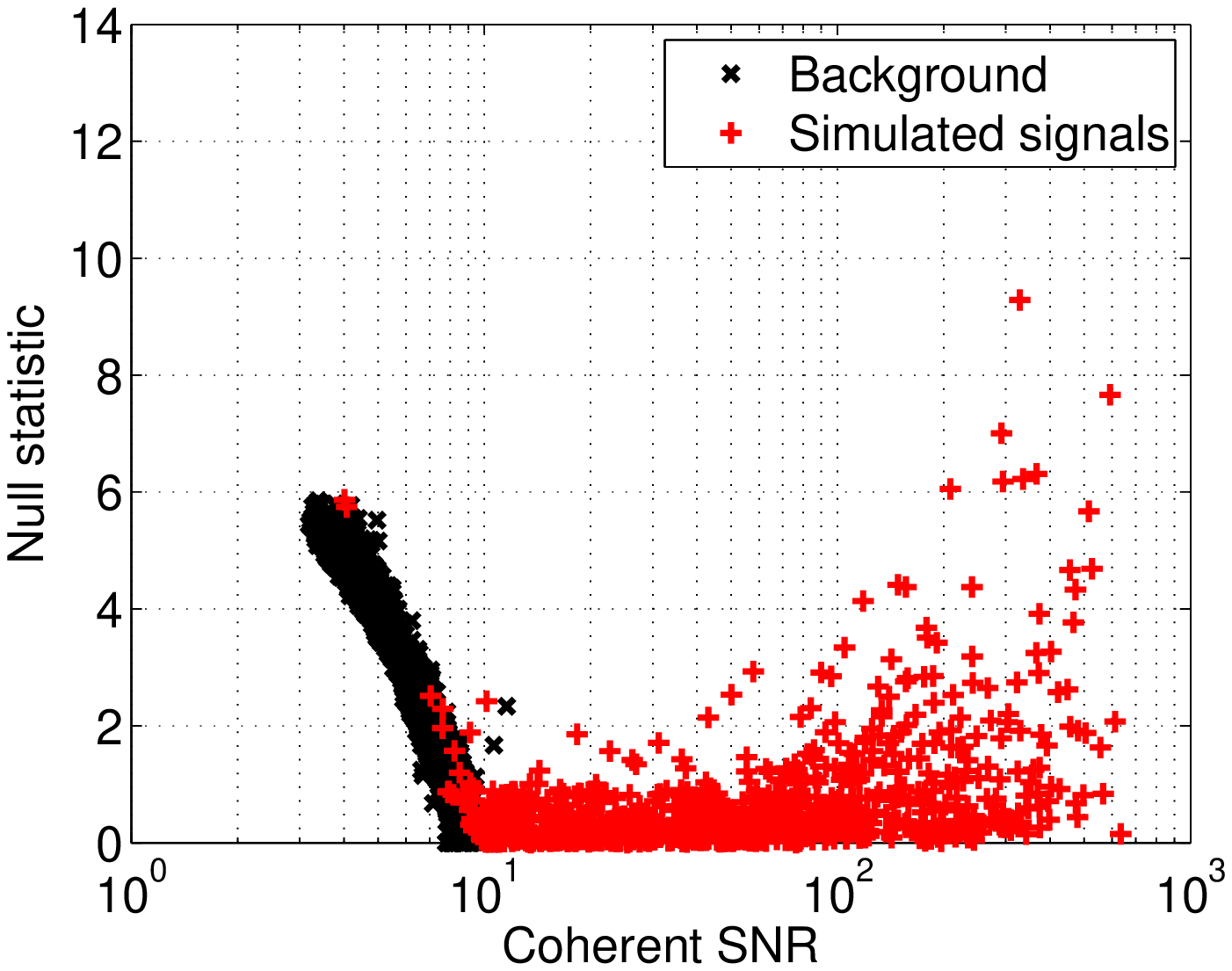}
\caption{
{\bf Left:} We plot the null-statistic for background and pN-NR simulated signals as a function of their coherent SNR. Most of the background triggers form a ``band'', which helps to separate signals from them. Notice that two of the loud background triggers prominently fall outside the band. These correspond to sky positions where the network does not resolve the GW polarizations very well \cite{Bose:2011km}.
In such a case, the trigger is vetoed. A better way to handle such triggers would be to not veto them but construct a more effective detection statistic by using a network's polarization resolving power at the trigger's sky position. This aspect will be explored elsewhere.
{\bf Right:} Here we show the injection triggers above the background triggers to highlight that for very weak injections, the null-statistic worsens and loses its discriminatory power. 
}
\label{fig:null}
\end{figure*}

Figure \ref{fig:null} presents the null-statistic distribution for simulations and background events as a function of their coherent SNR. For very strong signals the null statistic is large since even a slightly imperfect subtraction of two large SNR values can leave a moderately large residue. 
This is not a major concern since the distribution of loud injections on this plot is well separated from that of the background triggers. 
However, the null statistic can play a critical role in the detection of weak signals. Those occur in the region where their distribution on this plot mixes with that of the background triggers. Figure \ref{fig:null} shows that nearly all the background triggers fall on a relatively tight band and only a couple lie outside it. Those two outliers
are also the loudest background triggers in coherent SNR. Interestingly, these are triggers for which the sky positions were such that the LIGO-Virgo network could not resolve the putative signal polarizations. Just as for the injections, for these two triggers too the search pipeline used the coincident statistic as the detection statistic, the values of which were found to be well below 10.0.
Therefore, Fig.~\ref{fig:null} demonstrates that the null statistic is helpful in mitigating the impact of loud background triggers on a search pipeline's detection efficiency.

\begin{figure*}[tb]
\centering
\includegraphics[width=8.5cm]{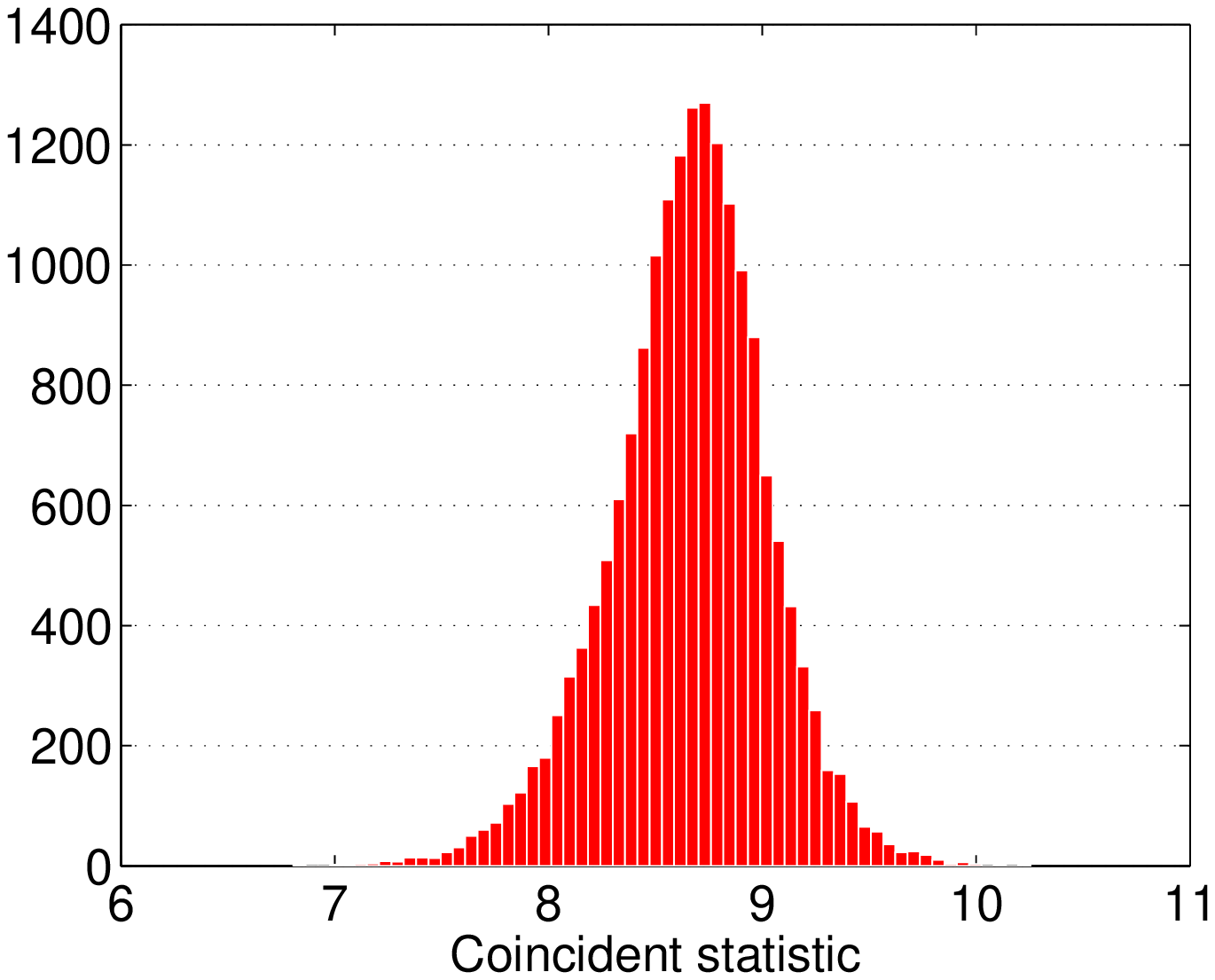}
\includegraphics[width=8.5cm]{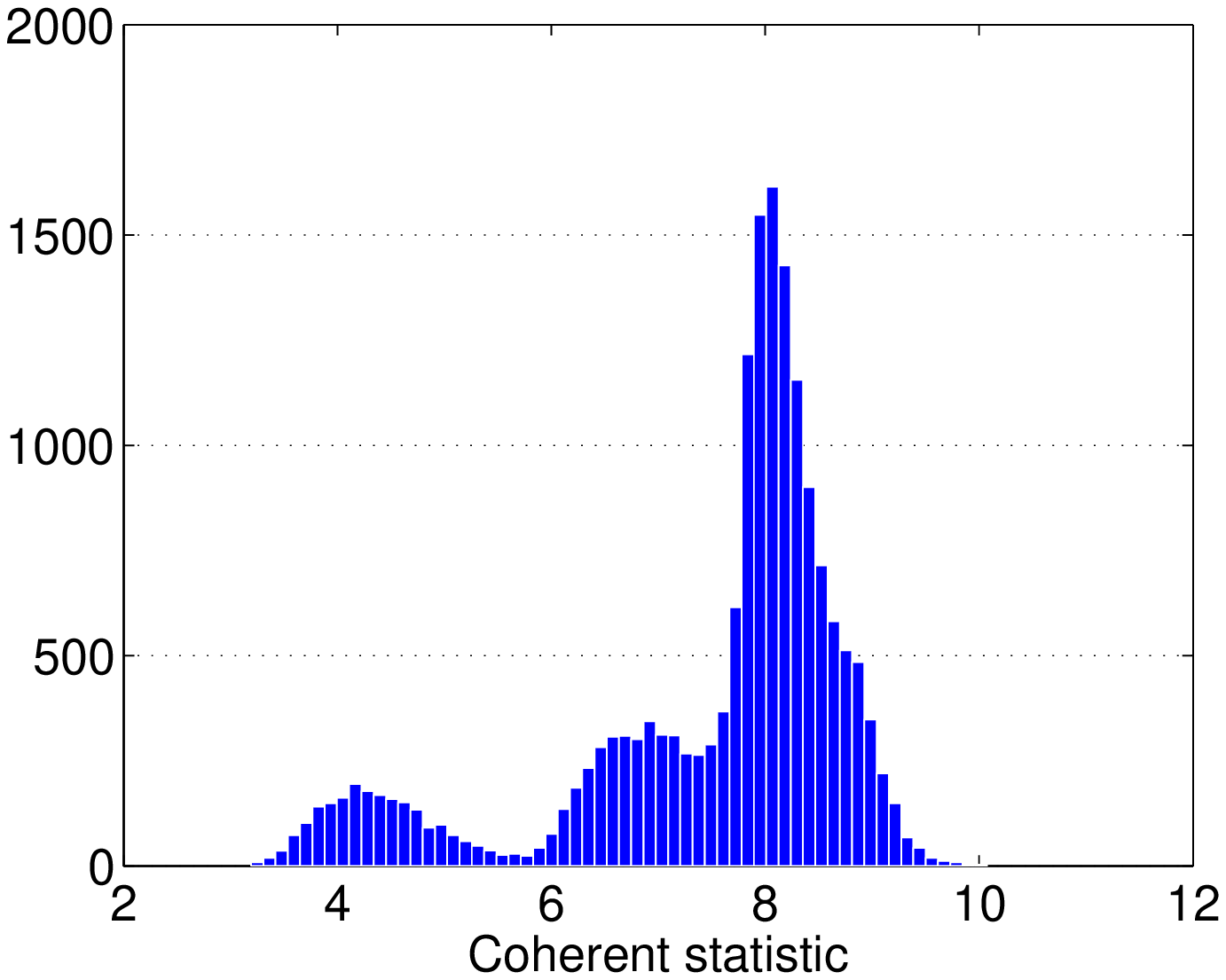}
\caption{
{\bf Left:} Distribution of background events in the coincident detection statistic used for the LIGO S6 - Virgo VSR2/3 high-mass search. {\bf Right:} Distribution of background events in the coherent detection statistic. 
}
\label{fig:background}
\end{figure*}

\begin{figure*}[tb]
\centering
\includegraphics[width=8.5cm]{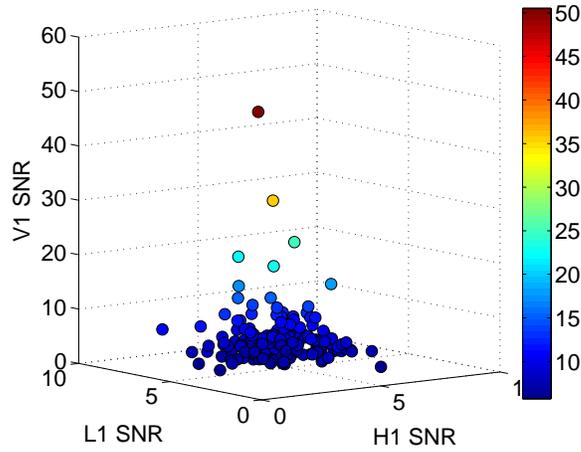}
\caption{
The coincident pipeline uses thresholds on matched-filter outputs from all detectors in a network to keep the number of background trigger coincidences at a manageable level so that the search remains computationally viable. An event has to be coincident in at least two detectors before it can be treated as a detection candidate.
While this requirement improves confidence in a detection, it also hurts detection efficiency by rejecting signals with marginally sub-threshold triggers in two of three detectors that would otherwise produce a candidate event with comparable detection confidence.
This plot shows the distribution of such events, with the colorbar displaying their coherent SNR. Out of 2000 simulated pN-NR signals, 48 were found with a zero false-alarm rate that have the SNR crossing the standard 5.5 threshold in only one of the three detectors. Also 54 events were found 
that had SNR crossing the standard 5.5 threshold in all three detectors.
}
\label{fig:sngl}
\end{figure*}

Although we will not be employing hierarchical search pipelines in ADE it is important to understand its issues in order to improve the efforts on developing fully coherent searches. As shown above, the thresholding criteria used in current coincident and hierarchical coherent pipelines can limit the detection efficiency of the search. In the current pipeline we only analyze triggers that cross the first matched-filtering stage with an SNR greater than 5.5. Additionally, we also require at least two such triggers in two different detectors within the window of light travel time between two sites to claim a gravitational-wave detection. It is possible to have gravitational-wave signals that only have threshold crossing trigger in one detector with significant coherent SNR. To investigate this in the same NR simulations, we searched for such events. Numerical relativity simulations plotted in Fig. \ref{fig:background} have only one threshold-crossing detector, yet their coherent SNRs are relatively strong: Out of a total of 2000 simulated pN-NR signal injections, we found that 52 had coherent SNR above 9.5. For each of these found injections only one of the three detectors in coincidence produced a single-detector SNR of above 5.5.\footnote{For a three-detector network with single-detector thresholds of 5.5 each, the minimum combined SNR for an event is $\sqrt{3}\times5.5 = 9.52$.} Furthermore, 35 of them had a coherent SNR higher than the loudest background trigger. Figure \ref{fig:sngl} reveals that some of these simulations had larger single-detector SNRs in Virgo compared to those in the two LIGO detectors. Due to their similar orientation and sky coverage, both LIGO detectors had very similar sensitivities in a large fraction of the sky. Additionally, their similar noise PSDs means that signals from the same source will have very similar SNRs in those detectors most of the time.
But Virgo's orientation is quite different from that of the LIGO detectors.
Therefore, in Fig. \ref{fig:sngl} some of the events have larger SNR contributions from Virgo than from any of the LIGO detectors.

\section{Parameter recovery}
\label{sec:estimation}

\begin{figure*}[tb]
\centering
\includegraphics[width=8.5cm]{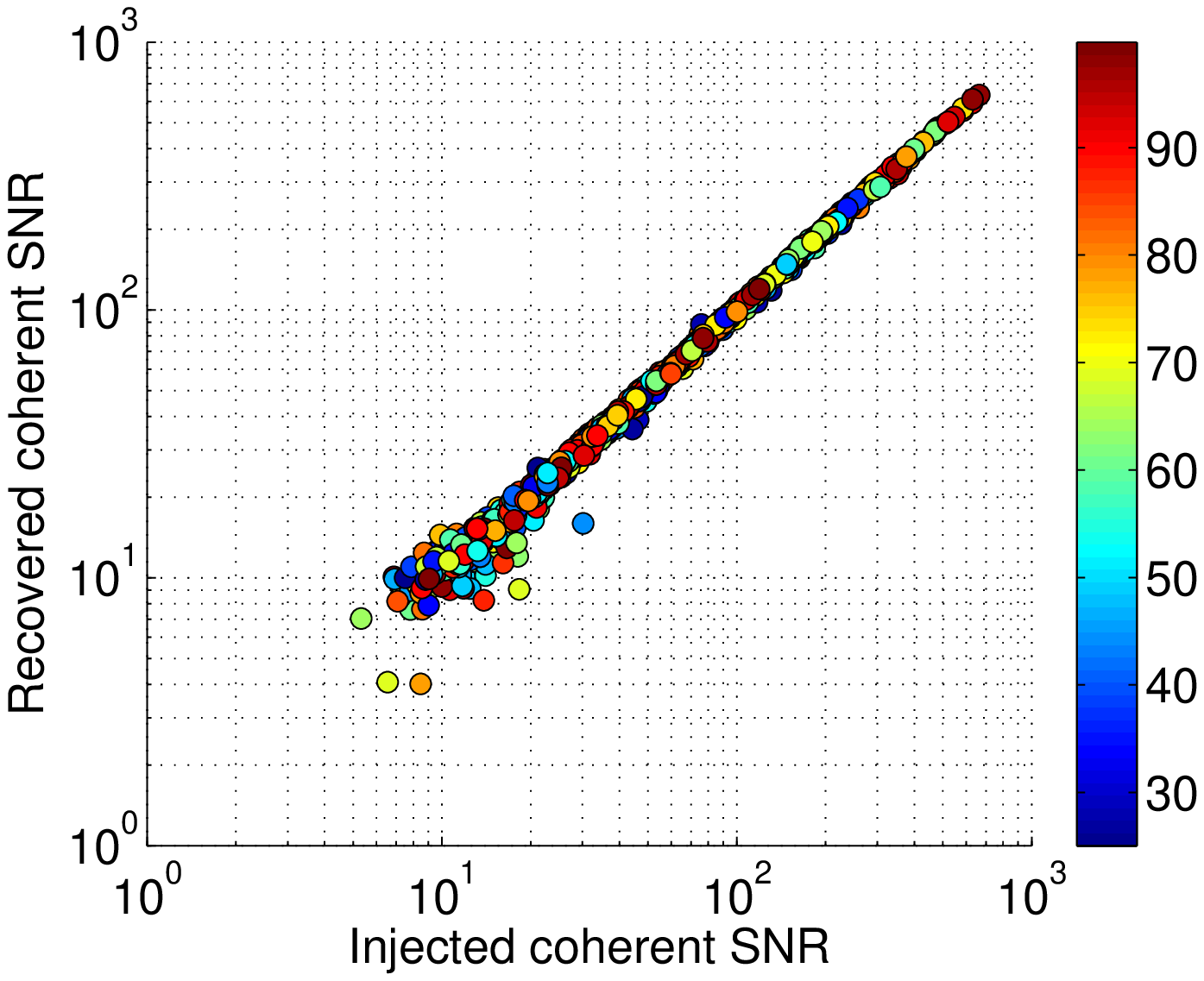}
\caption{
Recovered coherent SNR versus its injected value for NR simulated signals. The colorbar shows the total mass of the BBH systems whose pN-NR hybrid waveforms were injected. 
}
\label{fig:rec_cohsnr}
\end{figure*}

Parameter estimation will play an important role in making astrophysical statements about the origin of the GW signals we detect (see, e.g., Ref. \cite{Ajith:2009fz}). For instance, by measuring the masses of BBH components through GW observations we will be able to constrain stellar population synthesis models.
This section presents results from our EOBNR template searches of pN-NR simulated injections to educate us on how accurately one might be able to measure the parameters of real BBH signals.
In this study, parameters of those signals were estimated using the maximum likelihood method \cite{Helstrom}. Figure \ref{fig:rec_cohsnr} shows that the measured coherent SNR is within a few percent of the expected value when for signals the latter exceeds 20. Moreover, this observation holds across the whole range of total-mass values of the injected non-spinning BBH systems.

\begin{figure*}[tb]
\centering
\includegraphics[width=8.5cm]{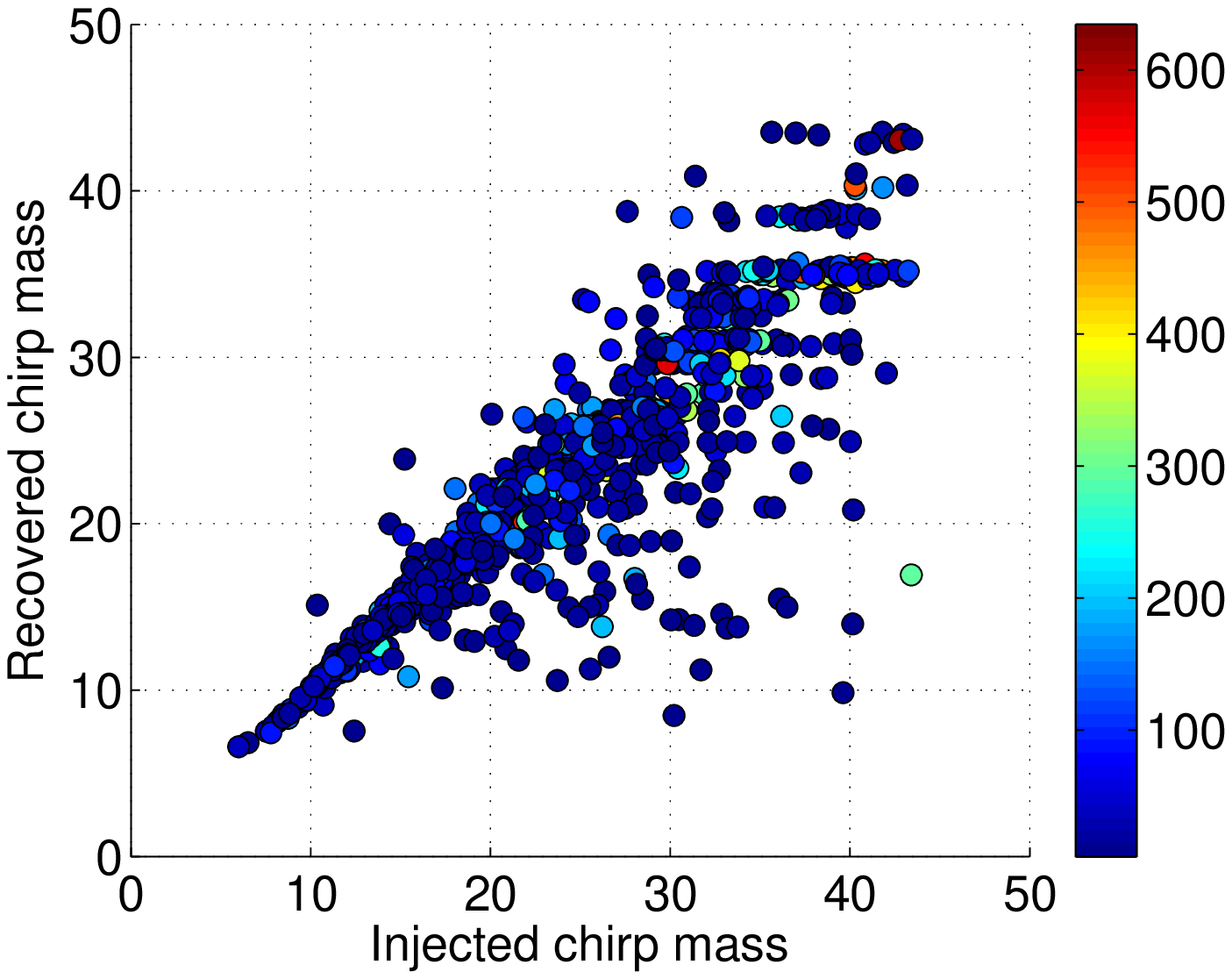}
\includegraphics[width=8.5cm]{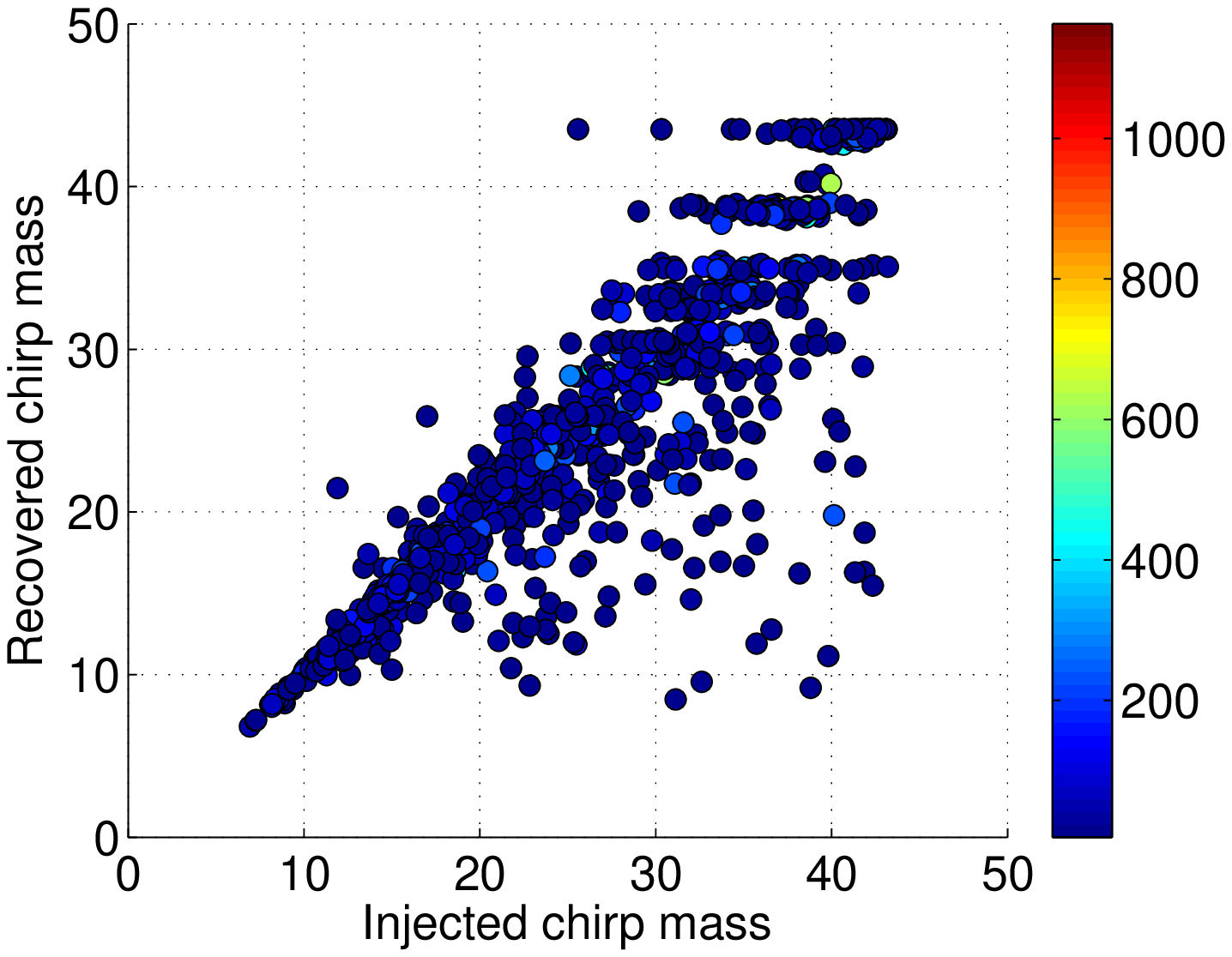}
\includegraphics[width=8.5cm]{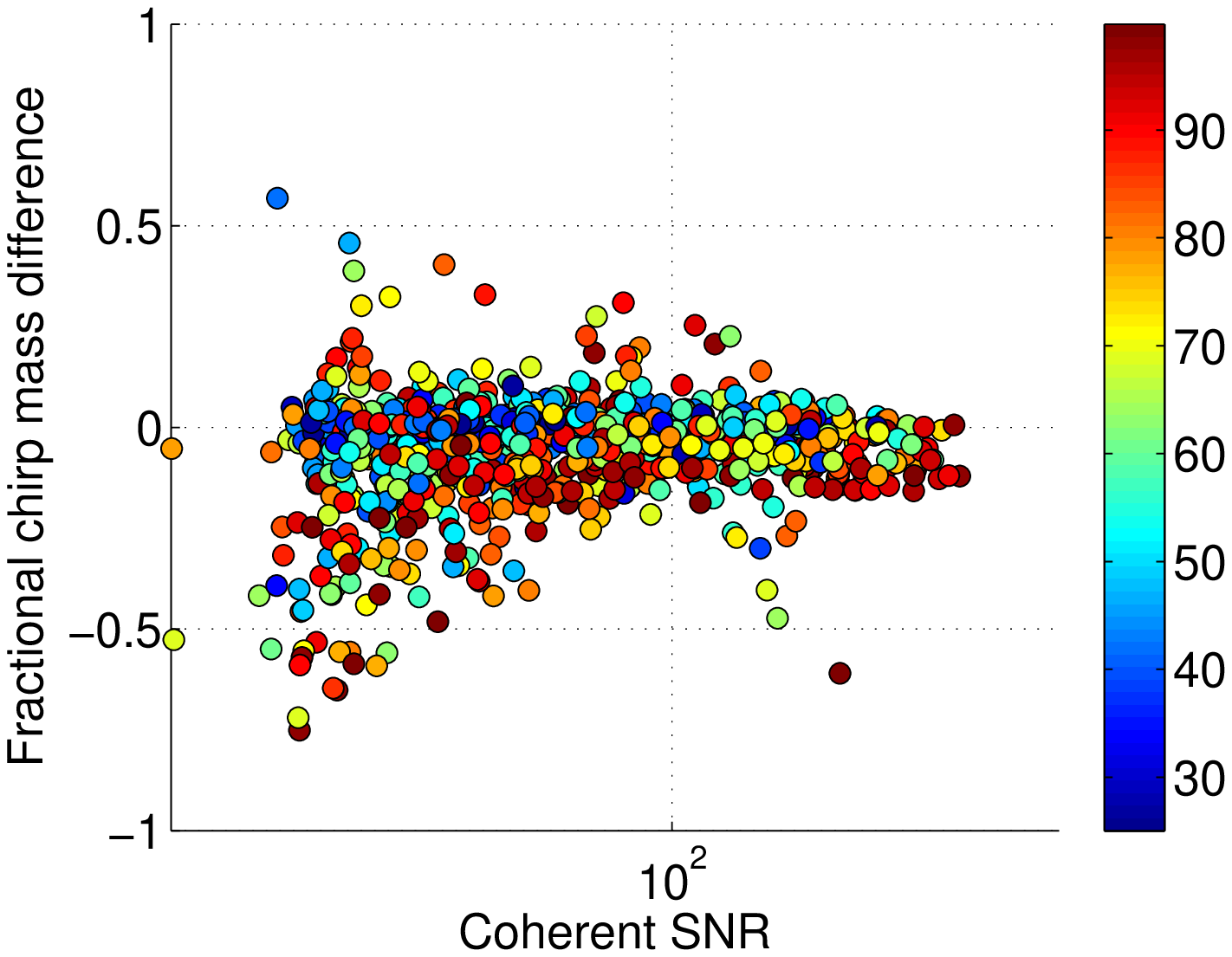}
\includegraphics[width=8.5cm]{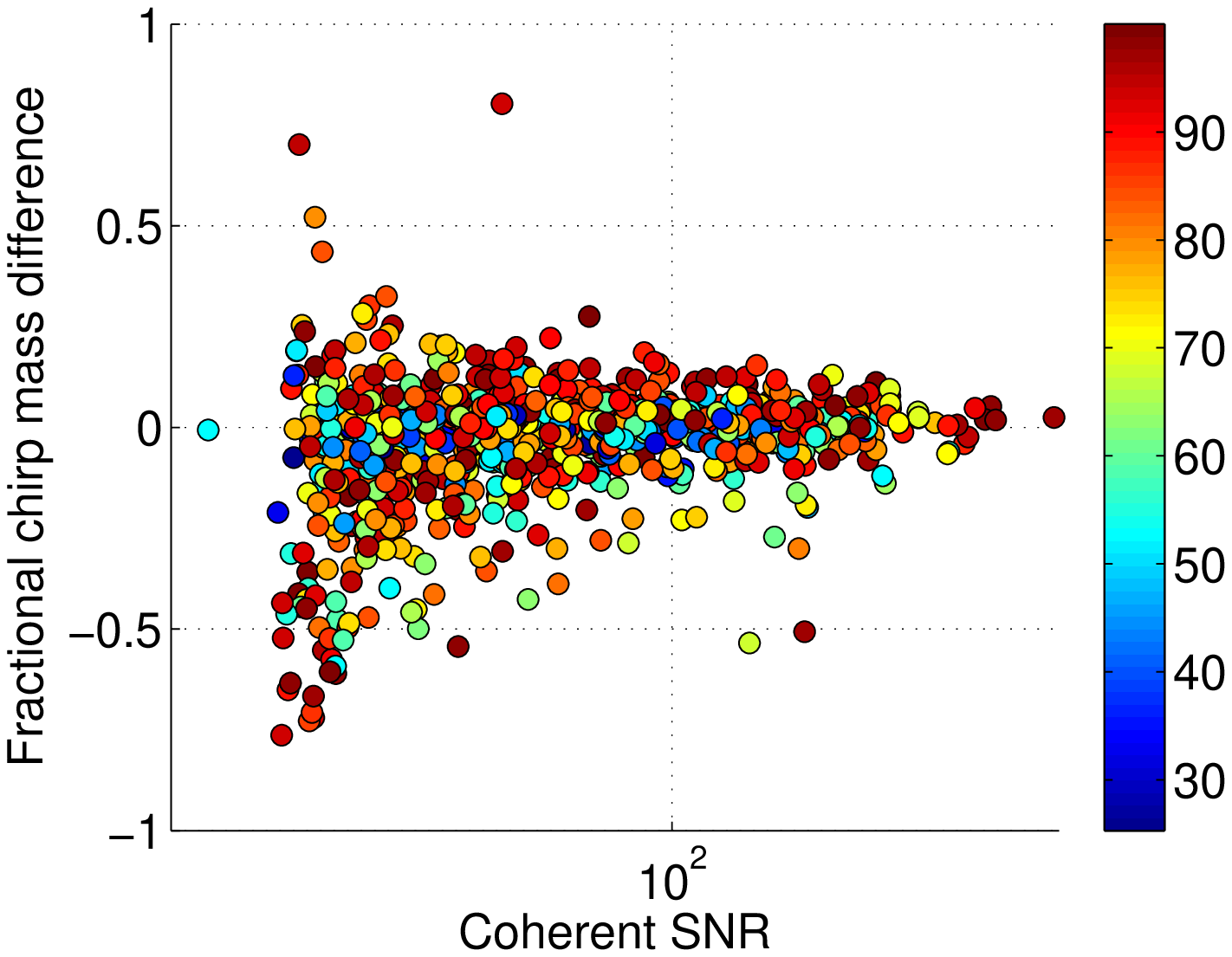}
\includegraphics[width=8.5cm]{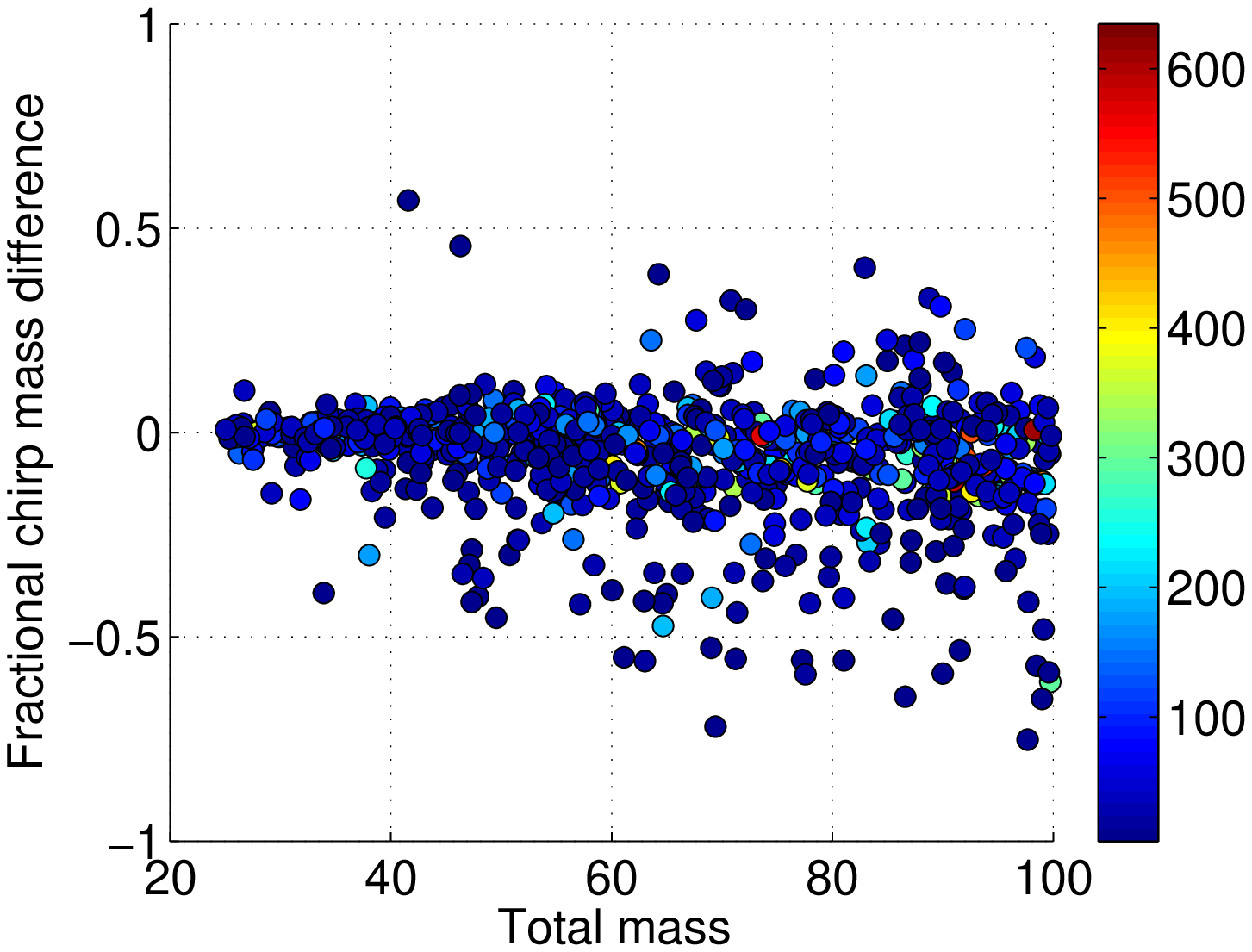}
\includegraphics[width=8.5cm]{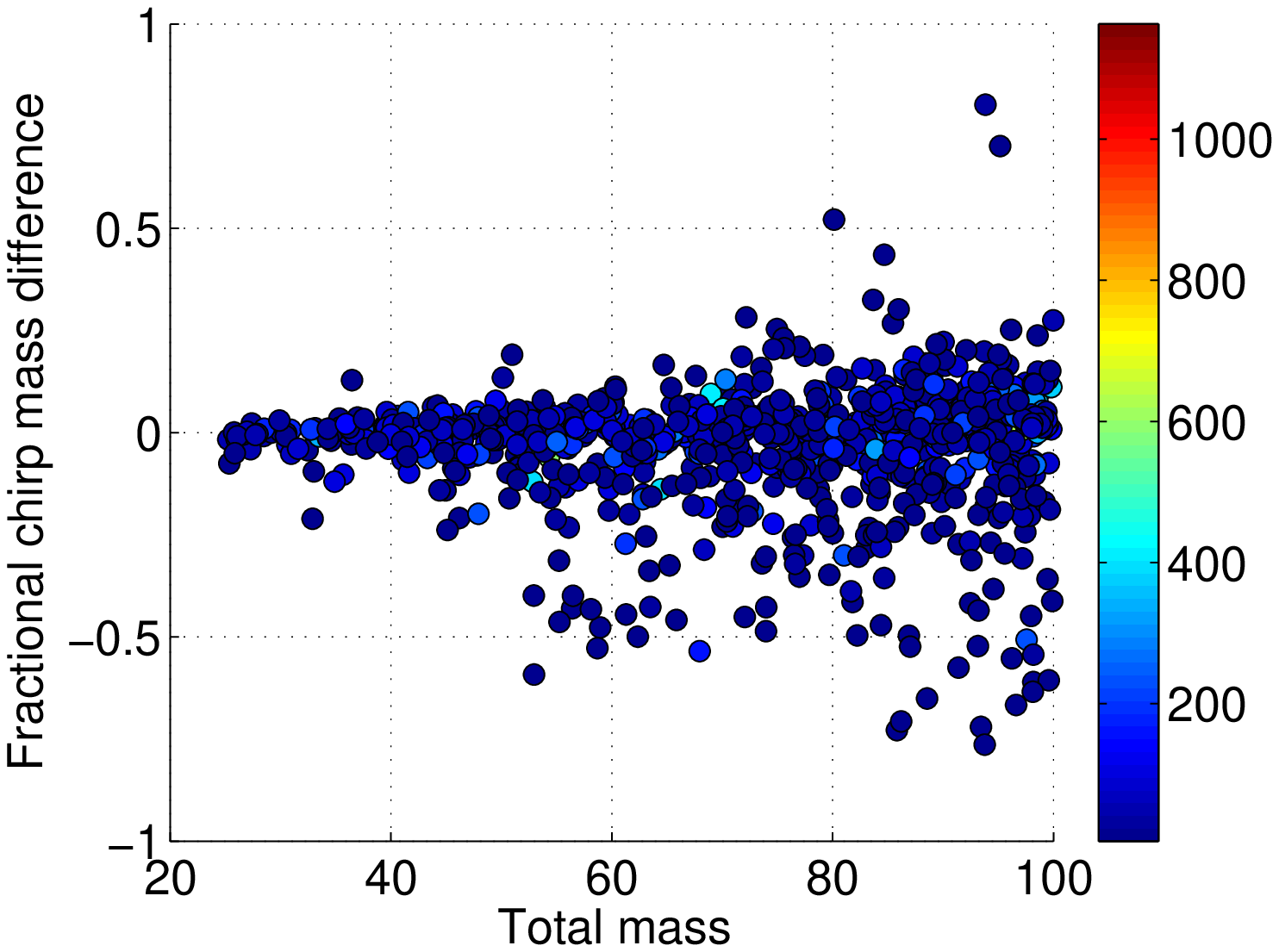}
\caption{
{\bf Left column:} Errors in the measurement of the chirp mass of pN-NR simulated signals. {\bf Right column:} Same as in the left column but for EOBNR simulated signals.
{\bf Top panel:} Recovered chirp mass as a function of injected chirp mass for pN-NR simulations (left) and EOBNR simulations (right), with the colorbar showing the coherent SNR. {\bf Middle panel:} Fractional chirp-mass difference as a function of coherent SNR with the colorbar displaying the total mass of the BBH system. {\bf Bottom panel:} Fractional chirp-mass difference as a function of the total mass of the BBH system, with the colorbar showing the coherent SNR. According to the bottom panel the overall chirp mass recovery is slightly better for EOBNR simulations than the pN-NR ones, due to better match with the templates in the former case.
}
\label{fig:mchirp_rec}
\end{figure*}

\begin{figure*}[tb]
\centering
\includegraphics[width=8.5cm]{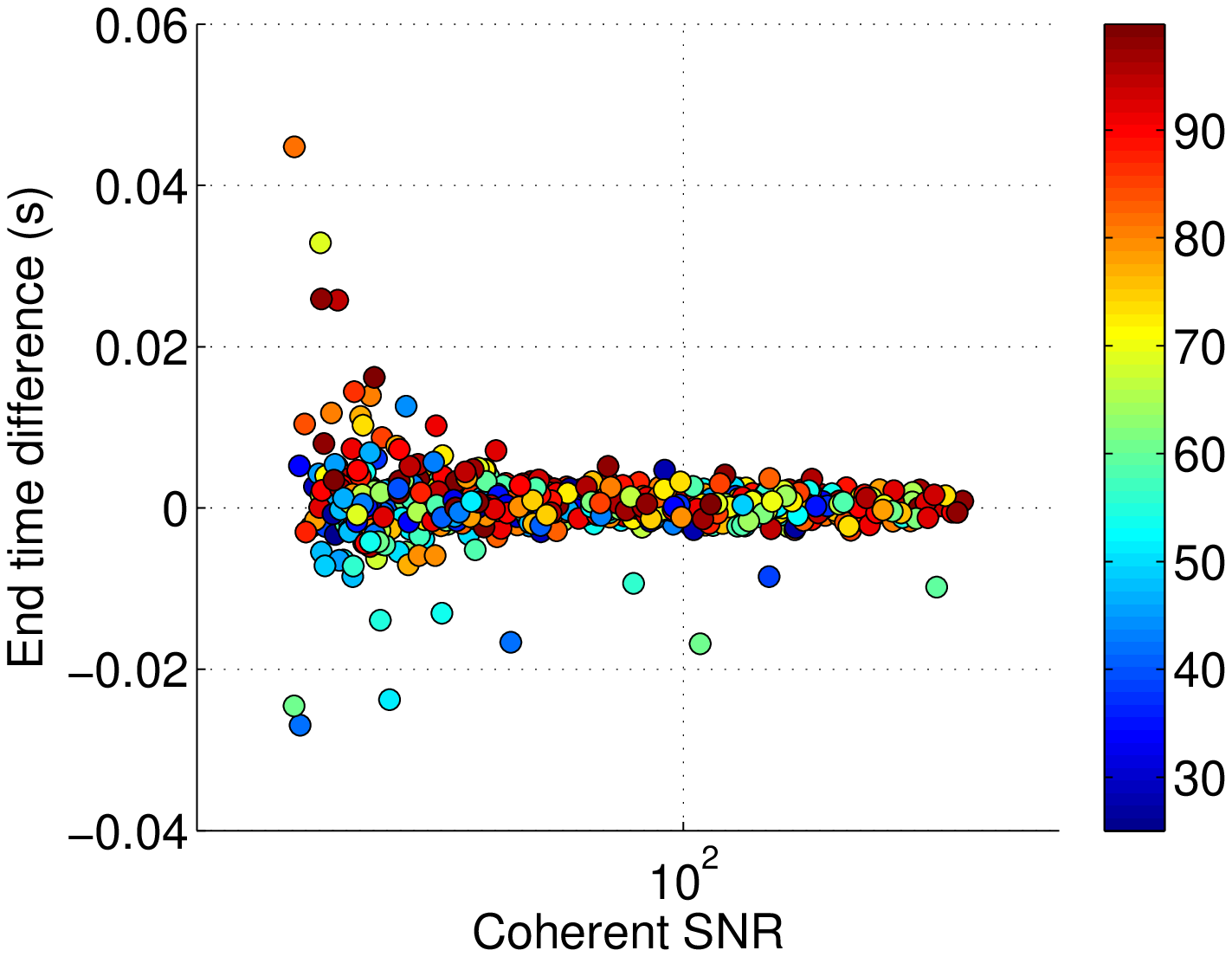}
\includegraphics[width=8.5cm]{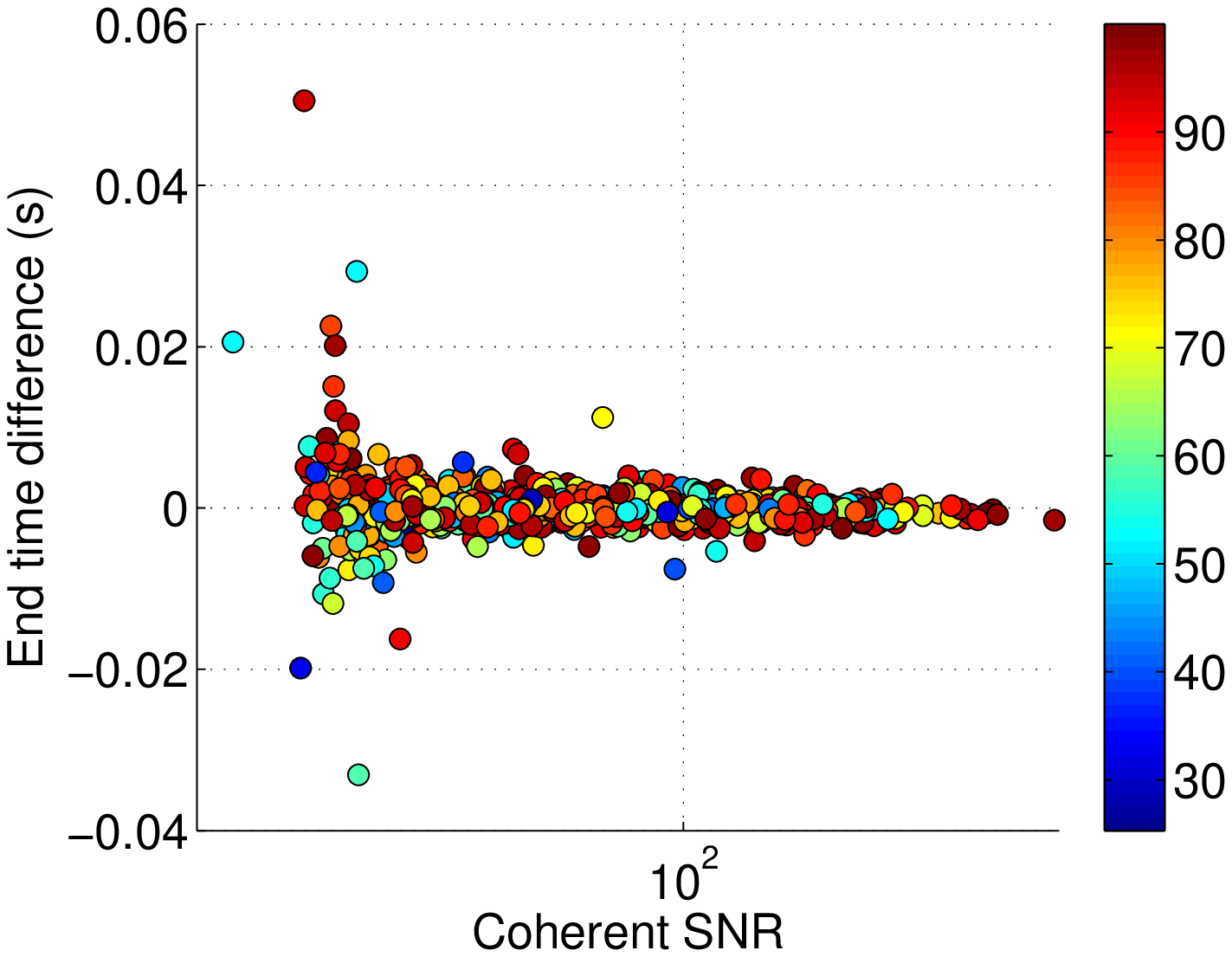}
\includegraphics[width=8.5cm]{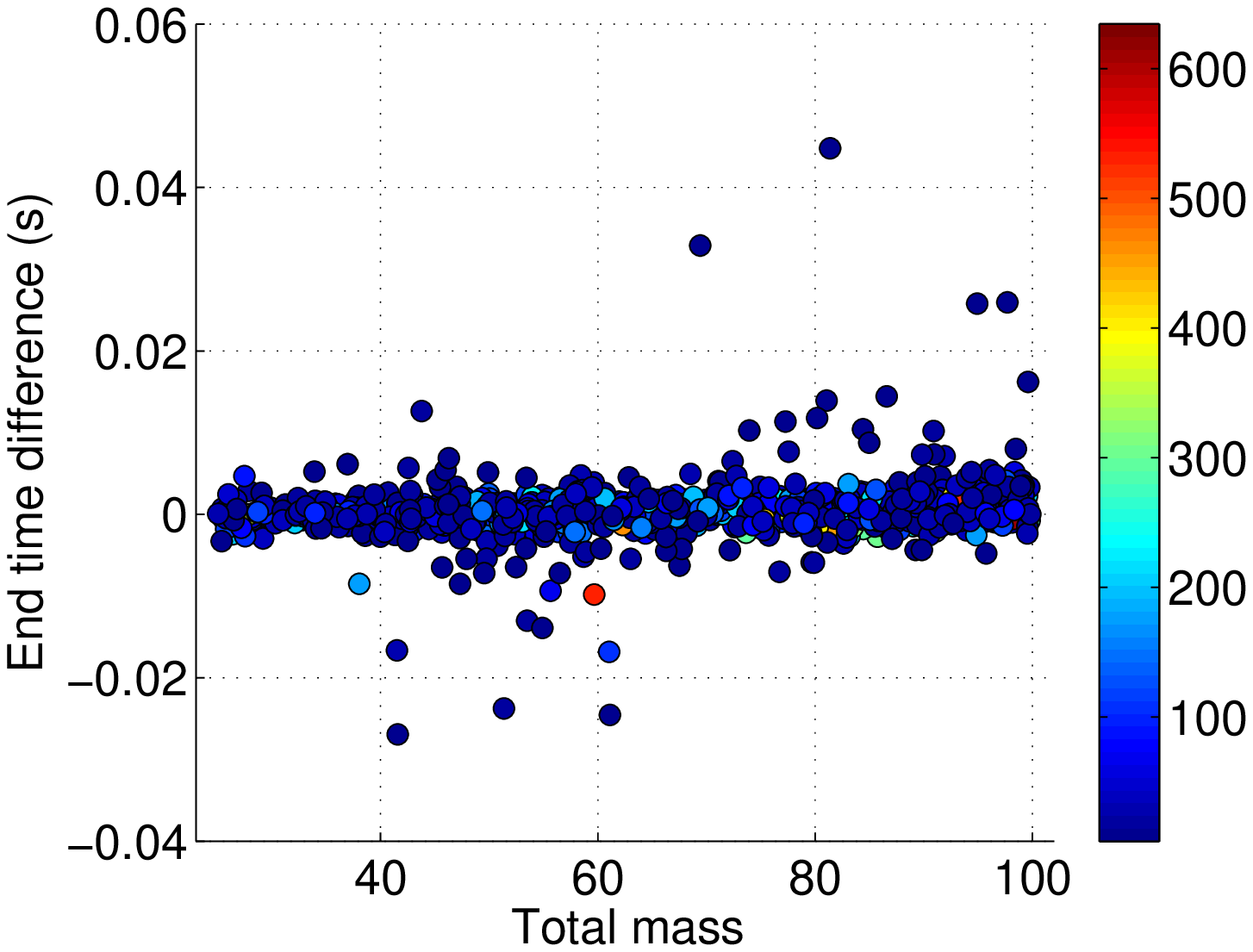}
\includegraphics[width=8.5cm]{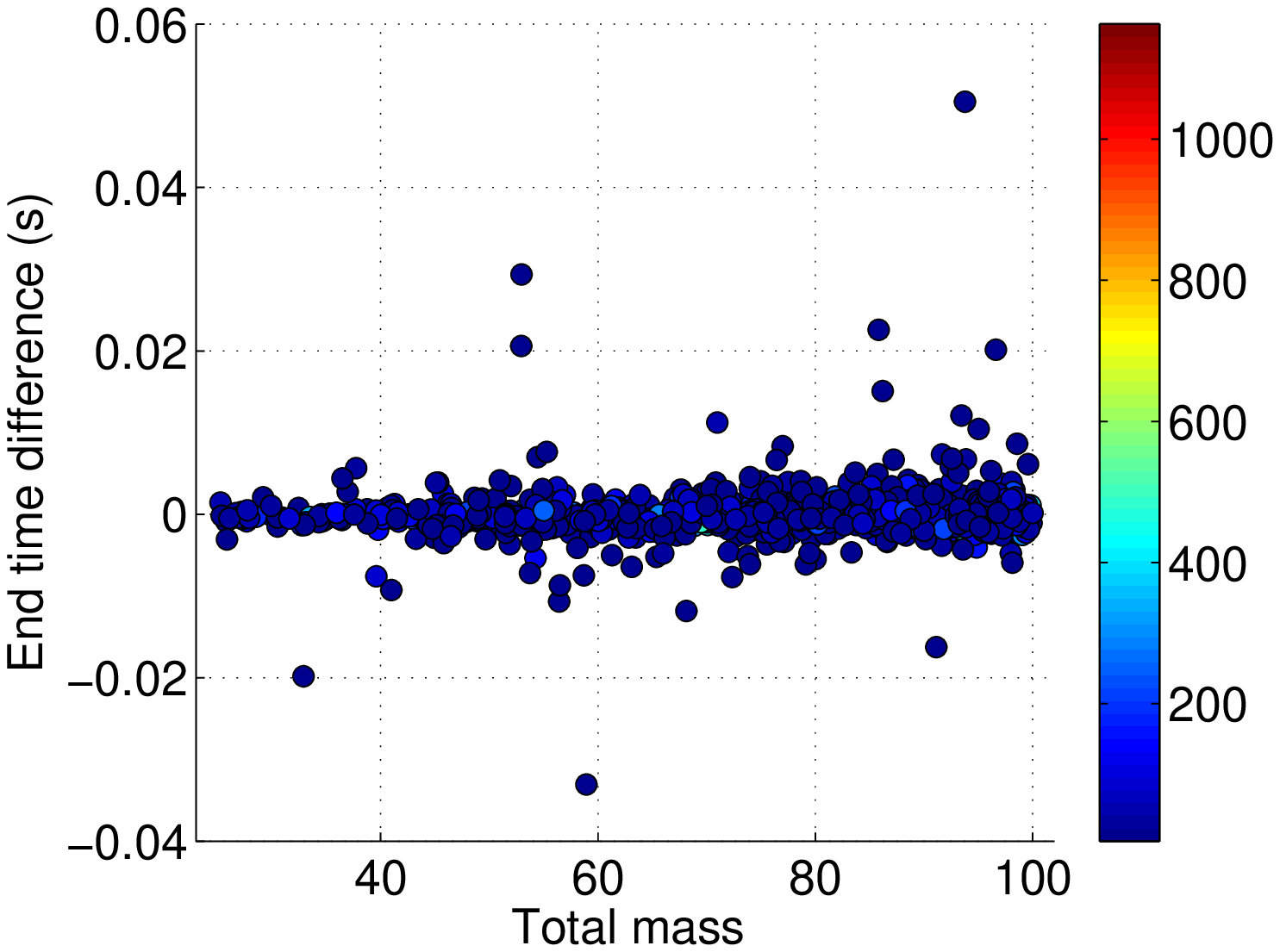}
\caption{
Difference between the measured and injected end time of detected simulated signals (in seconds). {\bf Top left:} The end-time difference is plotted as a function of the coherent SNR for pN-NR simulations. The colorbar represents the total mass of the binary system (in $M_\odot$). {\bf Top right:} The same quantity is plotted for EOBNR simulations. For pN-NR simulated signals a few were recovered with relatively high negative end-time difference, i.e., $\geq 0.01$ sec, in the total-mass range $40 M_\odot$ to $62 M_\odot$, at both high and low coherent SNRs. {\bf Bottom panel:} The end-time difference is plotted as a function of the total mass of the system.
}
\label{fig:end_time_diff}
\end{figure*}

A useful construct for describing the error in the measurement of the value of a dimensional parameter is:
\be
\mbox{fractional error} = \frac{\mbox{measured value - injected value}}{\mbox{injected value}}\,,
\ee
which we will use especially for studying the error in the measured or {\em recovered} chirp-mass values.
Comparing the injected and recovered parameter values helps us in determining if any systematics can develop in searching for 
pN-NR signals with a specific waveform family.
Consider, e.g., the chirp-mass recovered by EOBNR templates in Fig. \ref{fig:mchirp_rec}. There, in most cases, the injected chirp mass is greater than the recovered one. This can happen owing to the following possibilities:
(a) For some cases of pN-NR waveforms EOBNR templates with a smaller mass have the same or similar number of cycles. If this is true, then this is information that the hybridization schemes of those pN-NR waveforms can use to correct for this bias. (b) Templates that are longer in time and have more cycles than these pN-NR injections perform better at finding them. Longer templates would accrue noise from the cycles that do not overlap with any part of the signal and would, therefore, lose SNR. However a mismatch between an injection and the template, in amplitude or phase, can cause the projection of the signal on the template manifold to select a longer template as the best fit.
Indeed, the chirp-mass recovery and end-time recovery plots reveal that pN-NR injections tend to be found often with relatively higher chirp-mass EOBNR templates. Since higher chirp-mass templates will have a smaller duration than lower chirp-mass ones, all else being the same, one would expect that this bias should be accompanied by a systematic error whereby the template (or recovered) end-time occurs earlier than the injected (pN-NR) end-time. This expectation is confirmed in Fig. \ref{fig:end_time_diff}.

The end-time difference of recovered simulated signals in Fig. \ref{fig:end_time_diff} is computed by subtracting the injected end time from the measured end time. A negative end-time difference implies that the pN-NR signal is found at a later time than its injected value. A majority of the signals were recovered within 10 msec of the injected end time. A few outliers with a greater positive (smaller negative) end-time difference arise from systems with a large (small) total mass and relatively low coherent SNR. 

\begin{figure*}[tb]
\centering
\includegraphics[width=8.5cm]{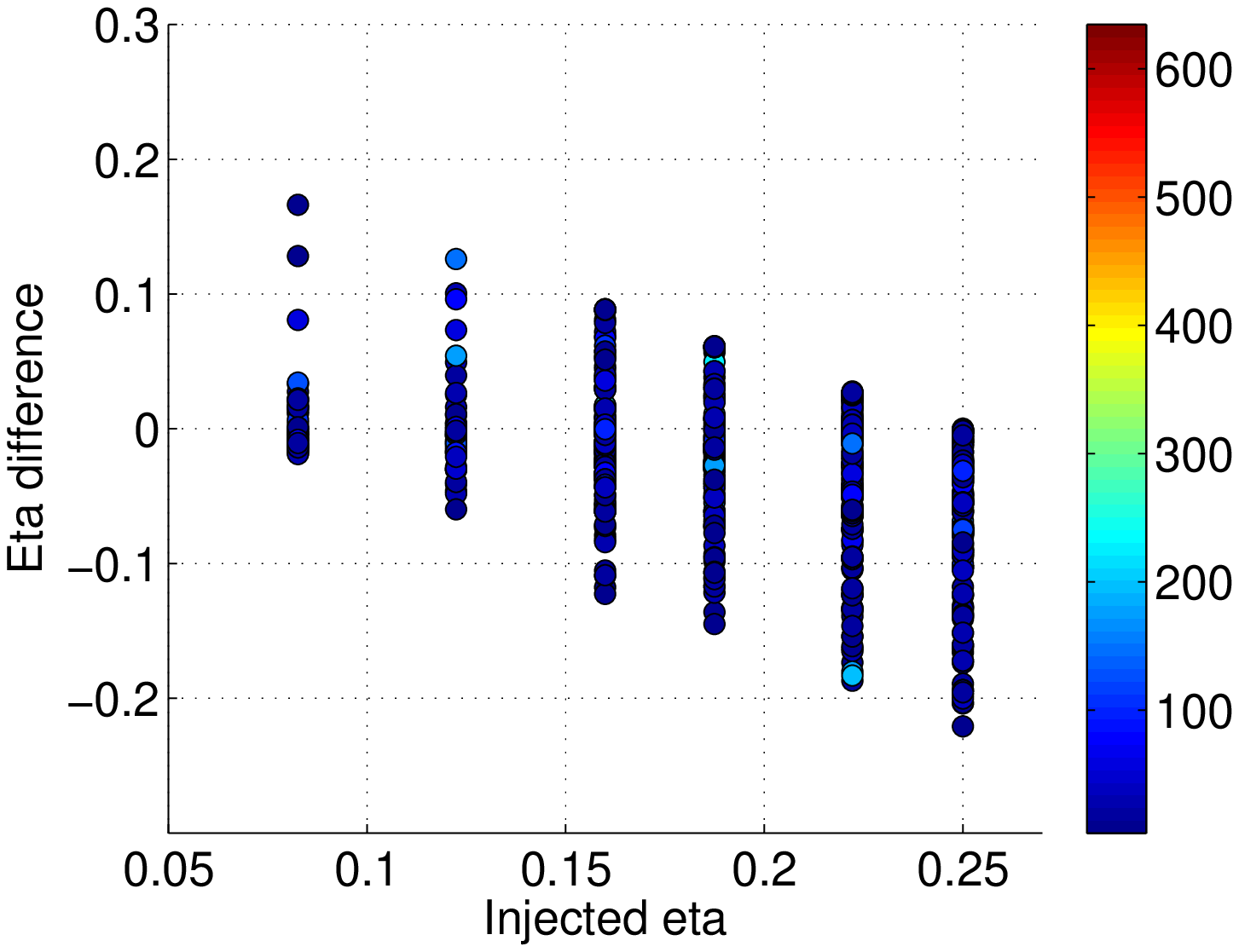}
\includegraphics[width=8.5cm]{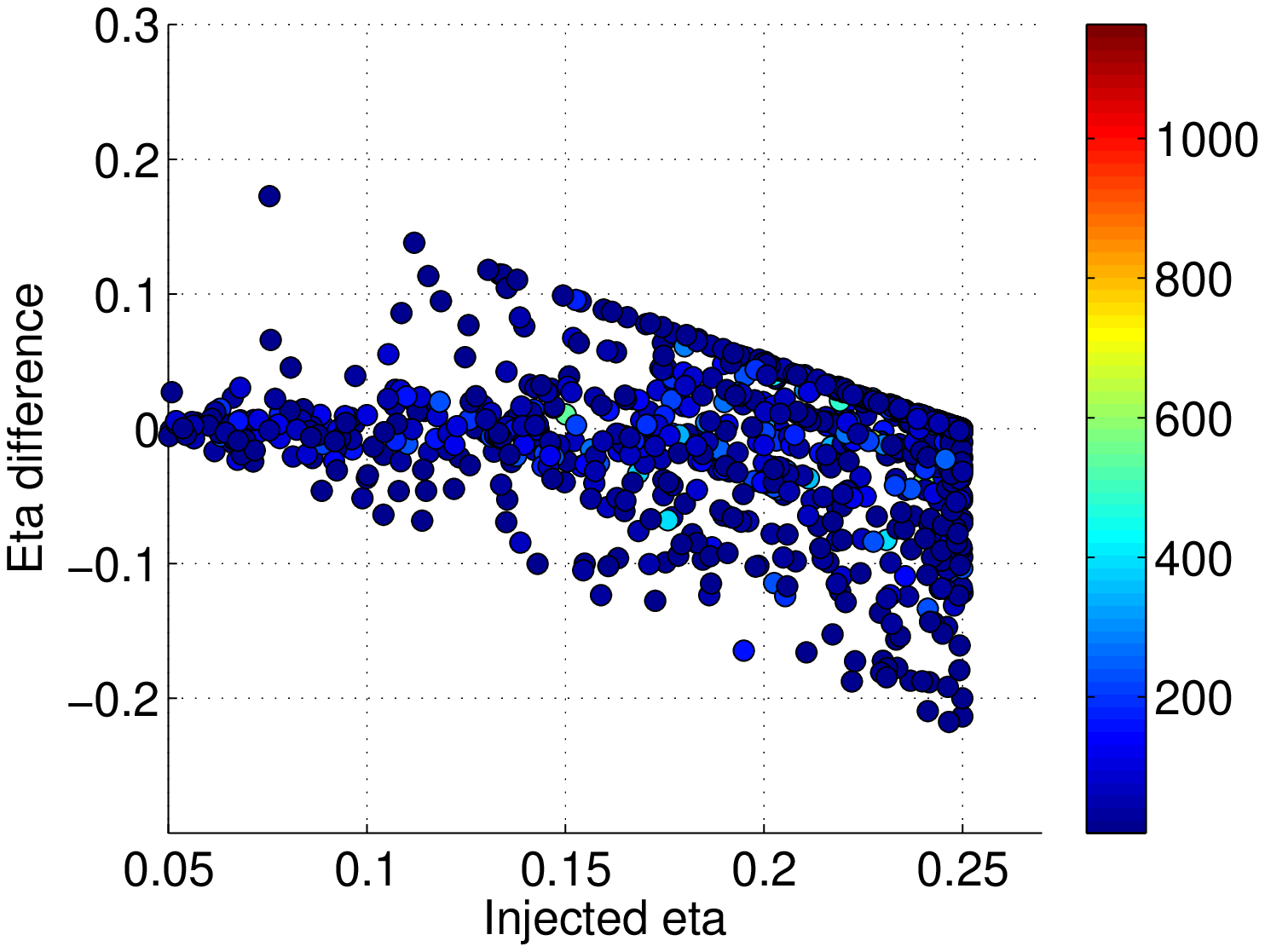}
\includegraphics[width=8.5cm]{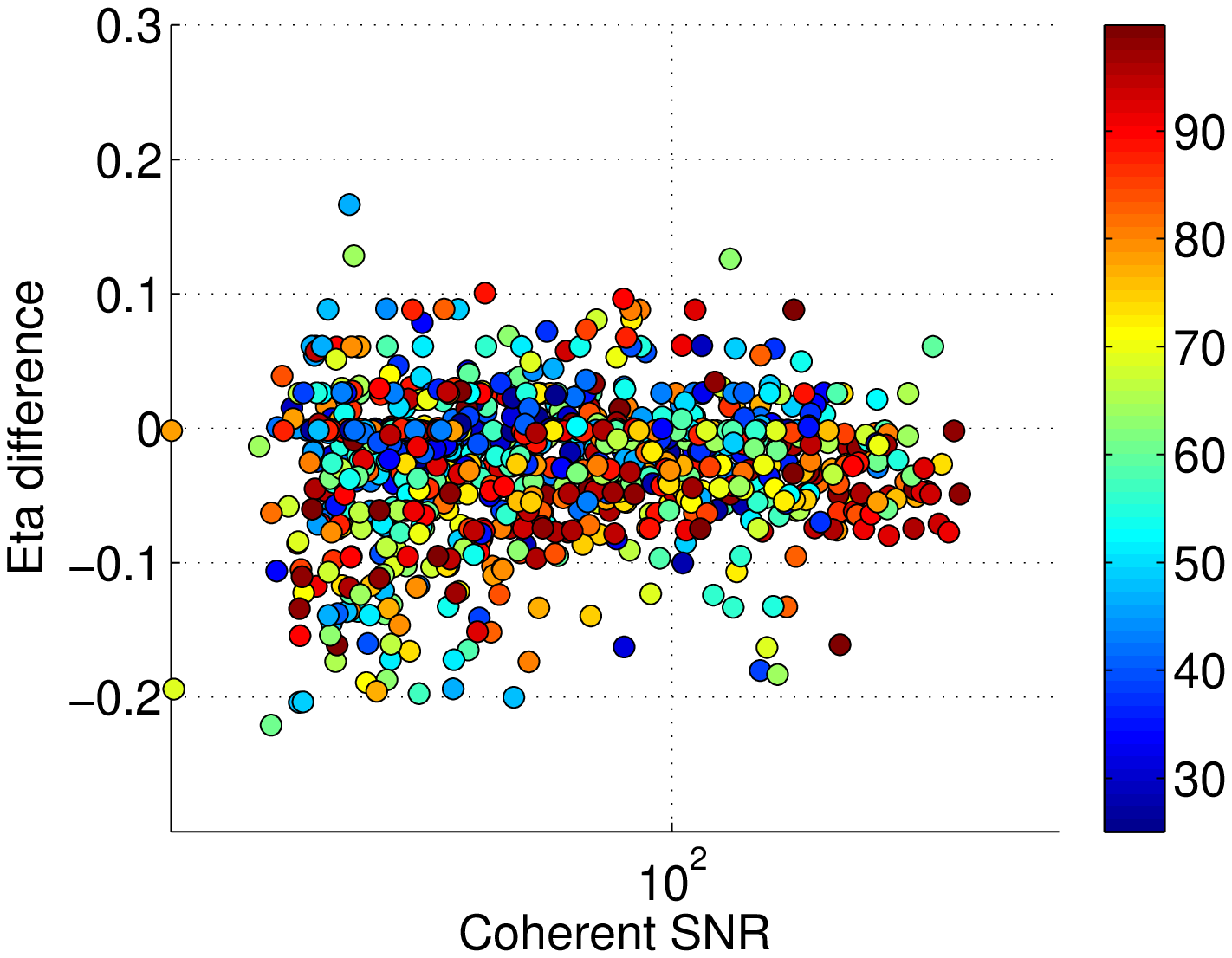}
\includegraphics[width=8.5cm]{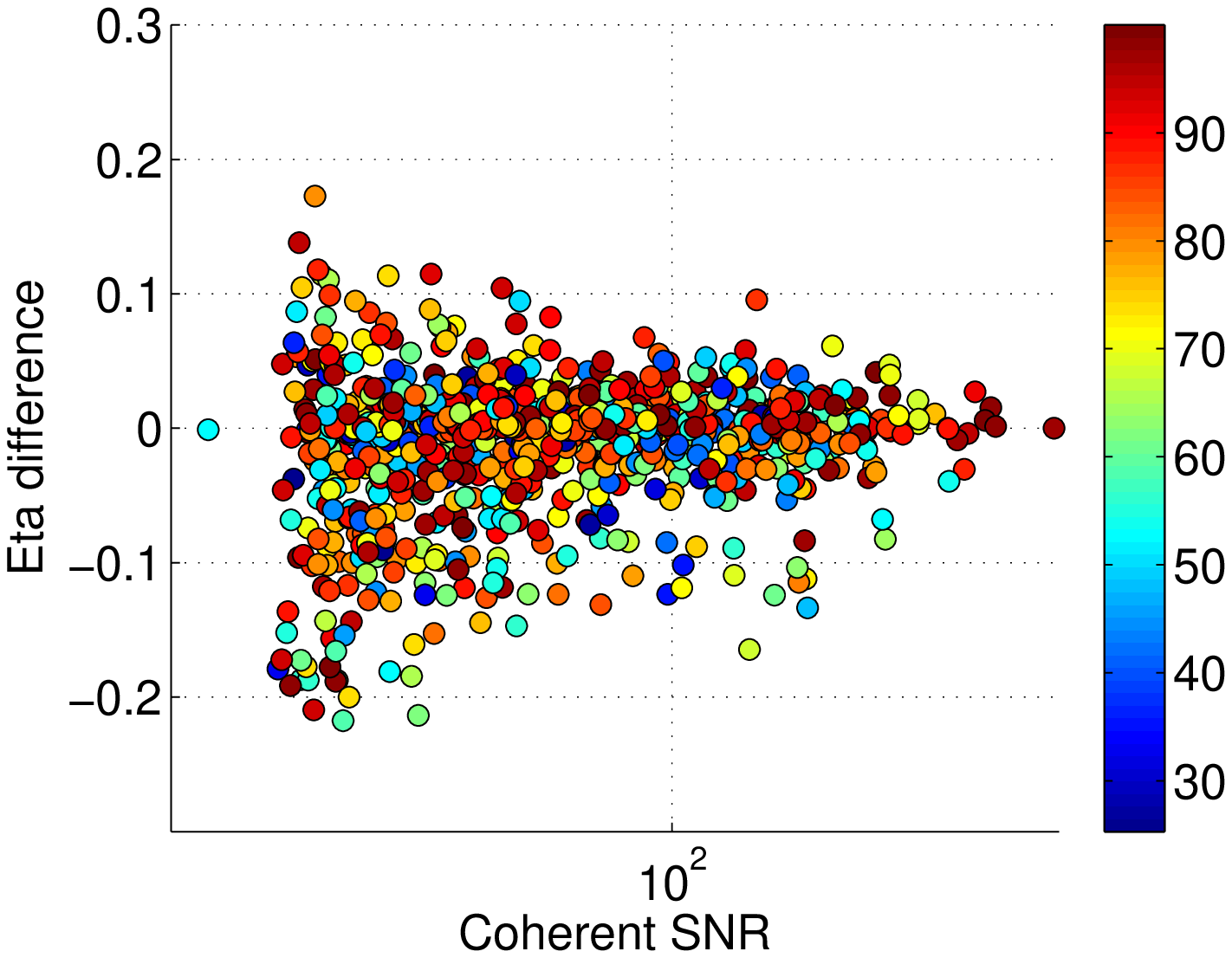}
\includegraphics[width=8.5cm]{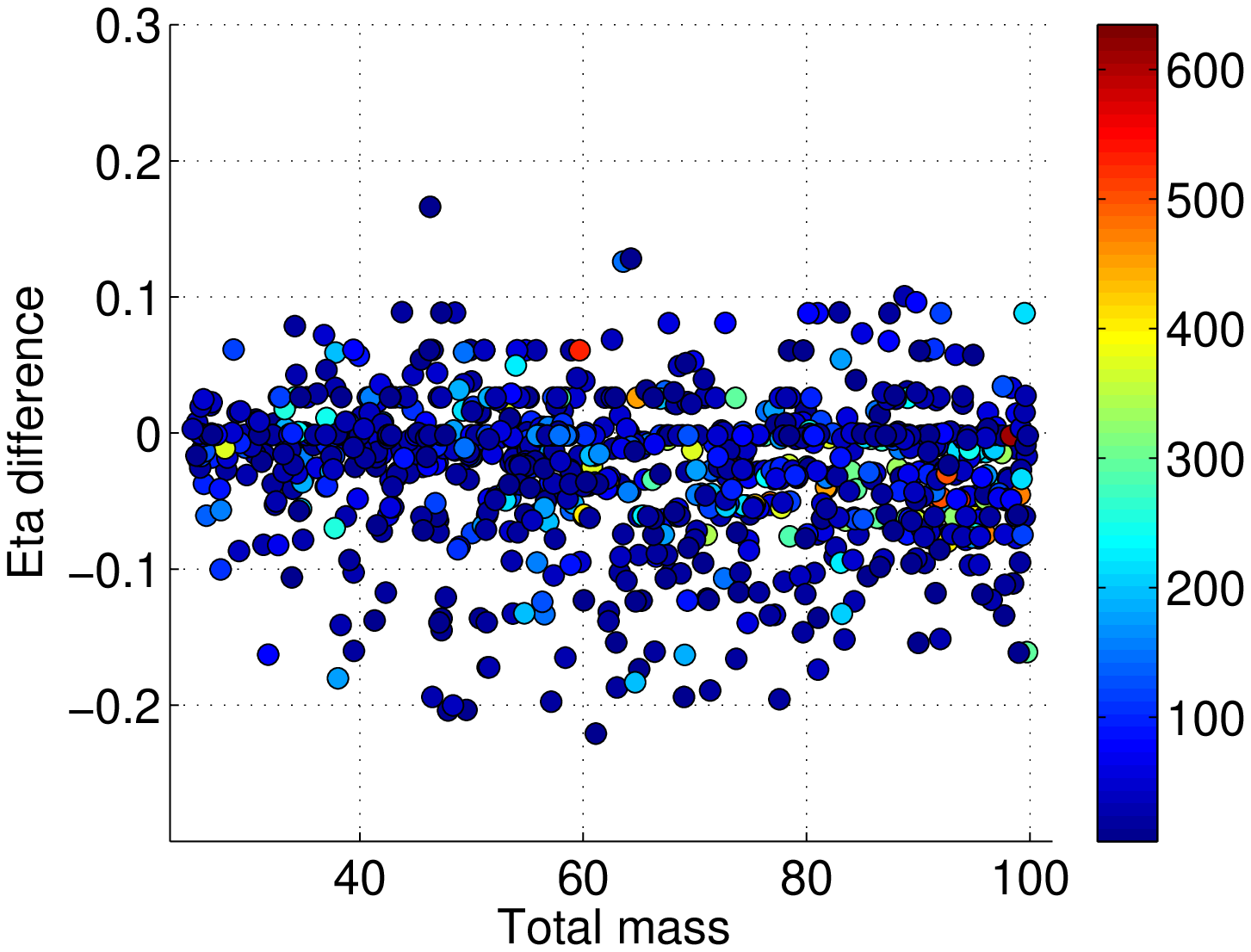}
\includegraphics[width=8.5cm]{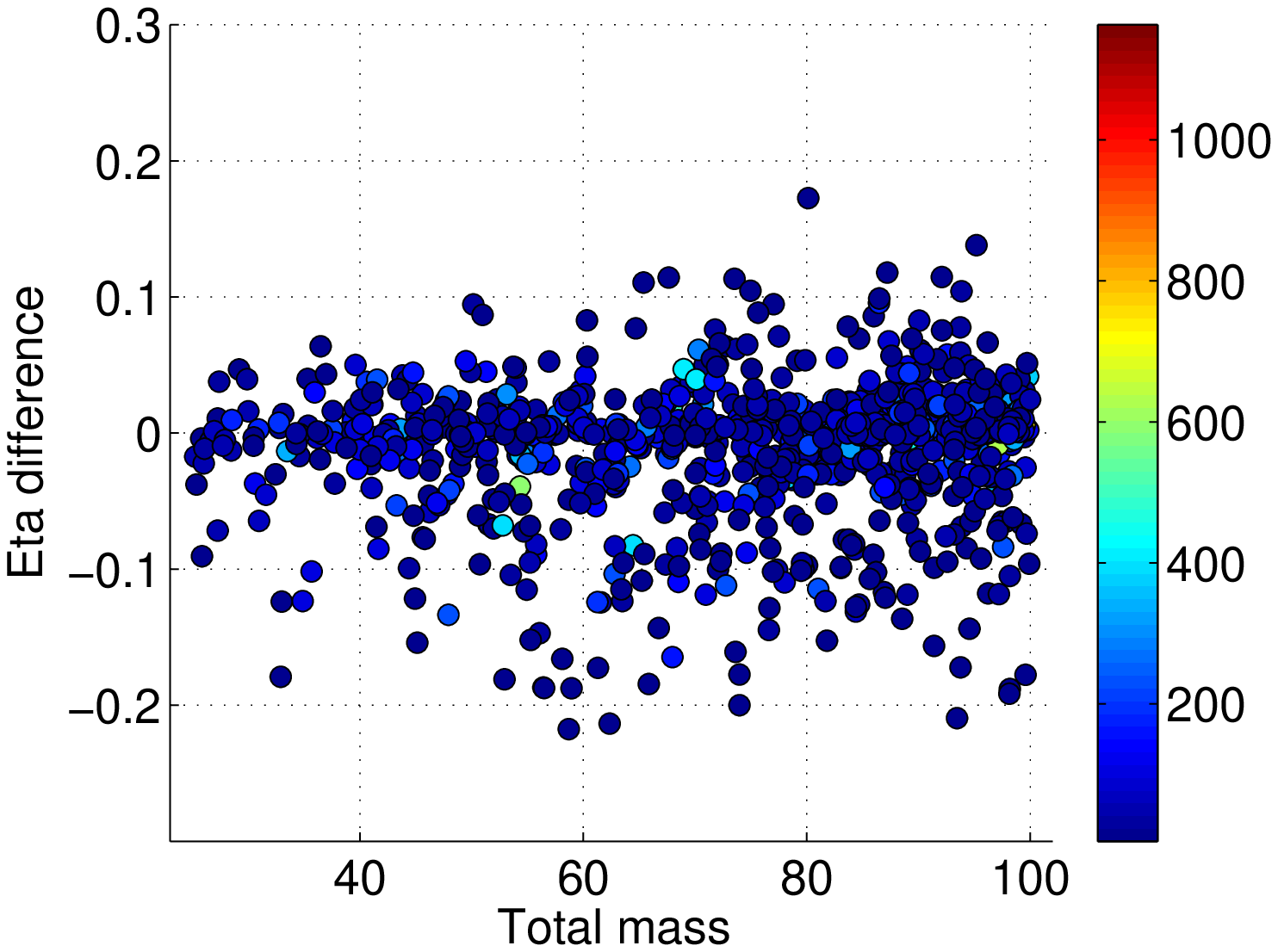}
\caption{
{\bf Left column:} Errors in the measurement of the symmetrized mass-ratio $\eta$ or ``Eta'' of pN-NR simulated signals. {\bf Right column:} Same as in the left column but for EOBNR simulated signals. {\bf Top panel:} This panel shows the $\eta$ difference as a function of injected $\eta$ values. Since pN-NR simulations are only available for a discrete set of $\eta$ values the left plot only has a few different values of injected $\eta$ compared to EOBNR counterpart, which is plotted in the right. Since $\eta \in (0,0.25]$, the maximum (minimum) $\eta$ difference is 0.25 (-0.25). {\bf Middle panel:} This panel shows that the $\eta$ difference of pN-NR hybrids is slightly worse compared to that of EOBNR simulations due to template mismatch. This is most apparent above at large coherent SNR values. {\bf Bottom panel:} Eta difference of two simulation categories is shown as a function of the total mass of the system. 
}
\label{fig:eta_rec}
\end{figure*}
\begin{figure*}[tb]
\centering
\includegraphics[width=8.5cm]{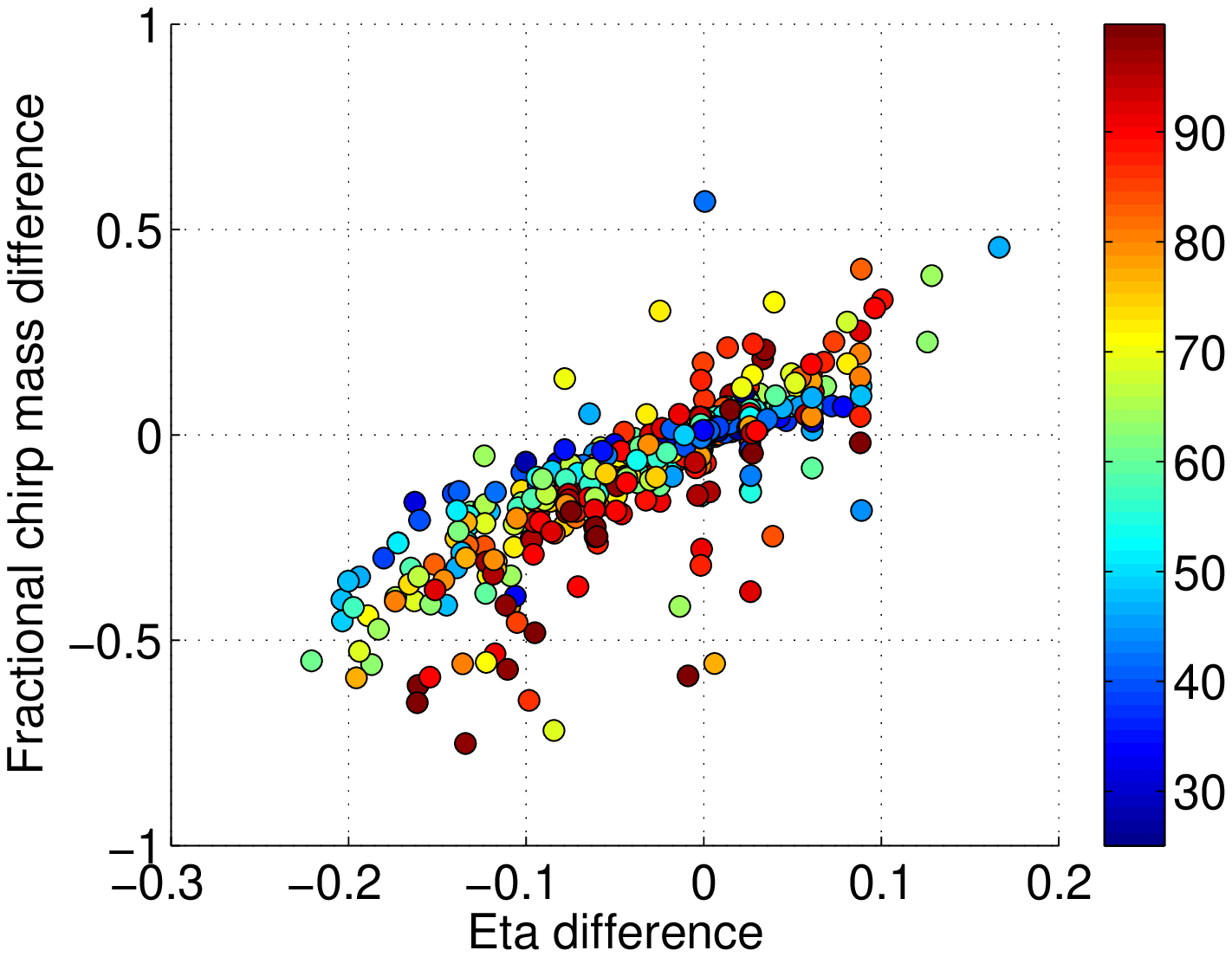}
\includegraphics[width=8.5cm]{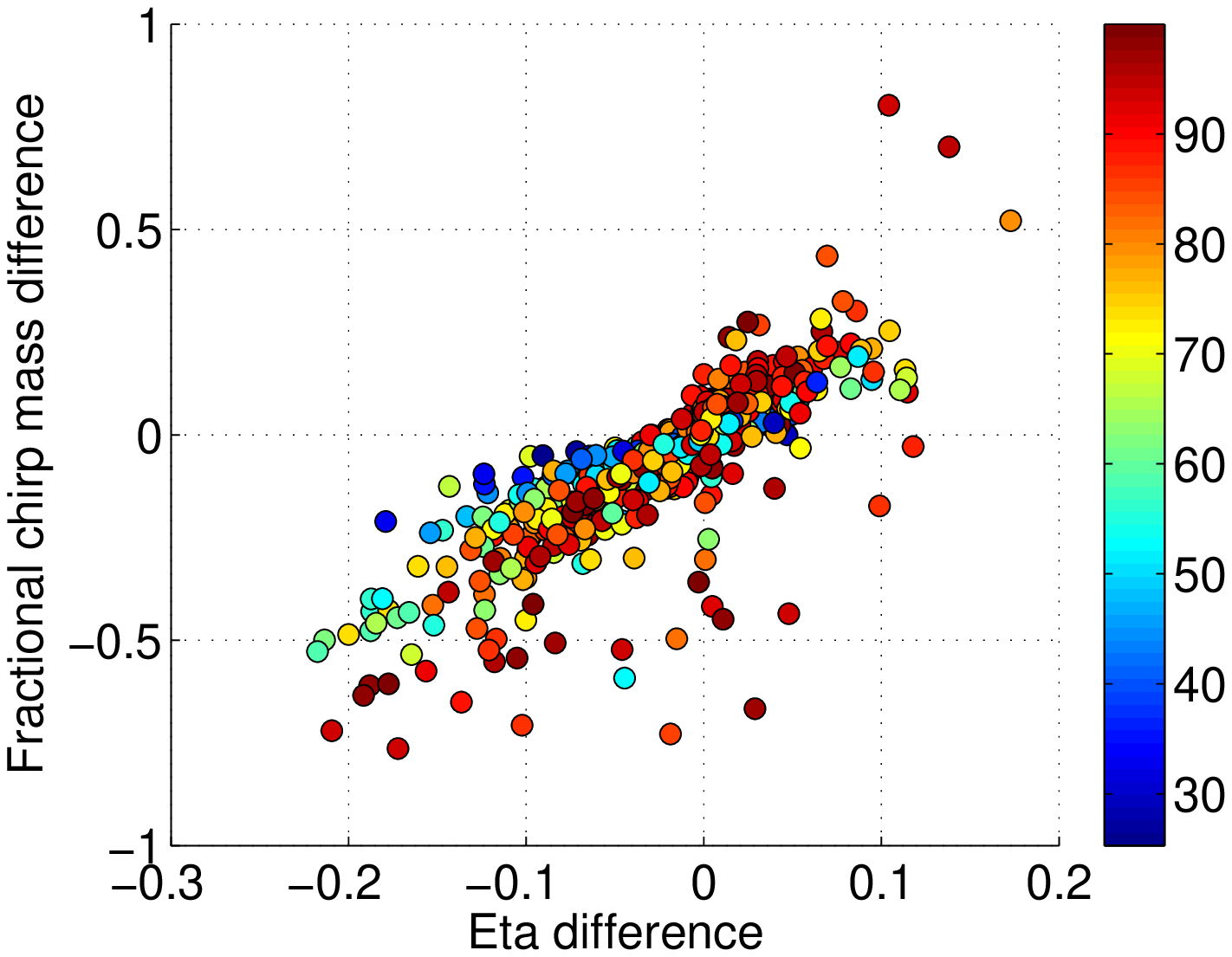}
\caption{
Covariance of chirp-mass error with $\eta$ error. The left figure shows this covariance for pN-NR injections and the right figure shows it for EOBNR injections. In both cases EOBNR templates were used to conduct the search. The colorbar shows the total mass of the BBH systems (in $M_\odot$).
}
\label{fig:chirpeta}
\end{figure*}

The results for the measurement of the symmetrized mass-ratio $\eta$ are presented in Fig. \ref{fig:eta_rec}. The end-time difference is computed by subtracting the injected value from the measured one. As shown in Ref. \cite{Ninja2Catalog:2012}, only six mass-ratio values were used to model and inject non-spinning pN-NR signals. 
For EOBNR signals, we produced a greater variety of mass ratios and, therefore, $\eta$, as can be seen in the top-right plot in that figure. The measured values of $\eta$ are always between 0.00 and 0.25 because those are the boundaries of the template bank that was employed, and correspond to mass ratios of infinity and 1, respectively. It is interesting to note that in a large fraction of cases small (large) $\eta$, or low (high) mass ratio, templates detect large (small) $\eta$ signals. 
This bias is, in fact, correlated with low (high) chirp-mass templates detecting high (low) chirp-mass signals, as confirmed by Fig. \ref{fig:chirpeta} for injections of pN-NR as well as EOBNR signals.
This covariance of errors occurs because templates with low (high) $\eta$ and low (high) $M_{\rm chirp}$ can have similar number of wave cycles and bandwidths as signals with high (low) $\eta$ and high (low) $M_{\rm chirp}$.\footnote{A similar correlation of errors in $\eta$ and the total-mass $M$ was found in Ref. \cite{Bose:2010zz} where the effect of using inspiral-only templates for searching inspiral-merger-ringdown signals on detection and parameter estimation was studied.}
In the same figure, also note how the degree of covariance changes somewhat as one goes from low total-mass systems (shown in blue) to high total-mass systems (shown in red). A similar behavior was observed in the Monte Carlo studies in Ref. \cite{Ajith:2009fz} of statistical errors in measuring BBH mass parameters with phenomenological waveforms \cite{Ajith:2007kx}.



\section{Discussion}
\label{sec:discussion}

Testing the preparedness of search pipelines to detect real signals in the ADE is one of the primary goals of NINJA. This study helped to quantify the sensitivity of one of the existing CBC search pipelines to numerical-relativity based BBH waveforms in early advanced detectors {\it albeit} in simulated Gaussian, stationary data. As we found here, the EOBNR waveforms, which employed only NASA-Goddard NR waveforms for calibration,
were able to detect pN-NR signals produced by a number of NR collaborations.
This is borne out by the fact that the ROC curve of EOBNR injections tracks that of the NR injections very closely.
Additionally, we studied some aspects of a coherent CBC search. Since a coherent search explores a larger dimensional parameter space than a coincident search it is more expensive, which makes the estimation of the background for the former type of search especially difficult. Therefore, we used the hierarchical coherent CBC search pipeline described in Ref. \cite{Bose:2011km} on the same NR based injections. Such an exercise is also useful to teach us about a subset of the potential issues we may face in a fully coherent search in the future. Here we demonstrated that the performance of that search conforms to expectations. Specifically, the characteristics of its background are consistent with theoretical predictions. Moreover, the coherent stage provides the null-stream statistic, which is a powerful multi-baseline signal consistency test, and can be employed to improve the performance of the search.
This test is especially useful for high-mass CBC searches where the chi-square test is less effective than low-mass ones, owing to fewer waveform cycles of high-mass signals in the detector band.

Finally, NINJA provided an important opportunity to test how well we might be able to measure the signal parameters. To address this question, we compared the maximum-likelihood estimates obtained by using the EOBNR family of templates. We focused our attention on only non-spinning injections. Here again the parameter accuracies of EOBNR injections are very similar to those of the pN-NR ones. The only small disagreement occurs for a few injections, mainly in the small total-mass region. For most of them, its cause was traced to the fact that compared to other injected waveforms these ones were {\it a priori} known to have a somewhat poorer match (by a few percent) \cite{Ninja2Catalog:2012} with the waveforms produced by the Spectral Einstein Code (SpEC). (See Ref. \cite{Boyle:2007ft} for details about SpEC.)
To study systematic errors stemming from signal-template mismatch in more detail, studies are ongoing with NR waveforms with spin and a variety of different mass ratios. Additionally, to study the impact of real data, which can be non-Gaussian and non-stationary, we plan to study the NINJA-2 waveforms injected in a recolored data set, where real data from past science runs will be scaled to have early aLIGO or aLIGO ASDs \cite{Ninja2BIC:2013}. 

\acknowledgments 

We would like to thank Duncan Brown, Alessandra Buonanno, Kipp Cannon, Collin Capano, Tom Dent, Steve Fairhurst, Ian Harry, Prayush Kumar, Satya Mohapatra, Greg Mendell, Larne Perkowsky, and Fred Raab for helpful discussions. We also thank the members of all the numerical relativity groups listed in Ref. \cite{Ninja2Catalog:2012} that contributed their waveforms for NINJA-2.
This work is supported in part by NSF grants PHY-0855679 and PHY-1206108. 

\bibliographystyle{plainnat}
\bibliographystyle{unsrt}
\bibliography{References}

\end{document}